%% file: DESY-13-245.tex
\documentclass[zpreprint,zbstpl]{zeus_paper}
\usepackage[english]{babel}
\usepackage{subfig}
\usepackage[mathscr]{euscript}
\input ./zeus_def.tex
\input ./def.tex
\begin{document}
%------------------------------------------------------------------------------
%       Title sheet
%------------------------------------------------------------------------------

\include{DESY-13-245-tit}
%------------------------------------------------------------------------------
%    List of authors
%--------------------------------------------------------------------------------
\clearpage

\input{DESY-13-245-authors}

%------------------------------------------------------------------------------
%       Text
%------------------------------------------------------------------------------
\include{DESY-13-245-txt}

%------------------------------------------------------------------------------
%       Bibliography
%------------------------------------------------------------------------------
%\include{highx-paper-ref}
%------------------------------------------------------------------------------
%       Tables
%------------------------------------------------------------------------------

\include{DESY-13-245-tab}
%------------------------------------------------------------------------------
%       Figures
%------------------------------------------------------------------------------
\include{DESY-13-245-fig}

%
%       ... that's it
%
\end{document}

%% file: zeus_def.tex
\newcommand{\Zsttdesc}[1]{%
The STT consisted of 48 sectors of two different sizes. Each sector
contained 192 (small sector) or 264 (large sector) straws of diameter
7.5 mm arranged into 3 layers. The sectors were trapezoidal in shape
and each subtended an azimuthal angle of $60^{\circ}$ -- 6 sectors
formed a so-called superlayer. A particle passing through the complete
detector traversed 8 superlayers, which were rotated around the beam
direction at angles of $30^{\circ}$ or $15^{\circ}$ to each other. The STT
covered the polar-angle region $5^{\circ}<\theta<23^{\circ}$.
}
\chardef\usc=95
\chardef\til=126
\catcode`\@=11 % @ signs are now treated as letters
\DeclareRobustCommand\xdotspace{\futurelet\@let@token\@xdotspace}
\def\@xdotspace{%
  \ifx\@let@token.\else
  \ifx\@let@token\bgroup.\else
  \ifx\@let@token\egroup.\else
  \ifx\@let@token\/.\else
  \ifx\@let@token\ .\else
  \ifx\@let@token~.\else
  \ifx\@let@token!.\else
  \ifx\@let@token,.\else
  \ifx\@let@token:.\else
  \ifx\@let@token;.\else
  \ifx\@let@token?.\else
  \ifx\@let@token/.\else
  \ifx\@let@token'.\else
  \ifx\@let@token).\else
  \ifx\@let@token-.\else
  \ifx\@let@token\@xobeysp.\else
  \ifx\@let@token\space.\else
  \ifx\@let@token\@sptoken.\else
   .\space
   \fi\fi\fi\fi\fi\fi\fi\fi\fi\fi\fi\fi\fi\fi\fi\fi\fi\fi}
\catcode`\@=12 % @ signs are no longer letters

\newcommand{\stru}[2]{%
   \relax\ifmmode\hbox{\vrule height#1 depth#2 width0pt}%
   \else\vrule height#1 depth#2 width0pt\fi}
\newcommand{\Ronum}[1]{\uppercase\expandafter{\romannumeral#1}}
\newcommand{\ronum}[1]{\expandafter{\romannumeral#1}}
\DeclareRobustCommand{\LaTeXZ}{%
  \LaTeX\kern-.05em4\kern-.1em
  {\raisebox{-0.2ex}{$\scriptstyle\text{ZEUS}$}}\xspace}
\DeclareMathAlphabet{\mathbf}{OT1}{cmr}{bx}{sl}
\newcommand{\eVdist}{\kern-0.06667em}

\newcommand{\gev}{{\,\text{Ge}\eVdist\text{V\/}}}

\newcommand{\mm}{\,\text{mm}}

\newcommand{\Tesla}{\,\text{T}}

\newcommand{\slashfrac}[2]{%
  \raisebox{0.5ex}{\ensuremath #1}\kern-0.12em/\kern-0.08em
  \raisebox{-.8ex}{\ensuremath #2}}
\newcommand{\sqr}[3]{%
    {\vcenter{\hrule height.#3ex\hbox{\vrule width.#2ex height#1ex
     \kern#1ex\vrule width.#3ex}\hrule height.#2ex}}}

\catcode`\@=11 % @ signs are now treated as letters
\newcommand{\parenbar}{\mathpalette\p@renb@r}
\def\p@renb@r#1#2{\vbox{%
  \ifx#1\scriptscriptstyle \dimen@.7em\dimen@ii.2em\else
  \ifx#1\scriptstyle \dimen@.8em\dimen@ii.25em\else
  \dimen@1em\dimen@ii.4em\fi\fi \offinterlineskip
  \ialign{\hfill##\hfill\cr
    \vbox{\hrule width\dimen@ii}\cr
    \noalign{\vskip-.3ex}%
    \hbox to\dimen@{$\mathchar300\hfil\mathchar301$}\cr
    \noalign{\vskip-.3ex}%
    $#1#2$\cr}}}
\catcode`\@=12 % @ signs are no longer letters

\newcommand{\IP}{{\rm I$\kern-0.01667em$P}\xspace}

\mathchardef\qsm=63
\mathchardef\pls=43
\mathchardef\mns=512
\mathchardef\plm=518
\mathchardef\eql=61
\mathchardef\smallleft=300
\mathchardef\smallright=301
\mathchardef\les=316
\mathchardef\gre=318
\mathchardef\leq=532
\mathchardef\grq=533
\catcode`\@=11 % @ signs are now treated as letters
\newcounter{pict@width}
\newcounter{pict@height}
\newlength{\pict@scale}
\newcommand{\psfigadd}[4]{%
\setcounter{pict@width}{1*\ratio{#2+\pict@scale/2}{\pict@scale}}
\setcounter{pict@height}{1*\ratio{#3+\pict@scale/2}{\pict@scale}}
\setlength{\unitlength}{\pict@scale}
\hbox to #2{\hspace{-\fill}\begin{picture}(\thepict@width,\thepict@height)
\put(0,0){\psfig{figure=#1,width=#2,height=#3,clip=}}
\SetScale{0.283466457}
\SetWidth{1.763889}
{#4}
\end{picture}}
}
\newcounter{pict@widthfst}
\newcounter{pict@widthscd}
\newcounter{pict@widthtot}
\newcommand{\psfigaddtwo}[7]{%
\setcounter{pict@widthfst}{1*\ratio{#2+\pict@scale/2}{\pict@scale}}
\setcounter{pict@widthscd}{1*\ratio{#2+#4+\pict@scale/2}{\pict@scale}}
\setcounter{pict@widthtot}{1*\ratio{#2+#4+#6+\pict@scale/2}{\pict@scale}}
\setcounter{pict@height}{1*\ratio{#3+\pict@scale/2}{\pict@scale}}
\setlength{\unitlength}{\pict@scale}
\hbox{\hspace{-\fill}\begin{picture}(\thepict@widthtot,\thepict@height)
\put(0,0){\psfig{figure=#1,width=#2,height=#3,clip=}}
\put(\thepict@widthscd,0){\psfig{figure=#5,width=#6,height=#3,clip=}}
\SetScale{0.283466457}
\SetWidth{1.763889}
{#7}
\end{picture}}
}
\newcommand{\psfigror}[4]{%
\setcounter{pict@width}{1*\ratio{#2+\pict@scale/2}{\pict@scale}}
\setcounter{pict@height}{1*\ratio{#3+\pict@scale/2}{\pict@scale}}
\setlength{\unitlength}{\pict@scale}
\hbox{\begin{picture}(\thepict@width,\thepict@height)
\put(0,\thepict@height){\psfig{figure=#1,width=#3,height=#2,clip=,angle=270}}
\SetScale{0.283466457}
\SetWidth{1.763889}
{#4}
\end{picture}}
}
\newcommand{\psfigrol}[4]{%
\setcounter{pict@width}{1*\ratio{#2+\pict@scale/2}{\pict@scale}}
\setcounter{pict@height}{1*\ratio{#3+\pict@scale/2}{\pict@scale}}
\setlength{\unitlength}{\pict@scale}
\hbox{\begin{picture}(\thepict@width,\thepict@height)
\put(0,0){\psfig{figure=#1,width=#3,height=#2,clip=,angle=90}}
\SetScale{0.283466457}
\SetWidth{1.763889}
{#4}
\end{picture}}
}
\catcode`\@=12 % @ signs are no longer letters
\newlength\listtextwidth

\catcode`\@=11 % @ signs are now treated as letters
\newlength{\@tabfninsert}
\newlength{\@tabfnwidth}
\newcommand{\tabfootnote}[2]{%
  \setlength{\@tabfninsert}{0.8em}
  \setlength{\@tabfnwidth}{\textwidth}
  \addtolength{\@tabfnwidth}{-\@tabfninsert}
  \addtolength{\@tabfnwidth}{-0.4em}
  \noindent\makebox[\@tabfninsert][r]{\footnotesize$^{#1}$\hfil}\hfill%
  \parbox[t]{\@tabfnwidth}{\footnotesize #2\hfill}}
\catcode`\@=12 % @ signs are no longer letters

%% file: def.tex
\newcommand\units{\,\mathrm}

\newcommand{\ddif}[3]{\frac{d^{2}#1}{d#2 d#3}}
\newcommand{\degree}{\ensuremath{^\circ}}

%% file: DESY-13-245-tit.tex
\title{
Measurement of neutral current \\
$\mathbf{e^\pm p}$ cross sections at high Bjorken $\mathbf{x}$ \\
with the ZEUS detector
}

\author{ZEUS Collaboration }
%\draftversion{3.5}
\prepnum{DESY--13--245}
\date{December 2013}

\abstract{
The neutral current $e^\pm p$ cross section has been measured up to 
values of Bjorken\,$x\cong 1$ with the ZEUS detector at HERA using an integrated luminosity of 187\,${\rm pb}^{-1}$ of $e^-p$ and 142\,${\rm pb}^{-1}$ of $e^+p$ collisions at $\surd s = 318 \gev$. Differential cross sections in $x$ and $Q^2$, the exchanged boson virtuality,  are presented for $Q^2 \geq 725 \gev^2$. An improved reconstruction method and greatly increased amount of data allows a finer binning in the high-$x$ region of the neutral current cross section and leads to a  measurement with much  improved precision compared to a similar earlier analysis.
%
%The analysis is focused on a fine-grained measurement of the high-$x$ region of the neutral current cross section, and has a much improved precision compared to a similar earlier analysis.
% with respect to previous measurements
%, which only used 16.7 $pb^{-1}$ of $e^-p$ and 65.1 $pb^{-1}$ of $e^+p$ collisions, is achieved, owing to larger data samples and improved kinematic reconstruction methods. 
The measurements are compared to Standard Model expectations based on a variety of recent parton distribution functions. 
}

\makezeustitle

%% file: DESY-13-245-authors.tex
%===================================================================
%
%  MEMBER NAME  AUTH181 (ZEUS)     M  TEX
%
%  JH.: transformed to a format, which is suited as input for
%       CONVERT, which automatically creates author-indices
%
%  Don't remove lines starting with a percent sign %,
%  CONVERT may need them urgently !
%  
%=====================================================================

%\documentstyle[12pt,twoside]{report}  

%\topmargin-1.cm
%\evensidemargin-0.3cm
%\oddsidemargin-0.3cm
%\textwidth 16.cm
%\textheight 680pt
%\parindent0.cm
%\parskip0.3cm plus0.05cm minus0.05cm
\def\3{\ss}
\pagenumbering{Roman}
                                    % this "%"s are for cosmetics only
%\begin{document}
                                                   %
\begin{center}
{                      \Large  The ZEUS Collaboration              }
\end{center}

{\small

%  members:

        {\raggedright
H.~Abramowicz$^{27, u}$, 
I.~Abt$^{21}$, 
L.~Adamczyk$^{8}$, 
M.~Adamus$^{34}$, 
R.~Aggarwal$^{4, a}$, 
S.~Antonelli$^{2}$, 
O.~Arslan$^{3}$, 
V.~Aushev$^{16, 17, o}$, 
Y.~Aushev,$^{17, o, p}$, 
O.~Bachynska$^{10}$
% ...
        }

\newpage

%       institutes:

\makebox[3em]{$^{1}$}
\begin{minipage}[t]{14cm}
{\it INFN Bologna, Bologna, Italy}~$^{A}$

\end{minipage}\\
\makebox[3em]{$^{2}$}
\begin{minipage}[t]{14cm}
{\it University and INFN Bologna, Bologna, Italy}~$^{A}$

\end{minipage}\\
\makebox[3em]{$^{3}$}
\begin{minipage}[t]{14cm}
{\it Physikalisches Institut der Universit\"at Bonn,
Bonn, Germany}~$^{B}$

\end{minipage}\\
\makebox[3em]{$^{4}$}
\begin{minipage}[t]{14cm}
{\it Panjab University, Department of Physics, Chandigarh, India}

\end{minipage}\\
\makebox[3em]{$^{5}$}
\begin{minipage}[t]{14cm}
{\it Calabria University,
Physics Department and INFN, Cosenza, Italy}~$^{A}$

\end{minipage}\\
\makebox[3em]{$^{6}$}
\begin{minipage}[t]{14cm}
{\it National Centre for Particle Physics, Universiti Malaya, 50603 Kuala Lumpur, Malaysia}~$^{C}$

\end{minipage}\\
\makebox[3em]{$^{7}$}
\begin{minipage}[t]{14cm}
{\it The Henryk Niewodniczanski Institute of Nuclear Physics, Polish Academy of \\
Sciences, Krakow, Poland}~$^{D}$

\end{minipage}\\
\makebox[3em]{$^{8}$}
\begin{minipage}[t]{14cm}
{\it AGH-University of Science and Technology, Faculty of Physics and Applied Computer
Science, Krakow, Poland}~$^{D}$

\end{minipage}\\
\makebox[3em]{$^{9}$}
\begin{minipage}[t]{14cm}
{\it Department of Physics, Jagellonian University, Cracow, Poland}

\end{minipage}\\
\makebox[3em]{$^{10}$}
\begin{minipage}[t]{14cm}
{\it Deutsches Elektronen-Synchrotron DESY, Hamburg, Germany}

\end{minipage}\\
\makebox[3em]{$^{11}$}
\begin{minipage}[t]{14cm}
{\it Deutsches Elektronen-Synchrotron DESY, Zeuthen, Germany}

\end{minipage}\\
\makebox[3em]{$^{12}$}
\begin{minipage}[t]{14cm}
{\it School of Physics and Astronomy, University of Glasgow,
Glasgow, United Kingdom}~$^{E}$

\end{minipage}\\
\makebox[3em]{$^{13}$}
\begin{minipage}[t]{14cm}
{\it Hamburg University, Institute of Experimental Physics, Hamburg,
Germany}~$^{F}$

\end{minipage}\\
\makebox[3em]{$^{14}$}
\begin{minipage}[t]{14cm}
{\it Institute of Particle and Nuclear Studies, KEK,
Tsukuba, Japan}~$^{G}$

\end{minipage}\\
\makebox[3em]{$^{15}$}
\begin{minipage}[t]{14cm}
{\it Institute of Physics and Technology of Ministry of Education and
Science of Kazakhstan, Almaty, Kazakhstan}

\end{minipage}\\
\makebox[3em]{$^{16}$}
\begin{minipage}[t]{14cm}
{\it Institute for Nuclear Research, National Academy of Sciences, Kyiv, Ukraine}

\end{minipage}\\
\makebox[3em]{$^{17}$}
\begin{minipage}[t]{14cm}
{\it Department of Nuclear Physics, National Taras Shevchenko University of Kyiv, Kyiv, Ukraine}

\end{minipage}\\
\makebox[3em]{$^{18}$}
\begin{minipage}[t]{14cm}
{\it Department of Physics, McGill University,
Montr\'eal, Qu\'ebec, Canada H3A 2T8}~$^{H}$

\end{minipage}\\
\makebox[3em]{$^{19}$}
\begin{minipage}[t]{14cm}
{\it Meiji Gakuin University, Faculty of General Education,
Yokohama, Japan}~$^{G}$

\end{minipage}\\
\makebox[3em]{$^{20}$}
\begin{minipage}[t]{14cm}
{\it Lomonosov Moscow State University, Skobeltsyn Institute of Nuclear Physics,
Moscow, Russia}~$^{I}$

\end{minipage}\\
\makebox[3em]{$^{21}$}
\begin{minipage}[t]{14cm}
{\it Max-Planck-Institut f\"ur Physik, M\"unchen, Germany}

\end{minipage}\\
\makebox[3em]{$^{22}$}
\begin{minipage}[t]{14cm}
{\it Department of Physics, University of Oxford,
Oxford, United Kingdom}~$^{E}$

\end{minipage}\\
\makebox[3em]{$^{23}$}
\begin{minipage}[t]{14cm}
{\it INFN Padova, Padova, Italy}~$^{A}$

\end{minipage}\\
\makebox[3em]{$^{24}$}
\begin{minipage}[t]{14cm}
{\it Dipartimento di Fisica dell' Universit\`a and INFN,
Padova, Italy}~$^{A}$

\end{minipage}\\
\makebox[3em]{$^{25}$}
\begin{minipage}[t]{14cm}
{\it Polytechnic University, Tokyo, Japan}~$^{G}$

\end{minipage}\\
\makebox[3em]{$^{26}$}
\begin{minipage}[t]{14cm}
{\it Dipartimento di Fisica, Universit\`a `La Sapienza' and INFN,
Rome, Italy}~$^{A}$

\end{minipage}\\
\makebox[3em]{$^{27}$}
\begin{minipage}[t]{14cm}
{\it Raymond and Beverly Sackler Faculty of Exact Sciences, School of Physics, \\
Tel Aviv University, Tel Aviv, Israel}~$^{J}$

\end{minipage}\\
\makebox[3em]{$^{28}$}
\begin{minipage}[t]{14cm}
{\it Department of Physics, Tokyo Institute of Technology,
Tokyo, Japan}~$^{G}$

\end{minipage}\\
\makebox[3em]{$^{29}$}
\begin{minipage}[t]{14cm}
{\it Universit\`a di Torino and INFN, Torino, Italy}~$^{A}$

\end{minipage}\\
\makebox[3em]{$^{30}$}
\begin{minipage}[t]{14cm}
{\it Universit\`a del Piemonte Orientale, Novara, and INFN, Torino,
Italy}~$^{A}$

\end{minipage}\\
\makebox[3em]{$^{31}$}
\begin{minipage}[t]{14cm}
{\it Department of Physics, University of Toronto, Toronto, Ontario,
Canada M5S 1A7}~$^{H}$

\end{minipage}\\
\makebox[3em]{$^{32}$}
\begin{minipage}[t]{14cm}
{\it Physics and Astronomy Department, University College London,
London, United Kingdom}~$^{E}$

\end{minipage}\\
\makebox[3em]{$^{33}$}
\begin{minipage}[t]{14cm}
{\it Faculty of Physics, University of Warsaw, Warsaw, Poland}

\end{minipage}\\
\makebox[3em]{$^{34}$}
\begin{minipage}[t]{14cm}
{\it National Centre for Nuclear Research, Warsaw, Poland}

\end{minipage}\\
\makebox[3em]{$^{35}$}
\begin{minipage}[t]{14cm}
{\it Department of Particle Physics and Astrophysics, Weizmann
Institute, Rehovot, Israel}

\end{minipage}\\
\makebox[3em]{$^{36}$}
\begin{minipage}[t]{14cm}
{\it Department of Physics, York University, Ontario, Canada M3J 1P3}~$^{H}$

\end{minipage}\\
\vspace{30em} \pagebreak[4]

%  references concerning institutes;

\makebox[3ex]{$^{ A}$}
\begin{minipage}[t]{14cm}
 supported by the Italian National Institute for Nuclear Physics (INFN) \
\end{minipage}\\
\makebox[3ex]{$^{ B}$}
\begin{minipage}[t]{14cm}
 supported by the German Federal Ministry for Education and Research (BMBF), under
 contract No. 05 H09PDF\
\end{minipage}\\
\makebox[3ex]{$^{ C}$}
\begin{minipage}[t]{14cm}
 supported by HIR grant UM.C/625/1/HIR/149 and UMRG grants RU006-2013, RP012A-13AFR and RP012B-13AFR from
 Universiti Malaya, and ERGS grant ER004-2012A from the Ministry of Education, Malaysia\
\end{minipage}\\
\makebox[3ex]{$^{ D}$}
\begin{minipage}[t]{14cm}
 supported by the National Science Centre under contract No. DEC-2012/06/M/ST2/00428\
\end{minipage}\\
\makebox[3ex]{$^{ E}$}
\begin{minipage}[t]{14cm}
 supported by the Science and Technology Facilities Council, UK\
\end{minipage}\\
\makebox[3ex]{$^{ F}$}
\begin{minipage}[t]{14cm}
 supported by the German Federal Ministry for Education and Research (BMBF), under
 contract No. 05h09GUF, and the SFB 676 of the Deutsche Forschungsgemeinschaft (DFG) \
\end{minipage}\\
\makebox[3ex]{$^{ G}$}
\begin{minipage}[t]{14cm}
 supported by the Japanese Ministry of Education, Culture, Sports, Science and Technology
 (MEXT) and its grants for Scientific Research\
\end{minipage}\\
\makebox[3ex]{$^{ H}$}
\begin{minipage}[t]{14cm}
 supported by the Natural Sciences and Engineering Research Council of Canada (NSERC) \
\end{minipage}\\
\makebox[3ex]{$^{ I}$}
\begin{minipage}[t]{14cm}
 supported by RF Presidential grant N 3920.2012.2 for the Leading Scientific Schools and by
 the Russian Ministry of Education and Science through its grant for Scientific Research on
 High Energy Physics\
\end{minipage}\\
\makebox[3ex]{$^{ J}$}
\begin{minipage}[t]{14cm}
 supported by the Israel Science Foundation\
\end{minipage}\\
\vspace{30em} \pagebreak[4]

%  references concerning mebers;

\makebox[3ex]{$^{ a}$}
\begin{minipage}[t]{14cm}
also funded by Max Planck Institute for Physics, Munich, Germany\
\end{minipage}\\
\makebox[3ex]{$^{ b}$}
\begin{minipage}[t]{14cm}
also funded by Max Planck Institute for Physics, Munich, Germany, now at Sri Guru Granth Sahib World University, Fatehgarh Sahib\
\end{minipage}\\
\makebox[3ex]{$^{ c}$}
\begin{minipage}[t]{14cm}
also at Agensi Nuklear Malaysia, 43000 Kajang, Bangi, Malaysia\
\end{minipage}\\
\makebox[3ex]{$^{ d}$}
\begin{minipage}[t]{14cm}
also at Cracow University of Technology, Faculty of Physics, Mathematics and Applied Computer Science, Poland\
\end{minipage}\\
\makebox[3ex]{$^{ e}$}
\begin{minipage}[t]{14cm}
partially supported by the Polish National Science Centre projects DEC-2011/01/B/ST2/03643 and DEC-2011/03/B/ST2/00220\
\end{minipage}\\
\makebox[3ex]{$^{ f}$}
\begin{minipage}[t]{14cm}
now at Rockefeller University, New York, NY 10065, USA\
\end{minipage}\\
\makebox[3ex]{$^{ g}$}
\begin{minipage}[t]{14cm}
now at University of Liverpool, United Kingdom\
\end{minipage}\\
\makebox[3ex]{$^{ h}$}
\begin{minipage}[t]{14cm}
also affiliated with University College London, UK\
\end{minipage}\\
\makebox[3ex]{$^{ i}$}
\begin{minipage}[t]{14cm}
supported by the Alexander von Humboldt Foundation\
\end{minipage}\\
\makebox[3ex]{$^{ j}$}
\begin{minipage}[t]{14cm}
now at CERN, Geneva, Switzerland\
\end{minipage}\\
\makebox[3ex]{$^{ k}$}
\begin{minipage}[t]{14cm}
also at Institute of Theoretical and Experimental Physics, Moscow, Russia\
\end{minipage}\\
\makebox[3ex]{$^{ l}$}
\begin{minipage}[t]{14cm}
Alexander von Humboldt Professor; also at DESY and University of Oxford\
\end{minipage}\\
\makebox[3ex]{$^{ m}$}
\begin{minipage}[t]{14cm}
also at University of Tokyo, Japan\
\end{minipage}\\
\makebox[3ex]{$^{ n}$}
\begin{minipage}[t]{14cm}
now at Kobe University, Japan\
\end{minipage}\\
\makebox[3ex]{$^{ o}$}
\begin{minipage}[t]{14cm}
supported by DESY, Germany\
\end{minipage}\\
\makebox[3ex]{$^{ p}$}
\begin{minipage}[t]{14cm}
member of National Technical University of Ukraine, Kyiv Polytechnic Institute, Kyiv, Ukraine\
\end{minipage}\\
\makebox[3ex]{$^{ q}$}
\begin{minipage}[t]{14cm}
now at DESY ATLAS group\
\end{minipage}\\
\makebox[3ex]{$^{ r}$}
\begin{minipage}[t]{14cm}
member of National University of Kyiv - Mohyla Academy, Kyiv, Ukraine\
\end{minipage}\\
\makebox[3ex]{$^{ s}$}
\begin{minipage}[t]{14cm}
now at BNL, USA\
\end{minipage}\\
\makebox[3ex]{$^{ t}$}
\begin{minipage}[t]{14cm}
now at LNF, Frascati, Italy\
\end{minipage}\\
\makebox[3ex]{$^{ u}$}
\begin{minipage}[t]{14cm}
also at Max Planck Institute for Physics, Munich, Germany, External Scientific Member\
\end{minipage}\\
\makebox[3ex]{$^{ v}$}
\begin{minipage}[t]{14cm}
now at Tokyo Metropolitan University, Japan\
\end{minipage}\\
\makebox[3ex]{$^{ w}$}
\begin{minipage}[t]{14cm}
also supported by DESY\
\end{minipage}\\
\makebox[3ex]{$^{ x}$}
\begin{minipage}[t]{14cm}
also at \L\'{o}d\'{z} University, Poland\
\end{minipage}\\
\makebox[3ex]{$^{ y}$}
\begin{minipage}[t]{14cm}
member of \L\'{o}d\'{z} University, Poland\
\end{minipage}\\
\makebox[3ex]{$^{ z}$}
\begin{minipage}[t]{14cm}
now at Polish Air Force Academy in Deblin\
\end{minipage}\\

}

%% file: DESY-13-245-txt.tex
\pagenumbering{arabic}
\pagestyle{plain}
% ----------------------------------------------------------------------------
%       Introduction
% ----------------------------------------------------------------------------
\section{Introduction}
\label{sec-int}

Measurements of  deep inelastic scattering (DIS) cross sections in lepton-proton collisions constitute an essential ingredient in probing the
structure of the proton, which is represented by the parton distribution functions (PDFs).
These PDFs are needed as input in perturbative quantum chromodynamics (pQCD) calculations 
to predict cross sections for hadronic processes with large momentum transfer. 
At large Bjorken $x$, uncertainties in the PDFs determination are often the dominant  uncertainty in the predictions. 

Most of the DIS data available for $x \ge 0.7$~\cite{pl:b223:485,pl:b282:475,jferson}
         have been obtained in fixed-target experiments in a kinematic range where the DGLAP~\cite{dglap} evolution equations, used for PDFs extraction, may not be
         fully applicable. The DIS cross-section measurements used for PDFs determination that originate from the HERA storage ring~\cite{Aaron:2009aa}, where $920\, \units{GeV}$ protons collided with $27.5\, \units{GeV}$ electrons or positrons, 
probe $Q^2$ values up to $40\,000 \units{GeV^2}$ but do not extend beyond $x$ of $0.65$. 
Therefore, in global pQCD fits of PDFs,  a parametrisation of the form  $(1-x)^\beta$ is assumed in order to extend PDFs to $x=1$. Such a behaviour is expected from counting rules~\cite{Brodsky:1973kr} with the power $\beta$ further constrained by the momentum-sum rule. 

As was observed by the ZEUS collaboration~\cite{epj:c49:523-544}, the kinematics of HERA and the design of the detectors allow extension of the measurements of the neutral current  (NC)  cross sections up to $x=1$.  
First results on the high-$x$ behaviour of the cross sections for $Q^2>600\, \units{GeV^2}$ were published~\cite{epj:c49:523-544}  using data taken in 1996\,--\,2000.

The NC cross-section measurements presented in this paper are based on data collected in 2004\,--\,2007, with integrated luminosities of $187 \units{pb^{-1}}$ for $e^-p$ and $142 \units{pb^{-1}}$ for $e^+p$ scattering, i.e. a factor ten and two, respectively, larger than in the previous high-$x$ ZEUS studies~\cite{epj:c49:523-544}. 
Given the much larger data sample and an improved analysis procedure, the uncertainties are substantially reduced. The data sets used for this analysis overlap with previously published ZEUS results~\cite{:2012bx,epj:c62:625-658}.  The finer binning in  the medium- and large-$x$ regions, as well as the extension of the cross-section measurements to $x\cong 1$,  are expected to provide  constraints on the  PDFs at high $x$, where the contribution of valence quarks is important. 
%A recommendation is made on how to use these measurements in the global QCD fits and avoid the overlap between the data from different ZEUS analyses.

\section{Kinematic variables and cross sections}
\label{sec:kin}

Inclusive deep inelastic lepton-proton scattering can be described in
terms of the kinematic variables $x$, $y$ and $Q^2$.
The variable $Q^2$ is defined as $Q^2 = -q^2 = -(k-k')^2$,
where $k$ and $k'$ are the four-momenta of the incoming and scattered lepton,
respectively.
Bjorken $x$ is defined as $x=Q^2/2P \cdot q$, where $P$ is
the four-momentum of the incoming proton.
The fraction of the lepton energy transferred to the proton in its rest frame
is given by $y = P \cdot q / P \cdot k$.
The variables $x$, $y$ and $Q^2$ are related by $Q^2=sxy$,
where $s$, the centre-of-mass-energy squared, is given
by $s=4E_e E_p$, and $E_{e}$ and $E_{p}$
are the initial energies of the electron and proton, respectively.

The electroweak Born-level cross section for the $e^ \pm p$ NC
interaction is given by
\begin{equation}
\ddif{\sigma(e^{\pm}p)}{x}{Q^{2}} =
\frac{2 \pi \alpha^{2} }{xQ^{4}}
[Y_{+} \tilde{F_{2}}(x,Q^{2})
\mp Y_{-} x\tilde{F_{3}}(x,Q^{2})
- y^{2}\tilde {F_{L}}(x,Q^{2})],
\label{eqn:unpol_xsec}
\end{equation}
where $\alpha$ is the fine-structure constant,
$Y_{\pm} = 1 \pm (1 - y)^{2}$, and
$\tilde{F_{2}}(x,Q^{2})$, $\tilde{F_{3}}(x,Q^{2})$ and
$\tilde{F_{L}}(x,Q^{2})$
are generalised structure functions~\cite{zfp:c24:151,devenish:2003:dis}. Next-to-leading-order (NLO)
pQCD calculations predict that the contribution of the longitudinal
structure function, $\tilde {F_L}$, to $d^2\sigma /dx dQ^2$ is approximately
$1.5\%$, averaged over the kinematic range considered.
%and therefore neglected in the discussion in this section.

Photon exchange dominates the cross section at low $Q^{2}$
and is described by $\tilde{F_{2}}$ and $\tilde{F_{L}}$.
It is only at $Q^{2}$ values comparable to $M_Z^2$ that
the $\gamma / Z$ interference and the $Z$ exchange terms become important
and the $\tilde{F_{3}}$ term contributes significantly to the cross section.
The sign of the $\tilde{F_3}$ term in Eq.~(\ref{eqn:unpol_xsec}) shows that
electroweak effects increase (decrease) the $e^- p$ ($e^+ p$) cross sections.

%Reduced cross sections, $\tilde{\sigma}$, for $e^-p$ and $e^+p$ scattering are defined as
%\begin{equation}
%\tilde{\sigma}^{e^{\pm} p}
%=
%\frac {xQ^{4}} {2 \pi \alpha^{2} }
%\frac {1} {Y_{+}}
%\ddif{\sigma(e^{\pm}p)}{x}{Q^{2}}.
%=
%\tilde{F_{2}}(x,Q^{2}) \mp \frac {Y_{-}} {Y_{+}} x \tilde{F_{3}}(x,Q^{2}).
%\label{eqn:red}
%\end{equation}

The generalised structure functions depend linearly on the longitudinal polarisation
of the electron\footnote{ Here and in the following the term electron denotes generically both the electron and the positron.} beam.
The extracted cross sections correspond to unpolarised lepton beams.

%In this paper, measurements of the differential and the reduced cross section as a function of $x$ and $Q^2$
%are presented using the full $e^\pm p$ datasets collected after the HERA luminosity upgrade. 
%The cross sections are given for the case of unpolarized lepton beams.

% ----------------------------------------------------------------------------
%       Experimental set-up
% ----------------------------------------------------------------------------
\section{Experimental set-up}
\label{sec-exp}

% ----------------------------------------------------------------------------
%       General detector blabla
% ----------------------------------------------------------------------------
A detailed description of the ZEUS detector can be found 
elsewhere~\cite{zeus:1993:bluebook}. A brief outline of the 
components that are most relevant for this analysis is given
below.\xspace

% ----------------------------------------------------------------------------
%       CTD/MVD/STT description, footnote on coordinate system is the argument
% ----------------------------------------------------------------------------
In the kinematic range of the analysis, charged particles were tracked
in the central tracking detector (CTD)~\cite{CTD}, the microvertex
detector (MVD)~\cite{MVD} and the straw-tube tracker (STT)~\cite{STT}. The CTD and the MVD
operated in a magnetic field of $1.43\Tesla$ provided by a thin
superconducting solenoid. The CTD drift chamber covered the
polar-angle\footnote{%
The ZEUS coordinate system is a right-handed Cartesian system, with the $Z$
axis pointing in the proton beam direction, referred to as the ``forward
direction'', and the $X$ axis pointing  towards the centre of HERA.
The coordinate origin is at the nominal interaction point.\xspace}
 region $15\degree<\theta<164\degree$. The MVD
silicon tracker consisted of a barrel (BMVD) and a forward (FMVD)
section. The BMVD provided polar-angle coverage for tracks with three
measurements from 30\degree \, to 150\degree. The FMVD extended the
polar-angle coverage in the forward region to 7\degree. The STT
covered the polar-angle region $5\degree<\theta<25\degree$.
For CTD-MVD tracks that pass
through all nine CTD superlayers, the transverse momentum, $p_T$, resolution was
$\sigma(p_{T})/p_{T} = 0.0029 p_{T} \oplus 0.0081 \oplus
0.0012/p_{T}$, with $p_{T}$ in \gev.

% ----------------------------------------------------------------------------
%       CAL description straight and simple
% ----------------------------------------------------------------------------
The high-resolution uranium--scintillator calorimeter (CAL)~\cite{CAL}
consisted of three parts: the forward (FCAL), the barrel (BCAL) and
the rear (RCAL) calorimeters. Each part was subdivided transversely
into towers and longitudinally into one electromagnetic section (EMC)
and either one (in RCAL) or two (in BCAL and FCAL) hadronic sections
(HAC). The smallest subdivision of the calorimeter was called a cell.
The CAL energy resolutions, as measured under test-beam conditions,
were $\sigma(E)/E=0.18/\sqrt{E}$ for electrons and
$\sigma(E)/E=0.35/\sqrt{E}$ for hadrons, with $E$ in \gev.

%\Zlumidesc{2.6~\%}

The luminosity was measured using the Bethe--Heitler reaction
$ep\,\rightarrow\, e\gamma p$ by a luminosity detector which consisted
of independent lead--scintillator calorimeter\cite{PCAL} and  magnetic
spectrometer\cite{SPECTRO} systems. The systematic
uncertainty on the measured luminosity was 1.8\%~\cite{Adamczyk:2013ewk }.

%Polarization
The analysed data were collected with longitudinally polarised $e^-$ and $e^+$ beams. The beam polarisation was measured using two independent polarimeters, a transverse~\cite{epj:c33:s1067,nim:a329:79} and a longitudinal polarimeter~\cite{nim:a479:334}. The luminosity-weighted polarisations for the $e^-$ beam were $+0.28$ and $-0.28$ and for the $e^+$ beam $+0.33$ and $-0.36$. Since the level of polarisation is very similar for the right-handed and left-handed lepton beams, the correction applied to obtain the unpolarised cross sections was very small.
 
\section{Monte Carlo samples}
\label{mc}
Monte Carlo (MC) simulations were used to determine the efficiency for selecting the events, the accuracy of kinematic reconstruction, to estimate the background rates and to extract the cross section in the kinematic region of interest. A sufficient  number of events were generated to ensure negligible statistical uncertainties compared  to data.

Neutral current DIS events were simulated including leading-order electroweak radiative effects, using the {\sc Heracles}~\cite{cpc:69:155} program with the {\sc Djangoh}~1.6~\cite{spi:www:djangoh11} interface to the fragmentation and hadronisation programs and using
% CTEQ5D PDF's\cite{epj:c12:375}.  
HERAPDF1.5 PDFs~\cite{herapdf1.5}.
To study the effect of the modelling of QCD radiation on the final results, two independent NC samples were generated, one with the {\sc Meps} model of {\sc Lepto} 6.5.1\cite{cpc:101:108} and the other with  the colour-dipole model of {\sc Ariadne} 4.12~\cite{cpc:71:15}. Both programs use the  Lund string model as implemented in {\sc  Jetset}~\cite{cpc:39:347} for hadronisation. A linear combination of the two MC samples was found to give the best description of the jet variables as discussed in Section~\ref{sec:mctune}. This combination was then used to extract the central values of the reduced cross sections.   

Hard photoproduction, a potential source of background, was simulated using {\sc Herwig}~5.9~\cite{cpc:67:465}. QED Compton events, another potential source of background, were generated using the {\sc Grape} MC generator~\cite{cpc:136:126}.

The ZEUS detector response was simulated using a program based on {\sc Geant}
3.21~\cite{tech:cern-dd-ee-84-1}. The generated events were passed through the
detector simulation, subjected to the same trigger requirements as the data and
processed by the same reconstruction programs.

\section{Event reconstruction}
\label{sec:evrec}

A typical NC high-$Q^{2}$ and high-$x$ event consists of the scattered electron and a high-energy collimated jet of particles in the direction of the  struck quark. The electron and the jet are balanced in transverse momentum. The proton remnant mostly disappears down the beam pipe. The $x$ and $Q^2$ of events, in which the jet is well contained in the detector, may be determined by various techniques, such as the double-angle (DA) method\cite{epj:c28:175,epj:c11:427}. However,  the maximum $x$ value that can be reached is limited by the fact that at the low values of $y$ typical of these events, the uncertainty on $x=Q^2/ys$  increases as $\Delta x\sim \Delta y/y^2$.

An improved $x$ reconstruction is achieved by observing that, in the limit of $x\rightarrow 1$, the energy of the struck quark represented by a collimated jet is $E_\mathrm{jet} \cong xE_p$. The  expression for $x$ is 
\begin{equation}
x = \frac {E_\mathrm{jet}(1+\cos \theta_\mathrm{jet})}{2 E_p \left( 1- \frac {E_\mathrm{jet}(1-\cos\theta_\mathrm{jet})}{2E_{e}} \right) } \, ,
\label{eq-xpt}
\end{equation}
where $\theta_\mathrm{jet}$ is the scattering angle of the jet in the detector. 

As $x$ increases and the jet associated with the struck quark disappears down the beam-pipe, the ability to reconstruct $x$ is limited by the energy loss. However, in these events, the cross section integrated  from a certain limit in $x$, $x_\mathrm{edge}$, up to $x=1$ is extracted. The value of $x_\mathrm{edge}$  below which the jet is fully contained in the detector depends on $Q^2$ and the higher the $Q^2$, the higher the value of $x_\mathrm{edge}$.   The above observation constitutes the essence of the methodology applied in the earlier ZEUS publication~\cite{epj:c49:523-544}. 

In the analysis presented here, two improvements were introduced~\cite{ronenphd,inderpalphd,rituphd}. Since, in these events, the scattered electron is very well measured in the detector, for the one-jet events, which statistically dominate the high-$Q^{2}$ and high-$x$ NC samples, the measured value of $E_\mathrm{jet}$ in Eq.~\eqref{eq-xpt} is replaced by
\begin{equation}
E_{\mathrm{jet}} =  E_{T}^{{e}} / \sin \theta _{\mathrm {jet}}  \, ,
\end{equation}
as the transverse energy of the (massless) jet is balanced by the transverse energy of the scattered electron, $E_{T}^{{e}}$, and the uncertainty on the electron-energy scale and resolution is smaller than that for the hadronic component. Hard QCD processes present in DIS, such as boson-gluon fusion or QCD Compton, may lead to extra jets in the events. The analysis is thus extended to include multi-jet events. For the latter, the smallest bias on the $x$ reconstruction was found for a modified Jacquet-Blondel (JB) method~\cite{proc:epfac:1979:391}, in which 
\begin{equation}
x=\frac{(E_T^\mathrm{jets})^2}{s (1-y_\mathrm{jets})y_\mathrm{jets}} \, ,
\end{equation}
where $E_T^\mathrm{jets}$ is the vector sum over the transverse-energy vectors of individual jets and $y_\mathrm{jets}=\sum_\mathrm{jets}E_i(1-\cos \theta_i)/2E_e$, where $E_i$ and $\theta_i$ denote the energy and scattering angle of jet $i$ and the sum runs over all jets. This approach is less sensitive to the contribution of particles that are not assigned to any jets.

The value of  $Q^2$ is calculated from the measured scattered-electron energy, $E^{'}_e$, and scattering angle, $\theta_e$, as
\begin{equation}
\label{q2formula}
  Q^{2}=2E_{e} E^{'}_{e} (1+\cos\theta_e) \, .
\end{equation}
This provides the best resolution for non-radiative events.

\subsection{Electron reconstruction}

Calorimeter and CTD information\cite{epj:c11:427} is used to identify the scattered electron and to determine its energy and scattering angle. The electron candidate is required to be isolated and matched to a well measured track if it  lies in the
% CTD acceptance 
range ($ 20^{\circ} < \theta_{e} < 150^{\circ}$). The energy of the electron is taken from the calorimeter information. The scattering angle is determined either from the track measurement,  for candidates within the CTD acceptance, or from the position reconstructed in the CAL and the interaction vertex. 

If the electron candidate is found in the BCAL, the measured energy  is corrected for non-uniformities in the response near the module and cell edges and for shower sampling variations. The absolute energy scale is determined using events from the kinematic peak, i.e.\,low-$y$ events for which the expected electron energy, $E_\mathrm{KP}$, can be inferred from the scattering angle. The linearity of the energy response is then checked by comparing $E^{'}_{e}$ with the energy predicted by the DA approach.  
The same procedure was applied on  data and  MC. The comparison of the distributions $E^{'}_{e}/E_\mathrm{KP}$ in the data and in the MC is shown in Fig.~\ref{fig:KP+jetscale} (a) and (b) for the JB estimator of $y$, $y_{\rm{JB}} <$ 0.1. Very good agreement is observed. The most probable values, obtained from  Gaussian fits around the peak to data and MC,  agree within 0.5\%, which is taken as an estimate of the systematic uncertainty on the electron energy scale.
%better then 0.5\%.
% which is then used as the absolute uncertainty on the electron energy scale. 

The electron energy resolution in the MC was found to be 5\% for $E^{'}_{e} > 15$ GeV, and improves to 3\% for electron energies above $50 \units{GeV}$. The widths of the  $E^{'}_{e}/E_\mathrm{KP}$ distributions in data and in MC agree within 10\%, which is taken as an estimate of the systematic uncertainty on the energy resolution.

The angular resolution is about 0.6 mrad in the forward region and increases to 0.7 mrad in the central region. The $Q^2$ resolution is fairly constant throughout the  range of the analysis and is dominated by the electron-energy resolution. 

The systematic uncertainties of the absolute energy scale, the energy resolution and the scattering angle were validated by a procedure based on the kinematic fit to the reconstructed NC events, for which the fit is over-constrained (see Section~\ref{sec:KF}). 

The track-matching efficiency was found to be slightly 
lower in the data than in the MC and was corrected for. For the kinematic range studied here and after applying the selection cuts described in Section~\ref{sec:eventsel}, the electron-detection efficiency is close to 100\%. 

\subsection{Jet reconstruction}

The jets were reconstructed in the laboratory frame using the inclusive $k_{T}$ algorithm with the radius parameter $R$ set to 1 in the massless scheme~\cite{np:b406:187}.  The algorithm was applied to the hadronic final state reconstructed using a combination of track and CAL information, excluding the cells and the track associated with the scattered electron. The selected tracks and CAL clusters were treated as massless energy-flow objects (EFOs)~\cite{thesis:briskin:1998}.
The jet variables are defined according to the
Snowmass convention \cite{proc:snowmass:1990:134},
\begin{alignat*}{2}
 E_T^\mathrm {jet} = \sum_i E_{T,i}\; ,\;\;\;
 \eta_{\rm {jet}} = \frac{\sum_i E_{T,i} \eta_i}{E_T^\mathrm {jet}},  
 \\ 
 \theta_{\rm {jet}} = 2\tan^{-1}(e^{-\eta_{\rm {jet}}})\; , \;\;\;
 E_{\rm {jet}} = \sum_i E_{i}, \nonumber
\end{alignat*}
where $E_{i}$, $E_{T,i}$ and $\eta_{i}$ are the energy, transverse energy 
and pseudorapidity of  the EFOs, respectively. 
A jet is required to have a minimum transverse energy of $10\units{GeV}$. 
Events for which no jet was found are called zero-jet events. 
%%% sentence below restored dur to referee's comments
The $x$ value for these events cannot be inferred and  they contribute to the large-$x$ integrated cross section.

The calibration of the jet-energy scale was performed through the transverse-momentum balance between the electron and the jet. Only 
one-jet events were selected for this procedure.  A comparison between data and MC is shown in Fig.~\ref{fig:KP+jetscale} (c)--(f), separately for jets found in the BCAL and in the FCAL. Very good agreement  is found. 
%Also the width of the distributions is well reproduced by the MC. 
This calibration procedure is estimated to yield a 1\% uncertainty on the jet-energy scale and a 5\% uncertainty on the jet-energy resolution.

For one-jet events, the expected angle of the jet, $\gamma_\mathrm{jet}$, may be inferred from the scattered-electron energy and angle.  
%The distribution of the ratio $\gamma_\mathrm{jet}/\theta_\mathrm{jet}$ in the data and in the MC is shown in Fig.~\ref{fig:jetangle} for the %$e^+p$ and $e^-p$ samples. 
A comparison between the distributions of the ratio $\gamma_\mathrm{jet}/\theta_\mathrm{jet}$ in  data and  in MC demonstrated 
a good agreement both in the most probable values (to better than 0.5\%) and in the widths (to better than 10\%).  
%An estimated $3 \units{mrad}$ uncertainty is assigned to the $\theta_\mathrm{jet}$ measurement. 
In the estimate of systematic uncertainties, the variation in $\theta_\mathrm{jet}$ is replaced by a possible misalignment between data and MC of $5 \units{mm}$ in the position of the jet on the face of the calorimeter.

As in the case of electron reconstruction, the final values for the jet-energy scale, the jet-energy resolution, their uncertainties, as well as the uncertainties on the scattering angle of the jets were validated by the kinematic-fit procedure.

The jet angular resolution and to a smaller degree its energy resolution were propagated to the resolution on the reconstructed $x$ value. 
For one-jet events, which constitute about 90\% of the sample, the $x$ resolution is about $0.01$ at $x\simeq 0.15$, increases to 0.03 at $x=0.35$ and remains constant at this value up to $x_\mathrm{edge}$.  The $x$ resolution for multi-jet events is about a factor of two worse. 

\subsection{Kinematic fit}
\label{sec:KF}
The information on the kinematics of each NC event is over-constrained by the fact that both the electron variables and the hadronic variables are available. 
An event-by-event kinematic fit using the BAT package~\cite{Caldwell:2008fw} was performed on the data and MC samples to study the description of the one-jet data by the MC simulation~\cite{rituphd}. The inputs to the fit were the measured energy and angle of the electron and the energy and angle of the jet. 

The output of the fit was the kinematic variables $x,y$ and the energy of a possible intial-state radiative (ISR) photon.   Bayesian inference was used,
$$ P(x,y,E_{\gamma}|\vec{D}) \propto P(\vec{D}|x,y,E_{\gamma})P_0(x,y,E_{\gamma}) \, , $$ 
where $E_{\gamma}$ represents the energy of the ISR photon and $\vec{D}$ is the set of measured quantities. Simple priors were chosen to reflect the basic features of the DIS cross-section dependence on $(x,y)$ and of the bremsstrahlung cross section on $E_{\gamma}$,
\begin{eqnarray}
P_0(x,y,E_{\gamma})&=&P_0(x,y)P_0(E_{\gamma})  \nonumber\\
P_0(x,y) &\propto&\frac{(1-x)^5}{x^2 y^2}  \nonumber\\
P_0(E_{\gamma}) &\propto& \frac{1+(1-E_{\gamma}/(E_e-E_{\gamma}))^2}{E_{\gamma}/(E_e-E_{\gamma})} \nonumber \; .
\end{eqnarray}
 The probability density $P(\vec{D}|x,y,E_{\gamma})$ was calculated using parametrisations of the resolutions as found from MC studies. The quantities $(x,y,E_{\gamma})$ at the global mode of the posterior were used to extract the kinematic-fit values for the energies and angles of the electron and hadronic system.
%and pull distributions between these values and the measured values were studied.   
The pulls were defined as the difference between the measured value and the fitted value, divided by the RMS extracted from the MC. Studies of these pull distributions allowed the determination of biases between the MC and the data, the variation of which were 
%. Furthermore, by varying the corrections to the MC, the level of agreement between the pull distributions in the data and the MC can be 
used to set limits on the systematic uncertainties. 
  
As an example, the pull distributions for data and MC are shown in Fig.~\ref{fig:batiep} for the $e^+p$ sample  for the different quantities entering the fit.  Excellent agreement is found between data and simulation. Similarly,  excellent agreement is observed for the $e^-p$ sample (not shown).
%The associated uncertainties are listed in table~\ref{tab:recsyst}.

\section{Event selection}
\label{sec:eventsel}

\subsection{Trigger}

ZEUS operated a three-level trigger system\cite{zeus:1993:bluebook,uproc:chep:1992:222,trig3}. At the first level, only coarse calorimeter and tracking information was available. Events were selected
using criteria based on an energy deposit in the CAL consistent with an isolated electron candidate.  In addition, events with high transverse energy in coincidence with a CTD track were accepted. 
At the second level, the value of 
\begin{equation}
  \delta \equiv \sum\limits_{i} (E-p_Z)_{i} = \sum\limits_{i} 
   ( E_i - E_i \cos
  \theta_{i} ) \; ,
%  \label{eq-Delta}
\end{equation}
  where the sum runs over all calorimeter energy deposits $E_i$ with
  polar angles $\theta_i$, was required to be higher than $29 \units{GeV}$ to ensure efficient selection of NC events, for which $\delta = 2E_e$ is expected when no detector effects are present.
%was used to select NC DIS
Timing information from the calorimeter was used to reject events inconsistent with the bunch-crossing time. At the third level, events were fully reconstructed and selected according to requirements similar to, but looser than, the offline cuts described below.

\subsection{Offline event selection}
\label{offlineevts}

Offline, events were selected according to the following criteria:
\begin{itemize}
\item  a scattered electron candidate 
%in the event was required. The scattered electrons were identified using an algorithm that combined information from the energy deposits in the CAL with tracks measured in the CTD~\cite{epj:c11:427}. To ensure high electron finding efficiency and to reject backgrounds, the identified electron was required to have an energy
 with $E^{'}_e> 15 \units{GeV}$ was required. A track matched to the energy deposit in the calorimeter was required for events in which the electron was found within the acceptance of the tracking detectors. This was done by requiring the distance of closest approach (DCA) between the track extrapolated to the calorimeter surface and the position of the energy cluster to be less than $10 \units{cm}$ and the electron track momentum, $p_e$,
to be larger than $5\units{GeV}$. A matched track was not required if the electron emerged in the FCAL with a polar angle outside the acceptance of the CTD. In this case, it was required to have $p^{e}_{T}>30 \units{GeV}$. An isolation requirement was imposed such that the energy not associated with the electron in an $\eta-\phi$-cone of radius 0.8 centred on the electron was less than $4 \units{GeV}$;

\item  fiducial-volume cuts were applied to the position of the scattered-electron candidate to ensure that it was within a region in which the experimental resolution is well understood. A cut excluded the electron candidates in the transition region between FCAL and BCAL ($0.52 < \theta< 0.65$). Events for which the $Z$ position of the electron in the CAL $Z_{e}< -95 \units{cm}$ were also excluded, as well as electrons in the BCAL within $1.4 \units{cm}$ of the module gap or $0.6 \units{cm }$ of the cell gap;

\item  up to three jets were allowed. Valid jets were required to have $E_{T}^{\mathrm{jet}} > 10 \units{GeV}$  and their reconstructed position had to be outside a box of 40$\times$40 cm$^{2}$ around the FCAL beam pipe.

\end{itemize}
The following set of cuts was applied to remove background events:

\begin{itemize}

\item  the value $\delta > 40 \units{GeV}$  was used to remove NC events with hard QED intial-state radiation and to reject photoproduction events (events in which the scattered electron escapes through the rear beam hole).  In addition, $\delta<65 \units{GeV}$ was required to remove events in which a
  normal DIS event coincided with additional energy deposits in the RCAL from beam-gas interactions. To remove residual photoproduction background, $y$ calculated from the electron-candidate variables was required to be less than 0.8; 

\item  to remove non-$ep$ background, events were required to have a reconstructed vertex within the range $-50 < Z_\mathrm{vtx} < 50 \units{cm}$;

\item  the net transverse momentum is expected to be small for NC events.
To remove cosmic-ray events and beam-related background events, the quantity $P_T / \sqrt{E_T}$ was required to be less than $4\sqrt{\units{GeV}}$. The variables $P_T$ and $E_T$ are defined by
\begin{eqnarray*}
  P_T^2 & = & P_X^2 + P_Y^2 = \left( \sum\limits_{i} E_i \sin \theta_i \cos
    \phi_i \right)^2+ \left( \sum\limits_{i} E_i \sin \theta_i \sin \phi_i \right)^2, \\
  E_T & = & \sum\limits_{i} E_i \sin \theta_i,
\end{eqnarray*}
  where the sums run over all calorimeter energy deposits, $E_i$, with
  polar and azimuthal angles $\theta_i$ and $\phi_i$ with respect to the
  event vertex, respectively;

\item  to ensure that zero-jet events originate from a large-$x$ scattering, a limit was imposed on $y_{\rm{JB}}$ such that  $y_{\rm {JB}} < 1.3 \cdot Q^2_{\rm {edge}}/(s \cdot x_{\rm {edge}})$, 
where $x_{\rm {edge}}$ and $Q^2_{\rm {edge}}$ are the lower $x$ and upper $Q^2$ edges of the
  bins defined for the integrated-cross-section measurements up to $x=1$.
\end{itemize}

The number of events remaining after all these selection cuts, and an additional requirement of $Q^2>550 \units{GeV^2}$, were 53\,099 for the $e^{-}p$ and 37\,361 for the $e^+p$ sample, of which 1\,877 and 1\,325 were found with zero jets, respectively.

\section{Data and MC comparison}

The MC simulation was used to correct for detector acceptance,  trigger and offline selection efficiencies and for smearing effects due to  finite resolution. Given that the measured distributions are steep functions of $x$ and $Q^2$, it is very important to have a good description of the underlying physics distributions and a good understanding of the resolution effects. 

\subsection{Treatment of the hadronic final state}\label{sec:mctune}

%After adjusting the energy scales and the energy resolutions for the scattered electron and for the jets in the MC to agree with those in the data as described in section~\ref{sec:evrec}, 
In order to assess the quality of modelling of the hadronic final state,
a detailed comparison of the data and the MC was performed in distributions such as the jet multiplicity, $N_\mathrm{jet}$, jet transverse energy, $E_T^{\mathrm{jet}}$, and the polar angle of the jets, $\theta_{\mathrm{jet}}$.  While the {\sc Ariadne} MC gave a better description of the $E_T^{\mathrm{jet}}$ distribution, {\sc Lepto} gave a better description of the $N_\mathrm{jet}$ distribution. Since both MC sets have identical generated $x$ and $Q^2$ distributions, a linear mixture of the two could be constructed, 
\begin{equation}
N_\mathrm{MC}(x,Q^2) = \lambda\cdot N_\mathrm{\sc Ariadne} (x,Q^2) + (1-\lambda) \cdot N_\mathrm{\sc Lepto}(x,Q^2) \, ,
\end{equation}
where $N$ denotes the number of generated events in a given $x$ and $Q^2$ bin and the subscripts denote the MC samples. The fraction $\lambda$ was determined by fitting the 3-dimensional distribution in $N_\mathrm{jet}$, $E_T^{\mathrm{jet}}$ and $\theta_{\mathrm{jet}}$ in the data. The best fit to the combined $e^-p$ and $e^+p$ data samples was obtained  
with $\lambda=0.3$ with a systematic uncertainty of 0.3. It was further checked that the description of the data improves for all relevant distributions, not only for those which were used in the fit. A comparison between data distributions and MC expectations with $\lambda=0.3$ for electron and jet variables is shown  for events with $Q^2>550 \units{GeV^2}$ and $y<0.8$, for the $e^-p$ sample in Fig.~\ref{fig:mcmixem} and for the  $e^+p$ sample in Fig.~\ref{fig:mcmixep}. The comparison between the distributions of $Q^2$,  $N_\mathrm{jet}$ and  $x$ for events with at least one reconstructed jet is shown in Fig.~\ref{fig:xq2mix}.  Very good agreement is achieved in all distributions. Subsequent comparisons use the linear combination of {\sc Ariadne} and  {\sc Lepto} with $\lambda=0.3$, which is denoted simply as $^\backprime$MC$^\prime$.

\section{Cross-section determination}
\label{binning}

 The bin widths in the $(x,Q^{2})$-plane are chosen to be sufficiently large compared to the resolution of the variables being measured so that the migrations between neighbouring bins are acceptable. The $x_{\mathrm{edge}}$, the lower $x$ edge of the zero-jet bin, was determined using $\theta_{\mathrm{jet}} = 0.11$ rad. The efficiency, defined as the number of events generated and reconstructed in a given bin after all selection cuts divided by the number of events generated in that bin, was typically 50\%, mainly driven by geometrical fiducial cuts. The efficiency was lower in low-$Q^{2}$ bins due to removal of electrons in RCAL and in the B/RCAL transition region. The purity, which is defined as the number of events reconstructed and generated in a particular bin after all selection cuts divided by the number of events reconstructed in that bin, was typically 60\%. 

%The efficiency and purity for simulations of 2004-06 $e^{-}p$ data are shown in Fig. \ref{effpur}. The %efficiency and purity values for zero jet bins are comparable to the mid-$x$ bins.\\

The double-differential cross section in bins of $Q^{2}$ and $x$ was determined as follows:
\begin{equation}
  \frac{d^2 \sigma(x, Q^2)}{dx dQ^2} = \frac{N_{\rm data}(\Delta x, \Delta Q^2)}
  {N_{\rm MC}(\Delta x, \Delta Q^2)}\frac{d^2
  \sigma_{\rm Born}^{\rm SM}(x,Q^2)}{dx dQ^2} ,
%  \label{eq-xsect}
\end{equation}
and the integrated cross section was determined as
\begin{equation}
  \int_{x_{\rm {edge}}}^{1} \frac{d^2 \sigma(x, Q^2)}{dx dQ^2} dx=
   \frac{N_{\rm data}(\Delta x, \Delta Q^2)}{N_{\rm MC}(\Delta x, \Delta Q^2)}
  \int_{x_{\rm {edge}}}^{1} \frac{d^2\sigma_{\rm Born}^{\rm SM}(x,Q^2)}{dx dQ^2}dx ,
%  \label{eq-xsecti}
\end{equation}
where $N_{\rm data}(\Delta x, \Delta Q^2)$ is the number of data events reconstructed
in a bin $(\Delta x, \Delta Q^2)$ and $N_{\rm MC}(\Delta x, \Delta Q^2)$
is the corresponding number of simulated events normalised to the data luminosity.
The SM prediction, $d^2 \sigma^{\rm SM}_{\rm Born}(x, Q^2)/dx dQ^2$,
was evaluated according to Eq.~(\ref{eqn:unpol_xsec}) with the same PDF and
electroweak parameters as used in the MC simulation.  This procedure
implicitly takes into account the acceptance, bin-centring and leading-order electroweak
radiative corrections from the MC simulation.

\section{Systematic uncertainties}
\label{sec:sysunc}

The systematic effects related to the uncertainties in the MC simulation were estimated by recalculating the cross section for variations of the parameters by their uncertainties (see Section~\ref{sec:evrec}). The positive and negative deviations from the nominal cross section were added separately in quadrature to obtain the total positive and negative systematic uncertainty. 
\subsection{Uncorrelated systematic uncertainties}

The following systematic uncertainties are either small or exhibit
no bin-to-bin correlations:
\begin{itemize}

 \item electron-energy resolution \,--\, the effect on the value of the cross section of changing the resolution of the
electron energy measured in the calorimeter by $\pm$3\%  was less than 1.5\% for almost all bins;

\item electron-isolation requirement \,--\, variation of the electron-isolation energy by $\pm2 \gev$ caused  the cross section to vary within 1.5\% for most of the bins, rising to a maximum of 3\% for a few bins;

\item track-matching efficiency \,--\, this correction was varied within the limits allowed by its statistical uncertainty; the resulting variation in the cross section was found to be less than $0.2$\% for all  bins; 
%\item track veto efficiency correction -  the applied track veto efficiency correction was varied by a value equal to $50$\% of it's %difference from one and the resulting systematic uncertainty in the cross sections was found to be less than $1$\%;
\item FCAL alignment \,--\, the  position of the jet on the face of the FCAL was varied by
  $\pm 5 \mm$ in both the $X$ and the $Y$ direction. The resulting
  changes in the cross sections were negligible in low-$Q^{2}$ bins, rising to a maximum of 5\% in high-$Q^{2}$ bins;
\item BCAL alignment \,--\,  the position of the electron on the face of the CAL along $Z$ and along the azimuthal direction was varied by $\pm$1 mm, respectively. The resulting changes in the cross sections were negligible;
%gives a very small variation  in the cross section;
%\item electron $Z$-position \,--\, the variation of the electron $Z$-position in BCAL by $\pm$1mm gives negligibly small variations in the cross section;
\item F/BCAL crack cut \,--\, the cut on the electron polar angle in this region was varied by $\pm 15 \units{mrad}$ and the resulting variation in the cross section was found to be negligible for almost all bins, except a few high-$Q^{2}$  bins, where the variation increased to $5$\%;
\item background \,--\, the estimated background from all sources was less than 0.015\% and therefore the associated contribution and systematic uncertainty were neglected.

\end{itemize}

\subsection{Correlated systematic uncertainties}

The significant correlated systematic uncertainties are listed below and labelled for further reference as follows:

\begin{itemize}
 \item $\{\delta_{1}\}$ electron-energy scale \,--\, the systematic uncertainty resulting from variation of the electron-energy scale by 0.5\%  resulted in typically less than 3\% variation in cross sections for almost all bins, except for some bins at high $Q^2$, where this value rose to $10$\%;

\item $\{\delta_{2}\}$ jet-energy scale \,--\, the uncertainty in the cross-section measurement due to variation of the jet-energy scale by 1\% had a negligible effect for almost all bins and was well within 1\%;

\item $\{\delta_{3}\}$ simulation of the hadronic final state \,--\, the analysis was repeated while varying the relative contribution of {\sc Ariadne} and  {\sc Lepto Meps} by $\pm 0.3$. The resulting uncertainty was well within $2$\% for low-$Q^2$ bins, increasing to $10$\% for the highest-$x$ bins at high $Q^2$;

\item $\{\delta_{4}\}$ PDF uncertainties \,--\, the MCs were reweighted to variants of HERAPDF1.5, representing the PDF uncertainties that were found to lead to the largest deviations from the central PDF set expectations for the NC cross sections. The resulting systematic uncertainties are very small with the exceptions of the highest $Q^2$ and $x$ bins, where the uncertainty reaches 3\%.  

\end{itemize}

In addition, there is a global uncertainty due to the luminosity measurement of 1.8\%.

%\end{description}

\section{Results}

The measured double-differential Born-level cross section as a function of $Q^2$ and $x$ is presented in Tables 1 and 2, for $e^-p$ and $e^+p$ scattering, respectively. For the highest integrated $x$ bin, the respective average cross sections, defined as
\begin{equation}
I(x) = \frac{1}{1-x_{\rm {edge}}}\int_{x_{\rm {edge}}}^{1}\frac{d^2\sigma(x,Q^2)}{dxdQ^2}dx
\;\; ,
\label{eqn-I(x)}
\end{equation}
are presented in Tables 3 and 4 as a function of $Q^2$. Also listed are the statistical and systematic uncertainties. The latter are given separately for the quadratically summed uncorrelated and correlated uncertainties. For bins populated by fewer than 50 events, the statistical uncertainties are quoted 
%the $\pm 34\%$ confidence limit derived from the appropriate Poisson distribution. 
 as the central 68\% probability interval calculated using the prior of Jeffreys~\cite{Jeffrey}.

The results are presented in Figs.~\ref{e-xsc} and~\ref{e+xsc}. The averaged integrated cross sections are plotted at $x=(x_\mathrm{edge}+1)/2$.    The measurements are compared to SM expectations obtained with the HERAPDF 1.5 PDFs~\cite{herapdf1.5}.  Within the quoted uncertainties, with statistical and systematic uncertainties added in quadrature, the agreement between measurements and expectations is good.

The ratio of the measured cross sections to those expected from HERAPDF1.5 are shown in Figs.~\ref{e-rxsc} and~\ref{e+rxsc}.  Note that for bins where no events are observed, the limit is quoted at $68$\% probability,  neglecting the systematic uncertainty. Also shown are the predictions from a number of other PDF sets (ABM11~\cite{abm11}, CT10~\cite{ct10}, MSTW2008~\cite{mstw2008}, NNPDF2.3~\cite{nnpdf2.3}), normalised to the predictions from HERAPDF1.5.  Within the quoted uncertainties, the agreement between measurements and expectations is good.
%though there is a tendency of the measured cross sections to lie above the expectations at high $x$ and high $Q^2$.  At lower $Q^2$, the measured cross sections at high $x$ tend to lie below the expectations.  The trend is more pronounced for $e^+p$ than for $e^-p$ interactions. 

%For convenience, the edges of the $Q^2$ bins in the vicinity of $Q^2 \simeq 600 \units{GeV^2}$ were chosen such that these measurements can be easily incorporated into PDF fits by discarding the existing ZEUS measurements~\cite{epj:c28:175,:2012bx} above $Q^2>600 \units{GeV^2}$. The advantage of the present measurements is a finer binning in $x$, an extension of the kinematic coverage up to $x=1$ and a better control of systematic uncertainties.
%These results are expected to have a significant impact in constraining the parton
%distributions at high $x$, where little data is available to
%date.

The measurement presented here is based on the same data set used for previous ZEUS publications~\cite{epj:c62:625-658,:2012bx}. The advantage of the present measurements is a finer binning in $x$ and an extension of the kinematic coverage up to $x \cong 1$. In the region of $Q^2 \geq 725 \units{GeV^2}$ and $y < 0.8$, the cross sections presented here could replace those previously published to assess their impact on the PDFs at high $x$,  where little data is available.

\section{Summary}

Neutral current  $e^-p$ and $e^+p$ DIS cross sections have been measured in the ZEUS detector as a function of $x$ and $Q^2$ for $Q^2 \geq 725 \units{GeV^2}$ and up to $x\cong 1$. The novel reconstruction method and the large volume of data available allowed a high precision, limited only by statistical uncertainties, to be achieved. The results are in good agreement with SM predictions from HERAPDF1.5 and several other commonly used sets of PDFs. The fine binning in $x$, the extension of the kinematic coverage up to $x \cong 1$ and the excellent control of the systematic uncertainties make these data an important input to fits constraining the PDFs in the valence-quark domain in a model-independent way.

%This confirms the validity of the commonly used parameterisation at high $x$.

% of the parton density functions in the proton. The measurements are presented in a form which makes them easy to incorporate in the global PDFs fits without need to use overlapping data samples. 

% ----------------------------------------------------------------------------
%       Mandatory acknowledgements. You may add your buddies to it.c
% ----------------------------------------------------------------------------
\section*{Acknowledgements}
\label{sec-ack}
We appreciate the contributions to the construction, maintenance and operation of the ZEUS detector made by many people who are not listed as authors. The HERA machine group and the DESY computing staff are especially acknowledged for their success in providing excellent operation of the collider and the data-analysis environment. We thank the DESY directorate for their strong support and encouragement.

%\Zacknowledge

\vfill\eject

%% file: DESY-13-245-tab.tex
\input{table_mine_eMp}
\clearpage
\input{table_mine}
\clearpage
\input{table_mine_eMp_int}
\clearpage
\input{table_mine_int}
\clearpage

%% file: table_mine_eMp.tex
\begin{table} 
\begin{center} 
\begin{tabular}{|c|l|c|c|c|c||c|c|c|c|c|}
\multicolumn{11}{l}{
{\normalsize}}\\
\hline
{$Q^2$} & 
\multicolumn{1}{c|}{$x$} & 
$N$& 
${d^2\sigma}/{dxdQ^2}$  &
$\delta_{\rm {stat}}$& 
$\delta_{\rm {sys}}$& 
$\delta_{\rm u}$& 
$\delta_1$ &
$\delta_2$ &
$\delta_3$ &
$\delta_4$ \\
{($\units{GeV^2}$)} & 
$$ &
$$ &
{($ \units{pb/GeV^{2}}$)} &
(\%) &
(\%) &
(\%) &
(\%) &
(\%) &
(\%) &
(\%)  \\
\hline \hline

$    725 $ &  $  0.06  $ & $ 743 $ &  $  3.39e+00 $ &       $^{   +3.7}_{   -3.7} $ &  $^{   +2.1}_{   -1.8} $ &  $^{   +1.2}_{   -1.2} $ &  $^{   +0.8}_{   -0.2} $ &  $^{   -0.2}_{   +0.1} $ &  $^{   +1.4}_{   -1.4} $ &  $^{   +0.1}_{   -0.1} $  \\ 
$    725 $ &  $  0.08  $ & $ 580 $ &  $  2.63e+00 $ &       $^{   +4.2}_{   -4.2} $ &  $^{   +1.5}_{   -2.3} $ &  $^{   +1.3}_{   -1.5} $ &  $^{   +0.7}_{   -1.7} $ &  $^{   -0.2}_{   +0.0} $ &  $^{   -0.1}_{   +0.1} $ &  $^{   +0.1}_{   -0.1} $  \\ 
$    725 $ &  $  0.10  $ & $ 441 $ &  $  1.89e+00 $ &       $^{   +4.8}_{   -4.8} $ &  $^{   +1.5}_{   -1.6} $ &  $^{   +1.5}_{   -1.5} $ &  $^{   +0.2}_{   -0.4} $ &  $^{   +0.0}_{   +0.1} $ &  $^{   +0.2}_{   -0.2} $ &  $^{   +0.1}_{   -0.1} $  \\ 
$    725 $ &  $  0.12  $ & $ 416 $ &  $  1.45e+00 $ &       $^{   +4.9}_{   -4.9} $ &  $^{   +3.2}_{   -2.3} $ &  $^{   +1.6}_{   -1.6} $ &  $^{   +2.1}_{   -1.6} $ &  $^{   -0.1}_{   +0.0} $ &  $^{   +0.5}_{   -0.5} $ &  $^{   +0.1}_{   -0.1} $  \\ 
$    725 $ &  $  0.16  $ & $ 283 $ &  $  1.05e+00 $ &       $^{   +5.9}_{   -5.9} $ &  $^{   +2.6}_{   -2.8} $ &  $^{   +1.9}_{   -2.1} $ &  $^{   +1.4}_{   -1.3} $ &  $^{   -0.0}_{   +0.2} $ &  $^{   +0.4}_{   -0.4} $ &  $^{   +0.1}_{   -0.1} $  \\ 
$    725 $ &  $  0.19  $ & $ 248 $ &  $  7.18e-01 $ &       $^{   +6.3}_{   -6.3} $ &  $^{   +2.7}_{   -2.9} $ &  $^{   +1.9}_{   -1.9} $ &  $^{   +1.7}_{   -1.8} $ &  $^{   -0.1}_{   -0.2} $ &  $^{   +0.3}_{   -0.3} $ &  $^{   +0.1}_{   -0.1} $  \\ 
$    725 $ &  $  0.23  $ & $ 227 $ &  $  5.88e-01 $ &       $^{   +6.6}_{   -6.6} $ &  $^{   +2.4}_{   -2.3} $ &  $^{   +2.1}_{   -2.2} $ &  $^{   +1.1}_{   -0.6} $ &  $^{   +0.1}_{   +0.1} $ &  $^{   -0.4}_{   +0.4} $ &  $^{   +0.1}_{   -0.1} $  \\ 
$    875 $ &  $  0.05  $ & $ 789 $ &  $  2.86e+00 $ &       $^{   +3.6}_{   -3.6} $ &  $^{   +1.1}_{   -1.2} $ &  $^{   +1.1}_{   -1.2} $ &  $^{   -0.3}_{   +0.2} $ &  $^{   -0.2}_{   +0.2} $ &  $^{   +0.1}_{   -0.1} $ &  $^{   +0.1}_{   -0.1} $  \\ 
$    875 $ &  $  0.07  $ & $ 681 $ &  $  2.05e+00 $ &       $^{   +3.8}_{   -3.8} $ &  $^{   +1.2}_{   -1.4} $ &  $^{   +1.1}_{   -1.2} $ &  $^{   -0.3}_{   +0.2} $ &  $^{   -0.1}_{   +0.2} $ &  $^{   -0.3}_{   +0.3} $ &  $^{   +0.1}_{   -0.1} $  \\ 
$    875 $ &  $  0.09  $ & $ 604 $ &  $  1.62e+00 $ &       $^{   +4.1}_{   -4.1} $ &  $^{   +1.4}_{   -1.7} $ &  $^{   +1.4}_{   -1.4} $ &  $^{   -0.3}_{   +0.1} $ &  $^{   -0.1}_{   +0.1} $ &  $^{   -0.4}_{   +0.4} $ &  $^{   +0.1}_{   -0.1} $  \\ 
$    875 $ &  $  0.11  $ & $ 493 $ &  $  1.16e+00 $ &       $^{   +4.5}_{   -4.5} $ &  $^{   +1.7}_{   -1.4} $ &  $^{   +1.5}_{   -1.4} $ &  $^{   +0.4}_{   +0.4} $ &  $^{   +0.1}_{   +0.1} $ &  $^{   -0.1}_{   +0.1} $ &  $^{   +0.1}_{   -0.1} $  \\ 
$    875 $ &  $  0.14  $ & $ 403 $ &  $  8.33e-01 $ &       $^{   +5.0}_{   -5.0} $ &  $^{   +1.7}_{   -2.2} $ &  $^{   +1.5}_{   -1.6} $ &  $^{   -0.7}_{   -0.4} $ &  $^{   -0.2}_{   +0.0} $ &  $^{   +0.8}_{   -0.8} $ &  $^{   +0.1}_{   -0.1} $  \\ 
$    875 $ &  $  0.17  $ & $ 385 $ &  $  6.13e-01 $ &       $^{   +5.1}_{   -5.1} $ &  $^{   +2.2}_{   -2.0} $ &  $^{   +1.8}_{   -1.8} $ &  $^{   -0.2}_{   +0.2} $ &  $^{   -0.2}_{   +0.1} $ &  $^{   +0.9}_{   -0.9} $ &  $^{   +0.1}_{   -0.1} $  \\ 
$    875 $ &  $  0.21  $ & $ 271 $ &  $  4.33e-01 $ &       $^{   +6.1}_{   -6.1} $ &  $^{   +2.0}_{   -2.4} $ &  $^{   +1.8}_{   -2.4} $ &  $^{   -0.5}_{   -0.0} $ &  $^{   -0.0}_{   -0.1} $ &  $^{   +0.8}_{   -0.8} $ &  $^{   +0.0}_{   -0.0} $  \\ 
$    875 $ &  $  0.25  $ & $ 258 $ &  $  3.47e-01 $ &       $^{   +6.2}_{   -6.2} $ &  $^{   +2.7}_{   -2.4} $ &  $^{   +2.0}_{   -2.0} $ &  $^{   -0.4}_{   +0.5} $ &  $^{   +0.0}_{   -0.0} $ &  $^{   -0.5}_{   +0.5} $ &  $^{   +0.0}_{   -0.0} $  \\ 
$   1025 $ &  $  0.05  $ & $ 598 $ &  $  2.14e+00 $ &       $^{   +4.1}_{   -4.1} $ &  $^{   +1.5}_{   -1.9} $ &  $^{   +1.3}_{   -1.3} $ &  $^{   -1.1}_{   +0.1} $ &  $^{   -0.2}_{   +0.1} $ &  $^{   +0.6}_{   -0.6} $ &  $^{   +0.1}_{   -0.1} $  \\ 
$   1025 $ &  $  0.07  $ & $ 489 $ &  $  1.46e+00 $ &       $^{   +4.5}_{   -4.5} $ &  $^{   +2.2}_{   -1.7} $ &  $^{   +1.4}_{   -1.4} $ &  $^{   -0.6}_{   +1.2} $ &  $^{   -0.2}_{   +0.3} $ &  $^{   +0.3}_{   -0.3} $ &  $^{   +0.1}_{   -0.1} $  \\ 
$   1025 $ &  $  0.09  $ & $ 450 $ &  $  1.11e+00 $ &       $^{   +4.7}_{   -4.7} $ &  $^{   +1.9}_{   -2.0} $ &  $^{   +1.4}_{   -1.5} $ &  $^{   -0.6}_{   +1.0} $ &  $^{   -0.0}_{   +0.1} $ &  $^{   +0.5}_{   -0.5} $ &  $^{   +0.1}_{   -0.1} $  \\ 
$   1025 $ &  $  0.11  $ & $ 402 $ &  $  8.58e-01 $ &       $^{   +5.0}_{   -5.0} $ &  $^{   +2.0}_{   -2.5} $ &  $^{   +1.5}_{   -1.5} $ &  $^{   -1.4}_{   +0.7} $ &  $^{   +0.2}_{   +0.1} $ &  $^{   +0.7}_{   -0.7} $ &  $^{   +0.1}_{   -0.1} $  \\ 
$   1025 $ &  $  0.14  $ & $ 342 $ &  $  6.18e-01 $ &       $^{   +5.4}_{   -5.4} $ &  $^{   +2.1}_{   -1.8} $ &  $^{   +1.7}_{   -1.7} $ &  $^{   -0.3}_{   +1.2} $ &  $^{   -0.4}_{   -0.1} $ &  $^{   -0.3}_{   +0.3} $ &  $^{   +0.1}_{   -0.1} $  \\ 
$   1025 $ &  $  0.17  $ & $ 231 $ &  $  4.39e-01 $ &       $^{   +6.6}_{   -6.6} $ &  $^{   +2.4}_{   -1.9} $ &  $^{   +1.9}_{   -1.9} $ &  $^{   +0.6}_{   +0.1} $ &  $^{   -0.1}_{   +0.2} $ &  $^{   -0.2}_{   +0.2} $ &  $^{   +0.1}_{   -0.1} $  \\ 
$   1025 $ &  $  0.20  $ & $ 224 $ &  $  3.13e-01 $ &       $^{   +6.7}_{   -6.7} $ &  $^{   +2.7}_{   -3.6} $ &  $^{   +2.1}_{   -2.2} $ &  $^{   -1.9}_{   +0.5} $ &  $^{   -0.0}_{   -0.1} $ &  $^{   -1.6}_{   +1.6} $ &  $^{   +0.0}_{   -0.0} $  \\ 
$   1025 $ &  $  0.27  $ & $ 363 $ &  $  2.12e-01 $ &       $^{   +5.2}_{   -5.2} $ &  $^{   +2.5}_{   -2.5} $ &  $^{   +1.6}_{   -1.7} $ &  $^{   -0.7}_{   +1.0} $ &  $^{   -0.0}_{   +0.1} $ &  $^{   +1.4}_{   -1.4} $ &  $^{   +0.0}_{   -0.0} $  \\ 

\hline

\end{tabular}
\end{center}
\renewcommand\thetable{1}
\caption[]
{The double-differential cross section, ${d^2\sigma}/{dxdQ^2}$, for NC $e^-p$ scattering at $\sqrt{s}=318 \units{GeV}$ as
a function of $Q^2$ and $x$. Also quoted are 
the number of events reconstructed and selected in 
the bin, $N$, the statistical uncertainty, $\delta_{\rm {stat}}$, the 
total systematic uncertainty,  $\delta_{\rm {sys}}$, the total uncorrelated
systematic uncertainty,  $\delta_{\rm u}$, followed by the bin-to-bin correlated systematic uncertainties, 
$\delta_1$--\,$\delta_4$, defined in Section~\ref{sec:sysunc}. 
%The upper limit on the cross section is given in case zero events are observed. 
The luminosity uncertainty of $1.8\%$ is not included.
This table has five continuations.}
\label{tab:e-pcros}
\end{table}

\begin{table} 
\begin{center} 
\begin{tabular}{|c|l|c|c|c|c||c|c|c|c|c|}
\multicolumn{11}{l}{
{\normalsize}}\\
\hline
{$Q^2$} & 
\multicolumn{1}{c|}{$x$} & 
$N$& 
${d^2\sigma}/{dxdQ^2}$  &
$\delta_{\rm{stat}}$& 
$\delta_{\rm{sys}}$& 
$\delta_{\rm{u}}$& 
$\delta_1$ &
$\delta_2$ &
$\delta_3$ &
$\delta_4$ \\
{($\units{GeV^2}$)} & 
$$ &
$$ &
{($ \units{pb/GeV^{2}}$)} &
(\%) &
(\%) &
(\%) &
(\%) &
(\%) &
(\%) &
(\%)  \\
\hline \hline

$   1200 $ &  $  0.06  $ & $ 552 $ &  $  1.46e+00 $ &       $^{   +4.3}_{   -4.3} $ &  $^{   +1.7}_{   -1.9} $ &  $^{   +1.4}_{   -1.4} $ &  $^{   -0.9}_{   +0.6} $ &  $^{   -0.3}_{   +0.4} $ &  $^{   -0.2}_{   +0.2} $ &  $^{   +0.1}_{   -0.1} $  \\ 
$   1200 $ &  $  0.07  $ & $ 454 $ &  $  1.02e+00 $ &       $^{   +4.7}_{   -4.7} $ &  $^{   +1.5}_{   -1.8} $ &  $^{   +1.4}_{   -1.4} $ &  $^{   -0.7}_{   +0.2} $ &  $^{   -0.1}_{   -0.0} $ &  $^{   +0.2}_{   -0.2} $ &  $^{   +0.1}_{   -0.1} $  \\ 
$   1200 $ &  $  0.09  $ & $ 412 $ &  $  7.52e-01 $ &       $^{   +4.9}_{   -4.9} $ &  $^{   +1.9}_{   -2.2} $ &  $^{   +1.5}_{   -1.5} $ &  $^{   -0.7}_{   +1.1} $ &  $^{   -0.3}_{   +0.0} $ &  $^{   +0.1}_{   -0.1} $ &  $^{   +0.1}_{   -0.1} $  \\ 
$   1200 $ &  $  0.12  $ & $ 430 $ &  $  5.39e-01 $ &       $^{   +4.8}_{   -4.8} $ &  $^{   +2.3}_{   -2.3} $ &  $^{   +1.5}_{   -1.4} $ &  $^{   -1.1}_{   +0.9} $ &  $^{   -0.1}_{   +0.1} $ &  $^{   +1.0}_{   -1.0} $ &  $^{   +0.1}_{   -0.1} $  \\ 
$   1200 $ &  $  0.14  $ & $ 326 $ &  $  4.22e-01 $ &       $^{   +5.5}_{   -5.5} $ &  $^{   +2.4}_{   -1.8} $ &  $^{   +1.9}_{   -1.7} $ &  $^{   -0.3}_{   +1.2} $ &  $^{   +0.0}_{   +0.1} $ &  $^{   -0.3}_{   +0.3} $ &  $^{   +0.1}_{   -0.1} $  \\
\hline

\end{tabular}
\end{center}
\renewcommand\thetable{1}
\caption[]
{Continuation 1.}
\label{tab:e-pcros_2}
\end{table}

\begin{table} 
\begin{center} 
\begin{tabular}{|c|l|c|c|c|c||c|c|c|c|c|}
\multicolumn{11}{l}{
{\normalsize}}\\
\hline
{$Q^2$} & 
\multicolumn{1}{c|}{$x$} & 
$N$& 
${d^2\sigma}/{dxdQ^2}$  &
$\delta_{\rm{stat}}$& 
$\delta_{\rm{sys}}$& 
$\delta_{\rm{u}}$& 
$\delta_1$ &
$\delta_2$ &
$\delta_3$ &
$\delta_4$ \\
{($\units{GeV^2}$)} & 
$$ &
$$ &
{($ \units{pb/GeV^{2}}$)} &
(\%) &
(\%) &
(\%) &
(\%) &
(\%) &
(\%) &
(\%)  \\
\hline \hline

$   1950 $ &  $  0.06  $ & $ 265 $ &  $  4.27e-01 $ &       $^{   +6.1}_{   -6.1} $ &  $^{   +2.2}_{   -2.4} $ &  $^{   +1.9}_{   -1.9} $ &  $^{   -1.1}_{   +0.8} $ &  $^{   -0.4}_{   +0.5} $ &  $^{   +0.4}_{   -0.4} $ &  $^{   +0.0}_{   -0.0} $  \\ 
$   1950 $ &  $  0.08  $ & $ 251 $ &  $  3.59e-01 $ &       $^{   +6.3}_{   -6.3} $ &  $^{   +2.4}_{   -2.2} $ &  $^{   +2.0}_{   -2.0} $ &  $^{   -0.8}_{   +0.9} $ &  $^{   -0.2}_{   +0.1} $ &  $^{   +0.5}_{   -0.5} $ &  $^{   +0.0}_{   -0.0} $  \\ 
$   1950 $ &  $  0.10  $ & $ 214 $ &  $  2.36e-01 $ &       $^{   +6.8}_{   -6.8} $ &  $^{   +2.2}_{   -2.4} $ &  $^{   +2.0}_{   -2.0} $ &  $^{   -0.8}_{   +0.2} $ &  $^{   -0.1}_{   +0.2} $ &  $^{   +0.7}_{   -0.7} $ &  $^{   +0.0}_{   -0.0} $  \\ 
$   1950 $ &  $  0.12  $ & $ 228 $ &  $  1.83e-01 $ &       $^{   +6.6}_{   -6.6} $ &  $^{   +2.3}_{   -2.3} $ &  $^{   +2.1}_{   -2.1} $ &  $^{   -0.9}_{   +0.8} $ &  $^{   -0.1}_{   +0.1} $ &  $^{   +0.3}_{   -0.3} $ &  $^{   +0.0}_{   -0.0} $  \\ 
$   1950 $ &  $  0.16  $ & $ 172 $ &  $  1.42e-01 $ &       $^{   +7.6}_{   -7.6} $ &  $^{   +2.6}_{   -3.0} $ &  $^{   +2.3}_{   -2.5} $ &  $^{   -1.4}_{   +0.8} $ &  $^{   -0.3}_{   +0.1} $ &  $^{   +0.1}_{   -0.1} $ &  $^{   +0.0}_{   -0.0} $  \\
\hline

\end{tabular}
\end{center}
\renewcommand\thetable{1}
\caption[]
{Continuation 2.}
\label{tab:e-pcros_3}
\end{table}

\begin{table} 
\begin{center} 
\begin{tabular}{|c|l|c|c|c|c||c|c|c|c|c|}
\multicolumn{11}{l}{
{\normalsize}}\\
\hline
{$Q^2$} & 
\multicolumn{1}{c|}{$x$} & 
$N$& 
${d^2\sigma}/{dxdQ^2}$  &
$\delta_{\rm{stat}}$& 
$\delta_{\rm{sys}}$& 
$\delta_{\rm{u}}$& 
$\delta_1$ &
$\delta_2$ &
$\delta_3$ &
$\delta_4$ \\
{($\units{GeV^2}$)} & 
$$ &
$$ &
{($ \units{pb/GeV^{2}}$)} &
(\%) &
(\%) &
(\%) &
(\%) &
(\%) &
(\%) &
(\%)  \\
\hline \hline

$   3000 $ &  $  0.10  $ & $ 104 $ &  $  9.11e-02 $ &       $^{   +9.8}_{   -9.8} $ &  $^{   +3.2}_{   -3.0} $ &  $^{   +2.9}_{   -2.8} $ &  $^{   -0.8}_{   +0.6} $ &  $^{   -0.7}_{   +0.4} $ &  $^{   +0.4}_{   -0.4} $ &  $^{   +0.0}_{   -0.0} $  \\ 
$   3000 $ &  $  0.12  $ & $ 106 $ &  $  6.44e-02 $ &       $^{   +9.7}_{   -9.7} $ &  $^{   +2.9}_{   -2.8} $ &  $^{   +2.7}_{   -2.7} $ &  $^{   -0.6}_{   +0.8} $ &  $^{   -0.0}_{   +0.2} $ &  $^{   -0.0}_{   +0.0} $ &  $^{   +0.0}_{   -0.0} $  \\ 
$   3000 $ &  $  0.16  $ & $ 77 $ &  $  4.65e-02 $ &       $^{  +11.4}_{  -11.4} $ &  $^{   +3.4}_{   -3.4} $ &  $^{   +3.4}_{   -3.2} $ &  $^{   -0.9}_{   +0.7} $ &  $^{   -0.1}_{   +0.2} $ &  $^{   -0.3}_{   +0.3} $ &  $^{   +0.0}_{   -0.0} $  \\ 
$   3000 $ &  $  0.19  $ & $ 82 $ &  $  3.81e-02 $ &       $^{  +11.0}_{  -11.0} $ &  $^{   +3.4}_{   -3.9} $ &  $^{   +3.1}_{   -3.2} $ &  $^{   -1.4}_{   +1.2} $ &  $^{   -0.2}_{   +0.1} $ &  $^{   +0.5}_{   -0.5} $ &  $^{   +0.0}_{   -0.0} $  \\ 
$   3000 $ &  $  0.23  $ & $ 68 $ &  $  3.15e-02 $ &       $^{  +12.1}_{  -12.1} $ &  $^{   +4.1}_{   -3.8} $ &  $^{   +3.8}_{   -3.8} $ &  $^{   +0.1}_{   +0.6} $ &  $^{   -0.4}_{   +0.3} $ &  $^{   -1.0}_{   +1.0} $ &  $^{   +0.0}_{   -0.0} $  \\
\hline

\end{tabular}
\end{center}
\renewcommand\thetable{1}
\caption[]
{Continuation 3.}
\label{tab:e-pcros_4}
\end{table}

\begin{table} 
\begin{center} 
\begin{tabular}{|c|l|c|c|c|c||c|c|c|c|c|}
\multicolumn{11}{l}{
{\normalsize}}\\
\hline
{$Q^2$} & 
\multicolumn{1}{c|}{$x$} & 
$N$& 
${d^2\sigma}/{dxdQ^2}$  &
$\delta_{\rm{stat}}$& 
$\delta_{\rm{sys}}$& 
$\delta_{\rm{u}}$& 
$\delta_1$ &
$\delta_2$ &
$\delta_3$ &
$\delta_4$ \\
{($\units{GeV^2}$)} & 
$$ &
$$ &
{($ \units{pb/GeV^{2}}$)} &
(\%) &
(\%) &
(\%) &
(\%) &
(\%) &
(\%) &
(\%)  \\
\hline \hline

$   5250 $ &  $  0.11  $ & $ 114 $ &  $  2.59e-02 $ &       $^{   +9.4}_{   -9.4} $ &  $^{   +3.4}_{   -3.2} $ &  $^{   +2.9}_{   -2.9} $ &  $^{   -0.6}_{   +1.1} $ &  $^{   -0.0}_{   +0.1} $ &  $^{   -1.3}_{   +1.3} $ &  $^{   +0.0}_{   -0.0} $  \\ 
$   5250 $ &  $  0.14  $ & $ 117 $ &  $  2.12e-02 $ &       $^{   +9.2}_{   -9.2} $ &  $^{   +3.6}_{   -3.4} $ &  $^{   +3.1}_{   -2.7} $ &  $^{   -1.8}_{   +1.6} $ &  $^{   -0.6}_{   +0.6} $ &  $^{   +0.4}_{   -0.4} $ &  $^{   +0.0}_{   -0.0} $  \\ 
$   5250 $ &  $  0.17  $ & $ 117 $ &  $  2.06e-02 $ &       $^{   +9.2}_{   -9.2} $ &  $^{   +3.2}_{   -3.4} $ &  $^{   +3.1}_{   -3.1} $ &  $^{   -1.2}_{   +0.5} $ &  $^{   -0.5}_{   +0.3} $ &  $^{   +0.2}_{   -0.2} $ &  $^{   +0.0}_{   -0.0} $  \\ 
$   5250 $ &  $  0.20  $ & $ 91 $ &  $  1.11e-02 $ &       $^{  +10.5}_{  -10.5} $ &  $^{   +3.3}_{   -3.1} $ &  $^{   +3.1}_{   -2.9} $ &  $^{   -0.5}_{   +1.0} $ &  $^{   +0.0}_{   +0.2} $ &  $^{   -0.7}_{   +0.7} $ &  $^{   +0.0}_{   -0.0} $  \\ 
$   5250 $ &  $  0.25  $ & $ 92 $ &  $  9.27e-03 $ &       $^{  +10.4}_{  -10.4} $ &  $^{   +3.6}_{   -4.2} $ &  $^{   +3.1}_{   -3.2} $ &  $^{   -1.1}_{   +1.5} $ &  $^{   -0.4}_{   +0.1} $ &  $^{   +0.4}_{   -0.4} $ &  $^{   +0.0}_{   -0.0} $  \\
\hline

\end{tabular}
\end{center}
\renewcommand\thetable{1}
\caption[]
{Continuation 4.}
\label{tab:e-pcros_5}
\end{table}

\begin{table} 
\begin{center} 
\begin{tabular}{|c|l|c|c|c|c||c|c|c|c|c|}
\multicolumn{11}{l}{
{\normalsize}}\\
\hline
{$Q^2$} & 
\multicolumn{1}{c|}{$x$} & 
$N$& 
${d^2\sigma}/{dxdQ^2}$  &
$\delta_{\rm{stat}}$& 
$\delta_{\rm{sys}}$& 
$\delta_{\rm{u}}$& 
$\delta_1$ &
$\delta_2$ &
$\delta_3$ &
$\delta_4$ \\
{($\units{GeV^2}$)} & 
$$ &
$$ &
{($ \units{pb/GeV^{2}}$)} &
(\%) &
(\%) &
(\%) &
(\%) &
(\%) &
(\%) &
(\%)  \\
\hline \hline

$  15500 $ &  $  0.43  $ & $ 120 $ &  $  3.22e-04 $ &       $^{   +9.1}_{   -9.1} $ &  $^{   +5.5}_{   -5.1} $ &  $^{   +3.0}_{   -3.0} $ &  $^{   -2.8}_{   +3.0} $ &  $^{   -0.8}_{   +0.7} $ &  $^{   -0.7}_{   +0.7} $ &  $^{   +0.1}_{   -0.1} $  \\ 
$  15500 $ &  $  0.80  $ & $ 8 $ &  $  3.39e-06 $ &       $^{  +41.8}_{  -29.4} $ &  $^{  +15.3}_{  -14.9} $ &  $^{  +12.9}_{  -12.9} $ &  $^{   -5.3}_{   +4.7} $ &  $^{   +0.0}_{   +0.4} $ &  $^{   +3.2}_{   -3.2} $ &  $^{   +3.0}_{   -3.0} $  \\

\hline

\end{tabular}
\end{center}
\renewcommand\thetable{1}
\caption[]
{Continuation 5.}
\label{tab:e-pcros_6}
\end{table}

%% file: table_mine.tex
\begin{table} 
\begin{center} 
\begin{tabular}{|c|l|c|c|c|c||c|c|c|c|c|}
\multicolumn{11}{l}{
{\normalsize}}\\
\hline
{$Q^2$} & 
\multicolumn{1}{c|}{$x$} & 
$N$& 
${d^2\sigma}/{dxdQ^2}$  &
$\delta_{\rm {stat}}$& 
$\delta_{\rm {sys}}$& 
$\delta_{\rm u}$& 
$\delta_1$ &
$\delta_2$ &
$\delta_3$ &
$\delta_4$ \\
{($\units{GeV^2}$)} & 
$$ &
$$ &
{($ \units{pb/GeV^{2}}$)} &
(\%) &
(\%) &
(\%) &
(\%) &
(\%) &
(\%) &
(\%)  \\
\hline \hline

$    725 $ &  $  0.06  $ & $ 567 $ &  $  3.31e+00 $ &       $^{   +4.2}_{   -4.2} $ &  $^{   +1.6}_{   -1.8} $ &  $^{   +1.3}_{   -1.3} $ &  $^{   +0.8}_{   -1.0} $ &  $^{   +0.0}_{   +0.0} $ &  $^{   -0.1}_{   +0.1} $ &  $^{   +0.1}_{   -0.1} $  \\ 
$    725 $ &  $  0.08  $ & $ 418 $ &  $  2.42e+00 $ &       $^{   +4.9}_{   -4.9} $ &  $^{   +1.9}_{   -2.0} $ &  $^{   +1.5}_{   -1.4} $ &  $^{   +1.1}_{   -0.9} $ &  $^{   -0.2}_{   +0.0} $ &  $^{   -0.3}_{   +0.3} $ &  $^{   +0.1}_{   -0.1} $  \\ 
$    725 $ &  $  0.10  $ & $ 321 $ &  $  1.81e+00 $ &       $^{   +5.6}_{   -5.6} $ &  $^{   +2.0}_{   -2.1} $ &  $^{   +1.7}_{   -1.7} $ &  $^{   +0.9}_{   -1.2} $ &  $^{   -0.0}_{   +0.4} $ &  $^{   -0.1}_{   +0.1} $ &  $^{   +0.1}_{   -0.1} $  \\ 
$    725 $ &  $  0.12  $ & $ 274 $ &  $  1.27e+00 $ &       $^{   +6.0}_{   -6.0} $ &  $^{   +2.5}_{   -2.8} $ &  $^{   +2.1}_{   -1.8} $ &  $^{   +1.4}_{   -1.8} $ &  $^{   -0.1}_{   -0.1} $ &  $^{   -0.1}_{   +0.1} $ &  $^{   +0.1}_{   -0.1} $  \\ 
$    725 $ &  $  0.16  $ & $ 207 $ &  $  1.04e+00 $ &       $^{   +7.0}_{   -7.0} $ &  $^{   +3.9}_{   -3.2} $ &  $^{   +2.4}_{   -2.1} $ &  $^{   +1.8}_{   -1.4} $ &  $^{   -0.1}_{   +0.2} $ &  $^{   +1.9}_{   -1.9} $ &  $^{   +0.1}_{   -0.1} $  \\
\hline

\end{tabular}
\end{center}
\renewcommand\thetable{2}
\caption[]
{The double-differential cross section, ${d^2\sigma}/{dxdQ^2}$, for NC $e^+p$ scattering at $\sqrt{s}=318 \units{GeV}$ as
    a function of $Q^2$ and $x$. Also quoted are 
  the number of events reconstructed and selected in 
    the bin, $N$, the statistical uncertainty, $\delta_{\rm{stat}}$, the 
     total systematic uncertainty, $\delta_{\rm{sys}}$, the total uncorrelated
    systematic uncertainty, $\delta_{\rm u}$, followed by the bin-to-bin correlated
    systematic uncertainties, $\delta_1$--\,$\delta_4$, defined in Section~\ref{sec:sysunc}. The upper limit on the cross section is given if no event is observed. The luminosity uncertainty of $1.8\%$ is not included.
This table has five continuations.}
\label{tab:e+pcros}
\end{table}

\begin{table} 
\begin{center} 
\begin{tabular}{|c|l|c|c|c|c||c|c|c|c|c|}
\multicolumn{11}{l}{
{\normalsize}}\\
\hline
{$Q^2$} & 
\multicolumn{1}{c|}{$x$} & 
$N$& 
${d^2\sigma}/{dxdQ^2}$  &
$\delta_{\rm{stat}}$& 
$\delta_{\rm{sys}}$& 
$\delta_{\rm{u}}$& 
$\delta_1$ &
$\delta_2$ &
$\delta_3$ &
$\delta_4$ \\
{($\units{GeV^2}$)} & 
$$ &
$$ &
{($ \units{pb/GeV^{2}}$)} &
(\%) &
(\%) &
(\%) &
(\%) &
(\%) &
(\%) &
(\%)  \\
\hline \hline

$   1200 $ &  $  0.06  $ & $ 377 $ &  $  1.35e+00 $ &       $^{   +5.2}_{   -5.2} $ &  $^{   +2.1}_{   -2.3} $ &  $^{   +1.7}_{   -1.7} $ &  $^{   -1.0}_{   +0.9} $ &  $^{   -0.5}_{   +0.5} $ &  $^{   -0.5}_{   +0.5} $ &  $^{   +0.1}_{   -0.1} $  \\ 
$   1200 $ &  $  0.07  $ & $ 314 $ &  $  9.50e-01 $ &       $^{   +5.6}_{   -5.6} $ &  $^{   +1.9}_{   -2.5} $ &  $^{   +1.7}_{   -1.7} $ &  $^{   -1.2}_{   +0.2} $ &  $^{   -0.0}_{   -0.0} $ &  $^{   +0.7}_{   -0.7} $ &  $^{   +0.1}_{   -0.1} $  \\ 
$   1200 $ &  $  0.09  $ & $ 285 $ &  $  6.97e-01 $ &       $^{   +5.9}_{   -5.9} $ &  $^{   +1.9}_{   -2.1} $ &  $^{   +1.7}_{   -1.7} $ &  $^{   -0.8}_{   +0.4} $ &  $^{   -0.2}_{   +0.2} $ &  $^{   -0.2}_{   +0.2} $ &  $^{   +0.1}_{   -0.1} $  \\ 
$   1200 $ &  $  0.12  $ & $ 325 $ &  $  5.33e-01 $ &       $^{   +5.5}_{   -5.5} $ &  $^{   +2.7}_{   -2.2} $ &  $^{   +1.9}_{   -1.8} $ &  $^{   -1.2}_{   +1.2} $ &  $^{   +0.3}_{   +0.0} $ &  $^{   -0.6}_{   +0.6} $ &  $^{   +0.1}_{   -0.1} $  \\ 
$   1200 $ &  $  0.14  $ & $ 213 $ &  $  3.76e-01 $ &       $^{   +6.9}_{   -6.9} $ &  $^{   +2.2}_{   -3.3} $ &  $^{   +2.0}_{   -2.1} $ &  $^{   -0.5}_{   -0.4} $ &  $^{   -0.5}_{   +0.3} $ &  $^{   -0.9}_{   +0.9} $ &  $^{   +0.1}_{   -0.1} $  \\
\hline

\end{tabular}
\end{center}
\renewcommand\thetable{2}
\caption[]
{Continuation 1.}
\label{tab:e-pcros_2}
\end{table}

\begin{table} 
\begin{center} 
\begin{tabular}{|c|l|c|c|c|c||c|c|c|c|c|}
\multicolumn{11}{l}{
{\normalsize}}\\
\hline
{$Q^2$} & 
\multicolumn{1}{c|}{$x$} & 
$N$& 
${d^2\sigma}/{dxdQ^2}$  &
$\delta_{\rm{stat}}$& 
$\delta_{\rm{sys}}$& 
$\delta_{\rm{u}}$& 
$\delta_1$ &
$\delta_2$ &
$\delta_3$ &
$\delta_4$ \\
{($\units{GeV^2}$)} & 
$$ &
$$ &
{($ \units{pb/GeV^{2}}$)} &
(\%) &
(\%) &
(\%) &
(\%) &
(\%) &
(\%) &
(\%)  \\
\hline \hline

$   1950 $ &  $  0.06  $ & $ 180 $ &  $  3.86e-01 $ &       $^{   +7.5}_{   -7.5} $ &  $^{   +2.9}_{   -2.6} $ &  $^{   +2.3}_{   -2.3} $ &  $^{   -1.0}_{   +1.3} $ &  $^{   -0.2}_{   +0.1} $ &  $^{   -0.6}_{   +0.6} $ &  $^{   +0.0}_{   -0.0} $  \\ 
$   1950 $ &  $  0.08  $ & $ 161 $ &  $  3.07e-01 $ &       $^{   +7.9}_{   -7.9} $ &  $^{   +2.6}_{   -2.7} $ &  $^{   +2.3}_{   -2.4} $ &  $^{   -1.0}_{   +0.8} $ &  $^{   -0.4}_{   +0.3} $ &  $^{   -0.6}_{   +0.6} $ &  $^{   +0.0}_{   -0.0} $  \\ 
$   1950 $ &  $  0.10  $ & $ 161 $ &  $  2.33e-01 $ &       $^{   +7.9}_{   -7.9} $ &  $^{   +2.8}_{   -2.8} $ &  $^{   +2.3}_{   -2.4} $ &  $^{   +0.0}_{   +0.8} $ &  $^{   -0.2}_{   +0.3} $ &  $^{   -1.1}_{   +1.1} $ &  $^{   +0.0}_{   -0.0} $  \\ 
$   1950 $ &  $  0.12  $ & $ 146 $ &  $  1.55e-01 $ &       $^{   +8.3}_{   -8.3} $ &  $^{   +2.5}_{   -2.6} $ &  $^{   +2.4}_{   -2.3} $ &  $^{   -1.0}_{   +0.6} $ &  $^{   -0.2}_{   +0.2} $ &  $^{   -0.4}_{   +0.4} $ &  $^{   +0.0}_{   -0.0} $  \\ 
$   1950 $ &  $  0.16  $ & $ 117 $ &  $  1.29e-01 $ &       $^{   +9.2}_{   -9.2} $ &  $^{   +3.1}_{   -2.9} $ &  $^{   +2.7}_{   -2.7} $ &  $^{   -0.9}_{   +1.2} $ &  $^{   -0.1}_{   +0.1} $ &  $^{   -0.3}_{   +0.3} $ &  $^{   +0.0}_{   -0.0} $  \\
\hline

\end{tabular}
\end{center}
\renewcommand\thetable{2}
\caption[]
{Continuation 2.}
\label{tab:e-pcros_3}
\end{table}

\begin{table} 
\begin{center} 
\begin{tabular}{|c|l|c|c|c|c||c|c|c|c|c|}
\multicolumn{11}{l}{
{\normalsize}}\\
\hline
{$Q^2$} & 
\multicolumn{1}{c|}{$x$} & 
$N$& 
${d^2\sigma}/{dxdQ^2}$  &
$\delta_{\rm{stat}}$& 
$\delta_{\rm{sys}}$& 
$\delta_{\rm{u}}$& 
$\delta_1$ &
$\delta_2$ &
$\delta_3$ &
$\delta_4$ \\
{($\units{GeV^2}$)} & 
$$ &
$$ &
{($ \units{pb/GeV^{2}}$)} &
(\%) &
(\%) &
(\%) &
(\%) &
(\%) &
(\%) &
(\%)  \\
\hline \hline

$   3000 $ &  $  0.10  $ & $ 75 $ &  $  8.89e-02 $ &       $^{  +11.5}_{  -11.5} $ &  $^{   +3.7}_{   -3.5} $ &  $^{   +3.5}_{   -3.5} $ &  $^{   -0.1}_{   +0.9} $ &  $^{   -0.5}_{   +0.7} $ &  $^{   +0.0}_{   -0.0} $ &  $^{   +0.0}_{   -0.0} $  \\ 
$   3000 $ &  $  0.12  $ & $ 99 $ &  $  8.21e-02 $ &       $^{  +10.1}_{  -10.1} $ &  $^{   +3.4}_{   -3.4} $ &  $^{   +3.3}_{   -3.3} $ &  $^{   -1.0}_{   +0.6} $ &  $^{   -0.2}_{   -0.0} $ &  $^{   +0.3}_{   -0.3} $ &  $^{   +0.0}_{   -0.0} $  \\ 
$   3000 $ &  $  0.16  $ & $ 79 $ &  $  6.51e-02 $ &       $^{  +11.3}_{  -11.3} $ &  $^{   +3.9}_{   -4.4} $ &  $^{   +3.7}_{   -3.8} $ &  $^{   -1.8}_{   +0.9} $ &  $^{   -0.1}_{   +0.1} $ &  $^{   +0.5}_{   -0.5} $ &  $^{   +0.0}_{   -0.0} $  \\ 
$   3000 $ &  $  0.19  $ & $ 61 $ &  $  3.67e-02 $ &       $^{  +12.8}_{  -12.8} $ &  $^{   +3.8}_{   -4.4} $ &  $^{   +3.6}_{   -3.7} $ &  $^{   -0.1}_{   +0.9} $ &  $^{   -0.1}_{   +0.1} $ &  $^{   -0.0}_{   +0.0} $ &  $^{   +0.0}_{   -0.0} $  \\ 
$   3000 $ &  $  0.23  $ & $ 48 $ &  $  3.03e-02 $ &       $^{  +14.4}_{  -14.4} $ &  $^{   +4.4}_{   -5.1} $ &  $^{   +4.3}_{   -4.3} $ &  $^{   -2.5}_{   +0.4} $ &  $^{   -0.3}_{   +0.4} $ &  $^{   +0.6}_{   -0.6} $ &  $^{   +0.0}_{   -0.0} $  \\
\hline

\end{tabular}
\end{center}
\renewcommand\thetable{2}
\caption[]
{Continuation 3.}
\label{tab:e-pcros_4}
\end{table}

\begin{table} 
\begin{center} 
\begin{tabular}{|c|l|c|c|c|c||c|c|c|c|c|}
\multicolumn{11}{l}{
{\normalsize}}\\
\hline
{$Q^2$} & 
\multicolumn{1}{c|}{$x$} & 
$N$& 
${d^2\sigma}/{dxdQ^2}$  &
$\delta_{\rm{stat}}$& 
$\delta_{\rm{sys}}$& 
$\delta_{\rm{u}}$& 
$\delta_1$ &
$\delta_2$ &
$\delta_3$ &
$\delta_4$ \\
{($\units{GeV^2}$)} & 
$$ &
$$ &
{($ \units{pb/GeV^{2}}$)} &
(\%) &
(\%) &
(\%) &
(\%) &
(\%) &
(\%) &
(\%)  \\
\hline \hline

$   5250 $ &  $  0.11  $ & $ 60 $ &  $  1.87e-02 $ &       $^{  +12.9}_{  -12.9} $ &  $^{   +4.1}_{   -4.8} $ &  $^{   +3.7}_{   -3.7} $ &  $^{   -1.1}_{   +1.1} $ &  $^{   -0.3}_{   -0.2} $ &  $^{   +1.4}_{   -1.4} $ &  $^{   +0.0}_{   -0.0} $  \\ 
$   5250 $ &  $  0.14  $ & $ 74 $ &  $  1.81e-02 $ &       $^{  +11.6}_{  -11.6} $ &  $^{   +4.7}_{   -3.9} $ &  $^{   +3.6}_{   -3.6} $ &  $^{   -1.3}_{   +1.6} $ &  $^{   -0.5}_{   +1.0} $ &  $^{   -0.5}_{   +0.5} $ &  $^{   +0.0}_{   -0.0} $  \\ 
$   5250 $ &  $  0.17  $ & $ 60 $ &  $  1.35e-02 $ &       $^{  +12.9}_{  -12.9} $ &  $^{   +4.5}_{   -4.0} $ &  $^{   +3.9}_{   -3.9} $ &  $^{   -0.8}_{   +1.4} $ &  $^{   -0.2}_{   +0.2} $ &  $^{   -0.3}_{   +0.3} $ &  $^{   +0.0}_{   -0.0} $  \\ 
$   5250 $ &  $  0.20  $ & $ 50 $ &  $  8.51e-03 $ &       $^{  +14.1}_{  -14.1} $ &  $^{   +4.2}_{   -4.5} $ &  $^{   +3.8}_{   -3.8} $ &  $^{   -1.7}_{   +1.0} $ &  $^{   -0.3}_{   +0.3} $ &  $^{   +1.0}_{   -1.0} $ &  $^{   +0.0}_{   -0.0} $  \\ 
$   5250 $ &  $  0.25  $ & $ 66 $ &  $  8.84e-03 $ &       $^{  +12.3}_{  -12.3} $ &  $^{   +4.2}_{   -4.1} $ &  $^{   +3.9}_{   -3.9} $ &  $^{   -0.9}_{   +1.4} $ &  $^{   -0.3}_{   +0.1} $ &  $^{   -0.3}_{   +0.3} $ &  $^{   +0.0}_{   -0.0} $  \\
\hline

\end{tabular}
\end{center}
\renewcommand\thetable{2}
\caption[]
{Continuation 4.}
\label{tab:e-pcros_5}
\end{table}

\begin{table} 
\begin{center} 
\begin{tabular}{|c|l|c|c|c|c||c|c|c|c|c|}
\multicolumn{11}{l}{
{\normalsize}}\\
\hline
{$Q^2$} & 
\multicolumn{1}{c|}{$x$} & 
$N$& 
${d^2\sigma}/{dxdQ^2}$  &
$\delta_{\rm{stat}}$& 
$\delta_{\rm{sys}}$& 
$\delta_{\rm{u}}$& 
$\delta_1$ &
$\delta_2$ &
$\delta_3$ &
$\delta_4$ \\
{($\units{GeV^2}$)} & 
$$ &
$$ &
{($ \units{pb/GeV^{2}}$)} &
(\%) &
(\%) &
(\%) &
(\%) &
(\%) &
(\%) &
(\%)  \\
\hline \hline

$  15500 $ &  $  0.43  $ & $ 41 $ &  $  1.57e-04 $ &       $^{  +15.6}_{  -15.6} $ &  $^{   +6.6}_{   -5.4} $ &  $^{   +4.8}_{   -4.5} $ &  $^{   -2.6}_{   +2.7} $ &  $^{   -0.6}_{   +0.6} $ &  $^{   -0.3}_{   +0.3} $ &  $^{   +0.1}_{   -0.1} $  \\ 
$  15500 $ &  $  0.80  $ & $ 3 $ &  $  1.72e-06 $ &       $^{  +75.7}_{  -42.3} $ &  $^{  +18.5}_{  -17.7} $ &  $^{  +16.7}_{  -16.7} $ &  $^{   -4.8}_{   +5.1} $ &  $^{   -0.6}_{   +0.2} $ &  $^{   -0.6}_{   +0.6} $ &  $^{   +2.4}_{   -2.4} $  \\

\hline

\end{tabular}
\end{center}
\renewcommand\thetable{2}
\caption[]
{Continuation 5.}
\label{tab:e-pcros_6}
\end{table}

%% file: table_mine_eMp_int.tex
\begin{table} 
\begin{center}
\begin{tabular}{|c|l|c|c|c|c||c|c|c|c|c|}
\multicolumn{11}{l}{
{\normalsize}}\\
\hline
{$Q^2$} & 
\multicolumn{1}{c|}{$x_{\rm edge}$} & 
$N$& 
$ I(x)$  &
$\delta_{\rm {stat}}$& 
$\delta_{\rm {sys}}$& 
$\delta_{\rm u}$& 
$\delta_1$ &
$\delta_2$ &
$\delta_3$ &
$\delta_4$ \\
{($\units{GeV^2}$)} & 
$$ &
$$ &
{($ \units{pb/GeV^{2}}$)} &
(\%) &
(\%) &
(\%) &
(\%) &
(\%) &
(\%) &
(\%)  \\
\hline \hline

 $    725 $ &  $  0.63  $ & $ 504 $ &   $  7.71e-02 $ &       $^{   +4.5}_{   -4.5} $ &  $^{   +2.8}_{   -3.2} $ &  $^{   +1.5}_{   -1.3} $ &  $^{   +1.3}_{   -2.0} $ &  $^{   +0.0}_{   -0.1} $ &  $^{   +1.9}_{   -1.9} $ &  $^{   +0.3}_{   -0.3} $  \\ 
 $    875 $ &  $  0.64  $ & $ 671 $ &   $  5.12e-02 $ &       $^{   +3.9}_{   -3.9} $ &  $^{   +2.3}_{   -1.9} $ &  $^{   +1.2}_{   -1.2} $ &  $^{   -0.1}_{   +1.0} $ &  $^{   +0.0}_{   -0.0} $ &  $^{   +1.3}_{   -1.3} $ &  $^{   +0.5}_{   -0.5} $  \\ 
 $   1025 $ &  $  0.66  $ & $ 414 $ &   $  2.75e-02 $ &       $^{   +4.9}_{   -4.9} $ &  $^{   +3.4}_{   -3.6} $ &  $^{   +1.5}_{   -1.5} $ &  $^{   -1.6}_{   +0.8} $ &  $^{   +0.0}_{   -0.0} $ &  $^{   +2.7}_{   -2.7} $ &  $^{   +0.6}_{   -0.6} $  \\ 
 $   1200 $ &  $  0.67  $ & $ 368 $ &   $  1.80e-02 $ &       $^{   +5.2}_{   -5.2} $ &  $^{   +3.7}_{   -2.9} $ &  $^{   +1.7}_{   -1.6} $ &  $^{   -1.5}_{   +2.5} $ &  $^{   +0.0}_{   -0.0} $ &  $^{   +1.4}_{   -1.4} $ &  $^{   +0.8}_{   -0.8} $  \\ 
 $   1400 $ &  $  0.68  $ & $ 202 $ &   $  1.04e-02 $ &       $^{   +7.0}_{   -7.0} $ &  $^{   +3.5}_{   -3.8} $ &  $^{   +2.1}_{   -2.0} $ &  $^{   -1.8}_{   +1.3} $ &  $^{   +0.0}_{   -0.0} $ &  $^{   +2.1}_{   -2.1} $ &  $^{   +1.0}_{   -1.0} $  \\
\hline

\end{tabular}
\end{center}
\renewcommand\thetable{3}
\caption[]
{The integrated cross section, $ I(x)$ (see Eq.(\ref{eqn-I(x)})),  for NC $e^-p$ scattering at $\sqrt{s}=318 \units{GeV}$ as a function of
     $Q^2$. Also quoted are the lower limit of integration, $x_{\rm edge}$,  the number of events reconstructed in
    the bin, $N$, the
    statistical uncertainty,  $\delta_{\rm {stat}}$, the
     total systematic uncertainty, $\delta_{\rm {sys}}$,  the total uncorrelated
    systematic uncertainty, $\delta_{\rm u}$, followed by the bin-to-bin correlated
    systematic uncertainties, $\delta_1$--\,$\delta_4$ defined in Section~\ref{sec:sysunc}. 
%In case zero events are observed, the 68\% upper limit is given.
    The luminosity uncertainty of $1.8\%$ is not included.}
\label{tab:2004-05cros}
\end{table}

%% file: table_mine_int.tex
\begin{table} 
\begin{center} 
\begin{tabular}{|c|l|c|c|c|c||c|c|c|c|c|}
\multicolumn{11}{l}{
{\normalsize}}\\
\hline
{$Q^2$} & 
\multicolumn{1}{c|}{$x_{\rm edge}$} & 
$N$& 
${ I(x) }$  &
$\delta_{\rm {stat}}$& 
$\delta_{\rm {sys}}$& 
$\delta_{\rm u}$& 
$\delta_1$ &
$\delta_2$ &
$\delta_3$ &
$\delta_4$ \\
{($\units{GeV^2}$)} & 
$$ &
$$ &
{($ \units{pb/GeV^{2}}$)} &
(\%) &
(\%) &
(\%) &
(\%) &
(\%) &
(\%) &
(\%)  \\
\hline \hline

  $    725 $ &  $  0.63  $ & $ 371 $ &   $  7.50e-02 $ &       $^{   +5.2}_{   -5.2} $ &  $^{   +2.4}_{   -2.3} $ &  $^{   +1.7}_{   -1.4} $ &  $^{   +1.2}_{   -1.3} $ &  $^{   +0.1}_{   -0.1} $ &  $^{   +0.8}_{   -0.8} $ &  $^{   +0.3}_{   -0.3} $  \\ 
 $    875 $ &  $  0.64  $ & $ 482 $ &   $  4.81e-02 $ &       $^{   +4.6}_{   -4.6} $ &  $^{   +1.9}_{   -1.7} $ &  $^{   +1.6}_{   -1.3} $ &  $^{   -0.5}_{   +0.6} $ &  $^{   +0.1}_{   -0.1} $ &  $^{   -0.1}_{   +0.1} $ &  $^{   +0.4}_{   -0.4} $  \\ 
 $   1025 $ &  $  0.66  $ & $ 281 $ &   $  2.50e-02 $ &       $^{   +6.0}_{   -6.0} $ &  $^{   +3.0}_{   -2.4} $ &  $^{   +1.7}_{   -1.6} $ &  $^{   -1.3}_{   +1.8} $ &  $^{   +0.0}_{   -0.0} $ &  $^{   +0.3}_{   -0.3} $ &  $^{   +0.5}_{   -0.5} $  \\ 
 $   1200 $ &  $  0.67  $ & $ 275 $ &   $  1.80e-02 $ &       $^{   +6.0}_{   -6.0} $ &  $^{   +3.1}_{   -2.9} $ &  $^{   +2.0}_{   -1.8} $ &  $^{   -1.5}_{   +1.8} $ &  $^{   +0.0}_{   -0.0} $ &  $^{   +0.9}_{   -0.9} $ &  $^{   +0.7}_{   -0.7} $  \\ 
 $   1400 $ &  $  0.68  $ & $ 146 $ &   $  9.99e-03 $ &       $^{   +8.3}_{   -8.3} $ &  $^{   +3.5}_{   -3.7} $ &  $^{   +2.4}_{   -2.3} $ &  $^{   -1.8}_{   +1.7} $ &  $^{   +0.0}_{   -0.0} $ &  $^{   +1.5}_{   -1.5} $ &  $^{   +0.8}_{   -0.8} $  \\ 
 $   1650 $ &  $  0.69  $ & $ 115 $ &   $  5.24e-03 $ &       $^{   +9.3}_{   -9.3} $ &  $^{   +3.5}_{   -3.5} $ &  $^{   +2.5}_{   -2.5} $ &  $^{   -1.9}_{   +1.7} $ &  $^{   +0.0}_{   -0.1} $ &  $^{   +0.7}_{   -0.7} $ &  $^{   +1.1}_{   -1.1} $  \\ 
 $   1950 $ &  $  0.71  $ & $ 62 $ &   $  2.86e-03 $ &       $^{  +12.7}_{  -12.7} $ &  $^{   +5.6}_{   -5.0} $ &  $^{   +3.8}_{   -3.4} $ &  $^{   -2.2}_{   +3.0} $ &  $^{   +0.1}_{   -0.0} $ &  $^{   +2.3}_{   -2.3} $ &  $^{   +1.4}_{   -1.4} $  \\ 
 $   2250 $ &  $  0.73  $ & $ 31 $ &   $  1.47e-03 $ &       $^{  +18.0}_{  -18.0} $ &  $^{   +5.6}_{   -5.9} $ &  $^{   +4.6}_{   -4.5} $ &  $^{   -2.5}_{   +2.1} $ &  $^{   +0.0}_{   -0.1} $ &  $^{   +1.6}_{   -1.6} $ &  $^{   +1.7}_{   -1.7} $  \\ 
 $   2600 $ &  $  0.75  $ & $ 27 $ &   $  9.47e-04 $ &       $^{  +19.2}_{  -19.2} $ &  $^{   +6.7}_{   -6.7} $ &  $^{   +5.4}_{   -5.3} $ &  $^{   -2.2}_{   +2.3} $ &  $^{   +0.0}_{   -0.1} $ &  $^{   +2.6}_{   -2.6} $ &  $^{   +1.9}_{   -1.9} $  \\ 
 $   3000 $ &  $  0.77  $ & $ 13 $ &   $  4.63e-04 $ &       $^{  +31.5}_{  -23.9} $ &  $^{  +10.4}_{  -10.7} $ &  $^{   +7.6}_{   -7.5} $ &  $^{   -3.1}_{   +2.4} $ &  $^{   +0.0}_{   +0.0} $ &  $^{   +6.3}_{   -6.3} $ &  $^{   +2.1}_{   -2.1} $  \\ 
 $   3500 $ &  $  0.79  $ & $ 11 $ &   $  2.61e-04 $ &       $^{  +34.7}_{  -25.7} $ &  $^{  +10.9}_{  -10.6} $ &  $^{   +8.7}_{   -8.7} $ &  $^{   -4.0}_{   +4.7} $ &  $^{   +0.1}_{   -0.3} $ &  $^{   +3.5}_{   -3.5} $ &  $^{   +2.4}_{   -2.4} $  \\ 
 $   4150 $ &  $  0.81  $ & $ 6 $ &   $  1.29e-04 $ &       $^{  +49.5}_{  -32.8} $ &  $^{  +13.5}_{  -14.8} $ &  $^{  +13.2}_{  -13.1} $ &  $^{   -3.2}_{   +0.7} $ &  $^{   +0.0}_{   +0.0} $ &  $^{   +0.0}_{   -0.0} $ &  $^{   +2.8}_{   -2.8} $  \\ 
 $   5250 $ &  $  0.85  $ & $ 3 $ &   $  2.99e-05 $ &       $^{  +75.7}_{  -42.3} $ &  $^{  +20.4}_{  -19.6} $ &  $^{  +17.1}_{  -16.7} $ &  $^{   -3.5}_{   +5.7} $ &  $^{   -0.2}_{   +0.0} $ &  $^{   +8.5}_{   -8.5} $ &  $^{   +3.7}_{   -3.7} $  \\ 
$   7000 $ &  $  0.87  $ & $ 0 $ &  $ < 9.68e-06 $ &  &  &  &  &  &   &\\ 
$   9500 $ &  $  0.89  $ & $ 0 $ &  $ < 2.10e-06 $ &  &  &  &  &  &   &\\

\hline

\end{tabular}
\end{center}
\renewcommand\thetable{4}
\caption[]
{The integrated cross section, $ I(x)$ (see Eq.(\ref{eqn-I(x)})),  for NC $e^+p$ scattering at $\sqrt{s}=318 \units{GeV}$ as a function of
     $Q^2$. Also quoted are the lower limit of integration, $x_{\rm edge}$,  the number of events reconstructed in
    the bin, $N$, the
    statistical uncertainty,  $\delta_{\rm {stat}}$, the
     total systematic uncertainty, $\delta_{\rm {sys}}$,  the total uncorrelated
    systematic uncertainty, $\delta_{\rm u}$, followed by the bin-to-bin correlated
    systematic uncertainties, $\delta_1$--\,$\delta_4$ defined in Section~\ref{sec:sysunc}. If no event is observed, the 68\% upper limit is given.
    The luminosity uncertainty of $1.8\%$ is not included.}

\label{tab:2006-07crosINT}
\end{table}

%% file: DESY-13-245-fig.tex
% calib plots plots eMp

\captionsetup[subfloat]{font={bf}, captionskip=-0.4cm, margin=3.7cm}

\begin{figure}[h!]
\begin{center}
\vspace{-1cm}
\includegraphics[height=0.06\textheight]{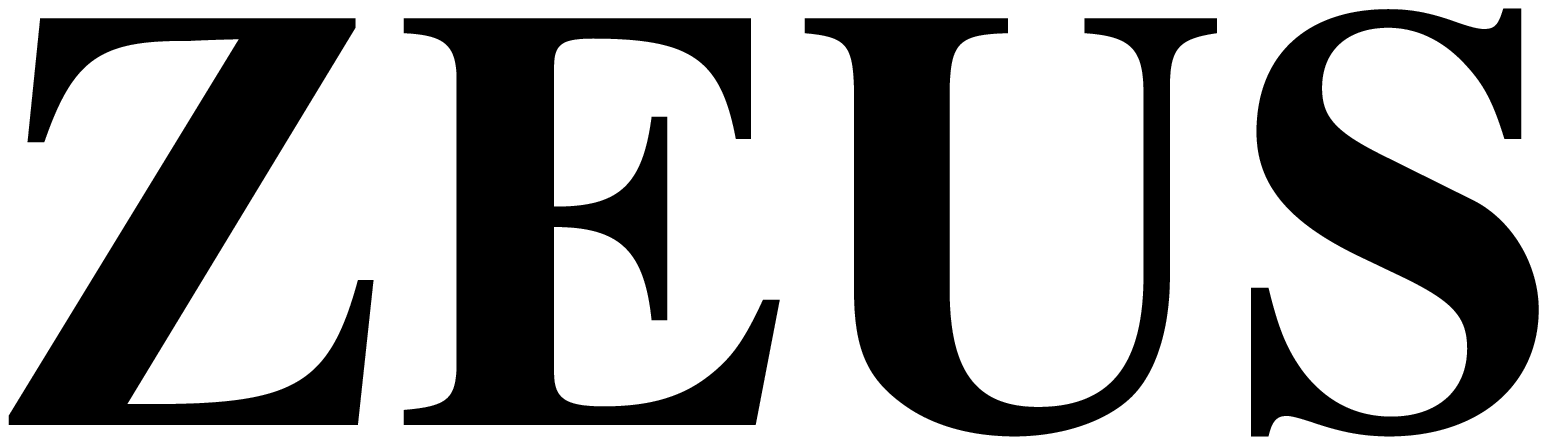}\\
\showcaptionsetup{subfloat} \subfloat[]{\label{fig:KP+jetscale:a}\includegraphics[height=0.25\textheight]{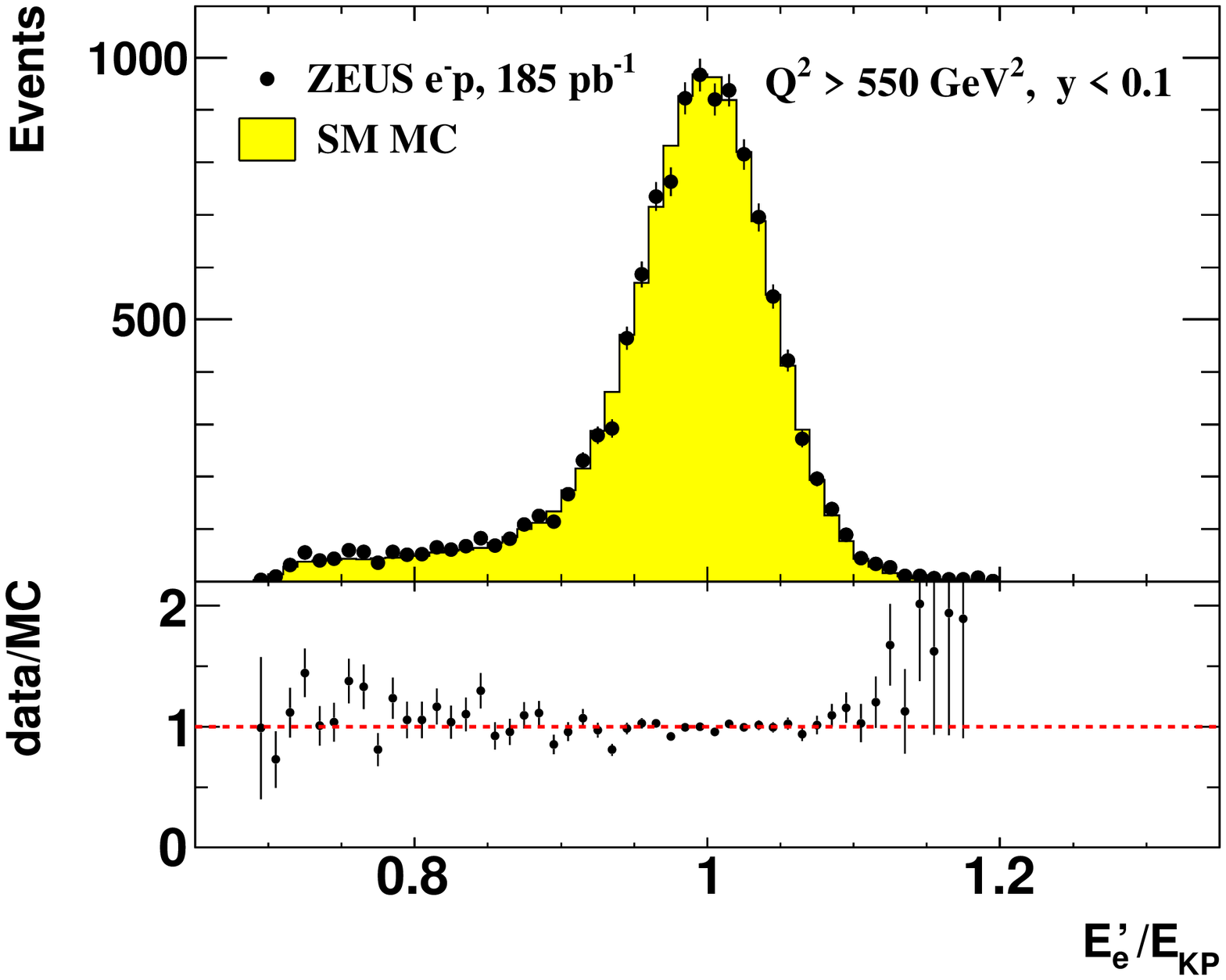}}%\\
\subfloat[]{\includegraphics[height=0.25\textheight]{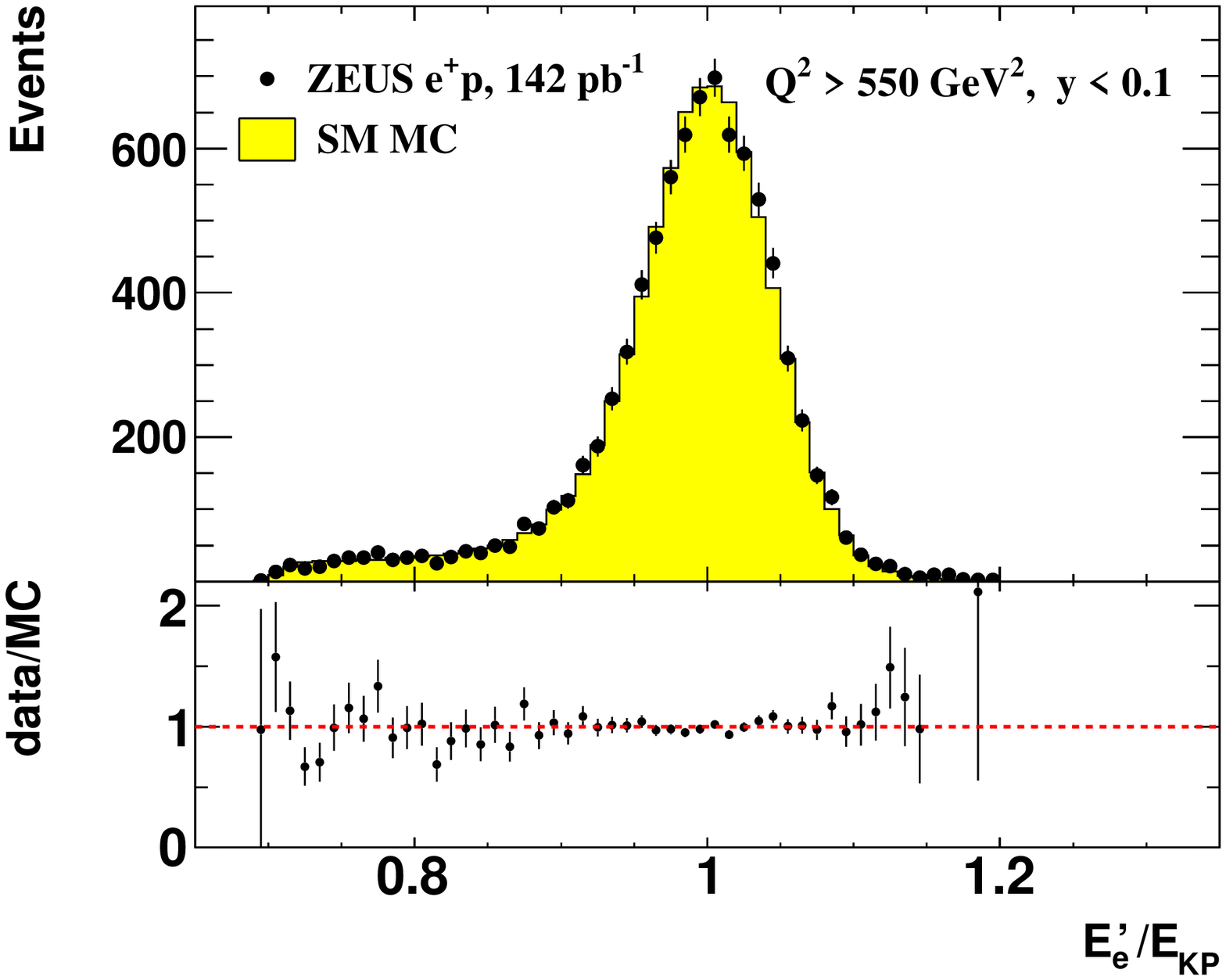}}\\
\subfloat[]{\includegraphics[height=0.25\textheight]{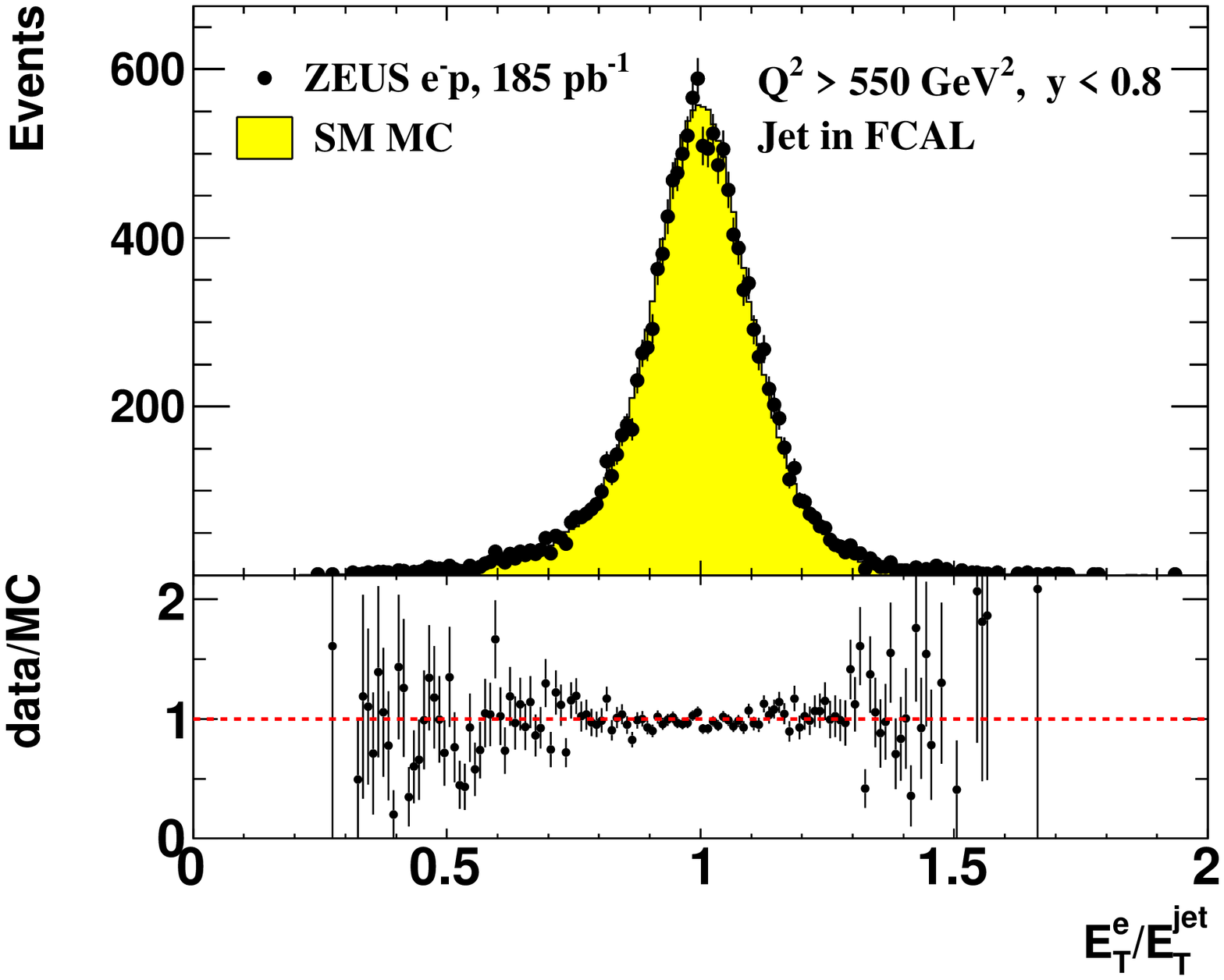}}
\subfloat[]{\includegraphics[height=0.25\textheight]{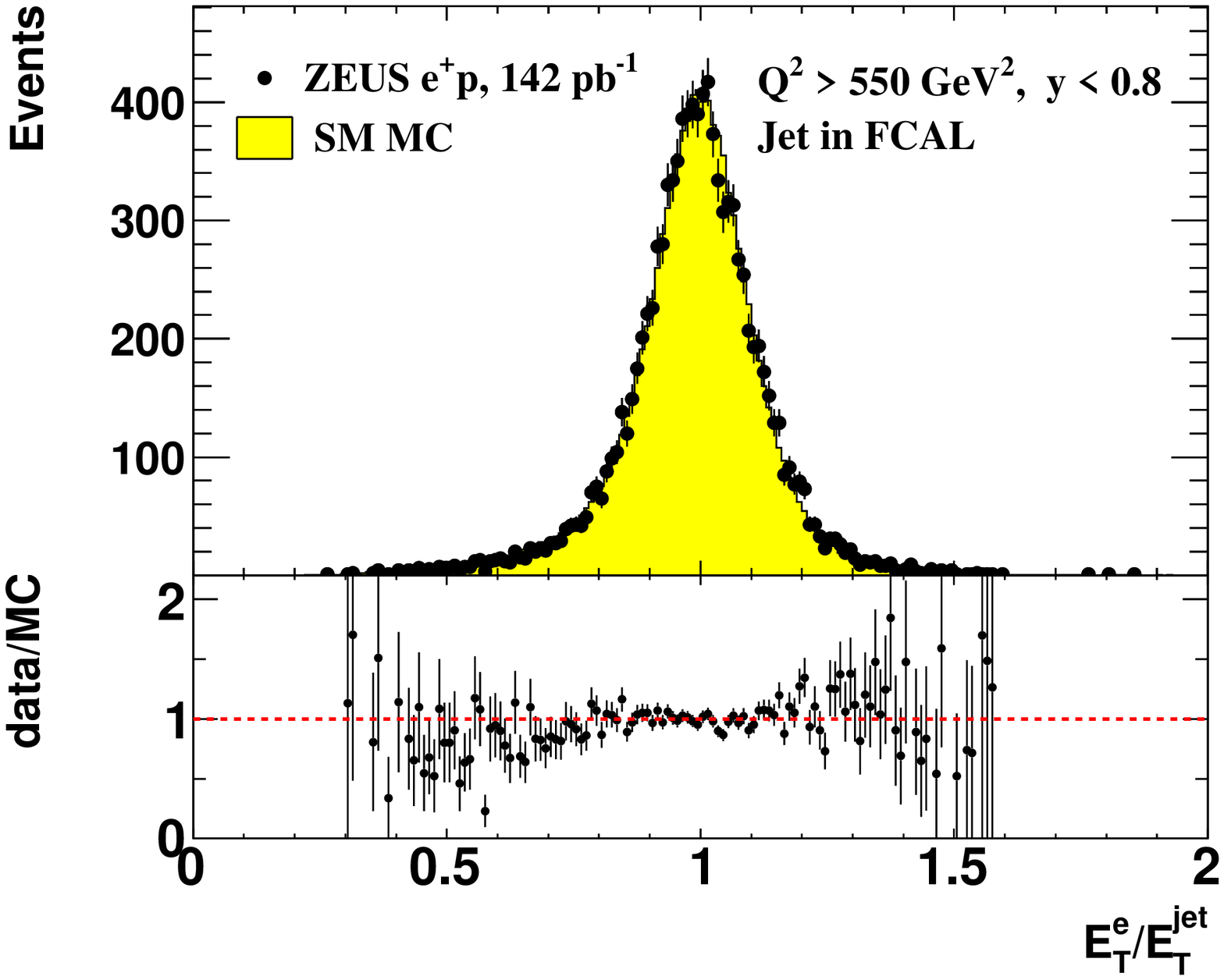}}\\
\subfloat[]{\includegraphics[height=0.25\textheight]{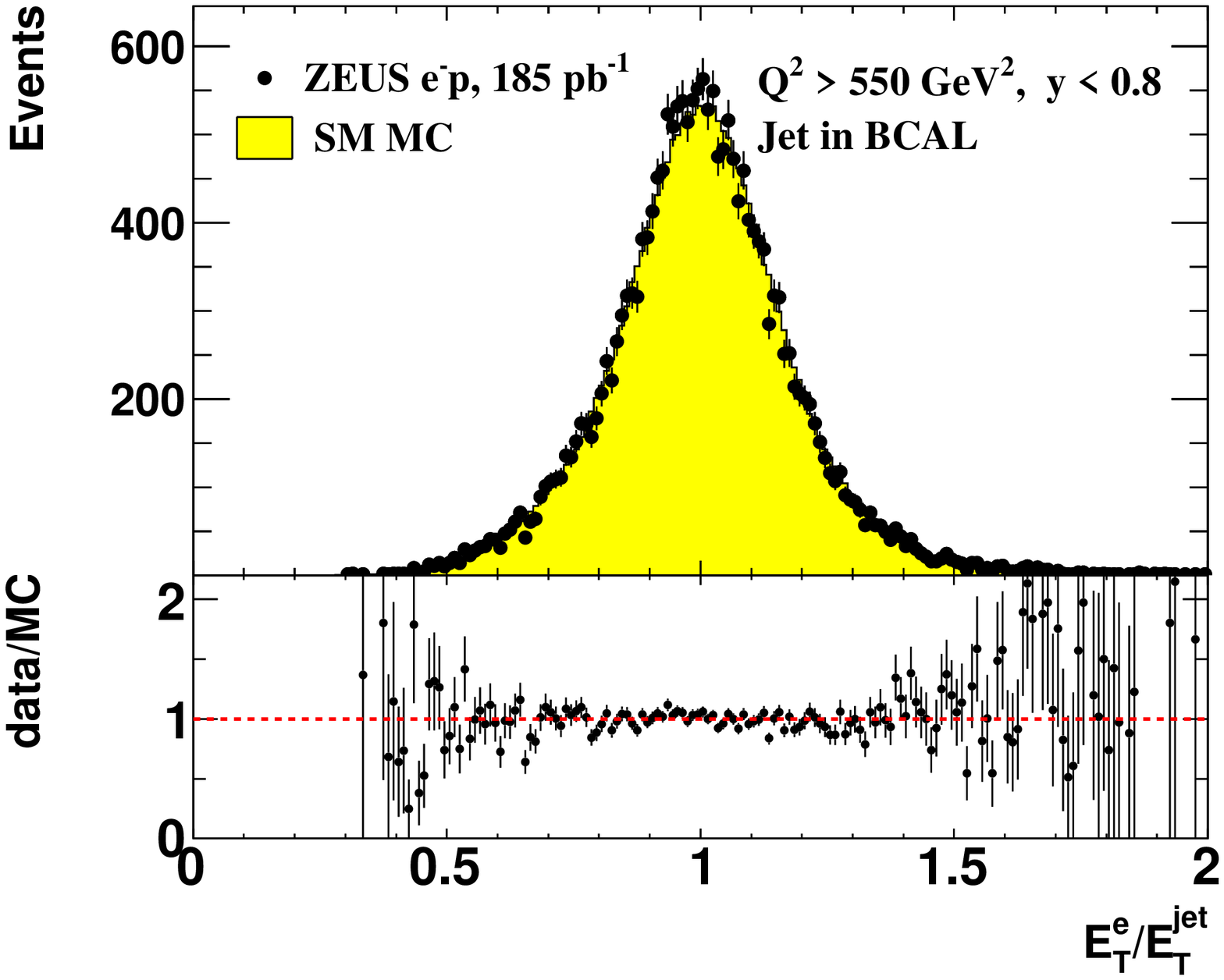}}
\subfloat[]{\includegraphics[height=0.25\textheight]{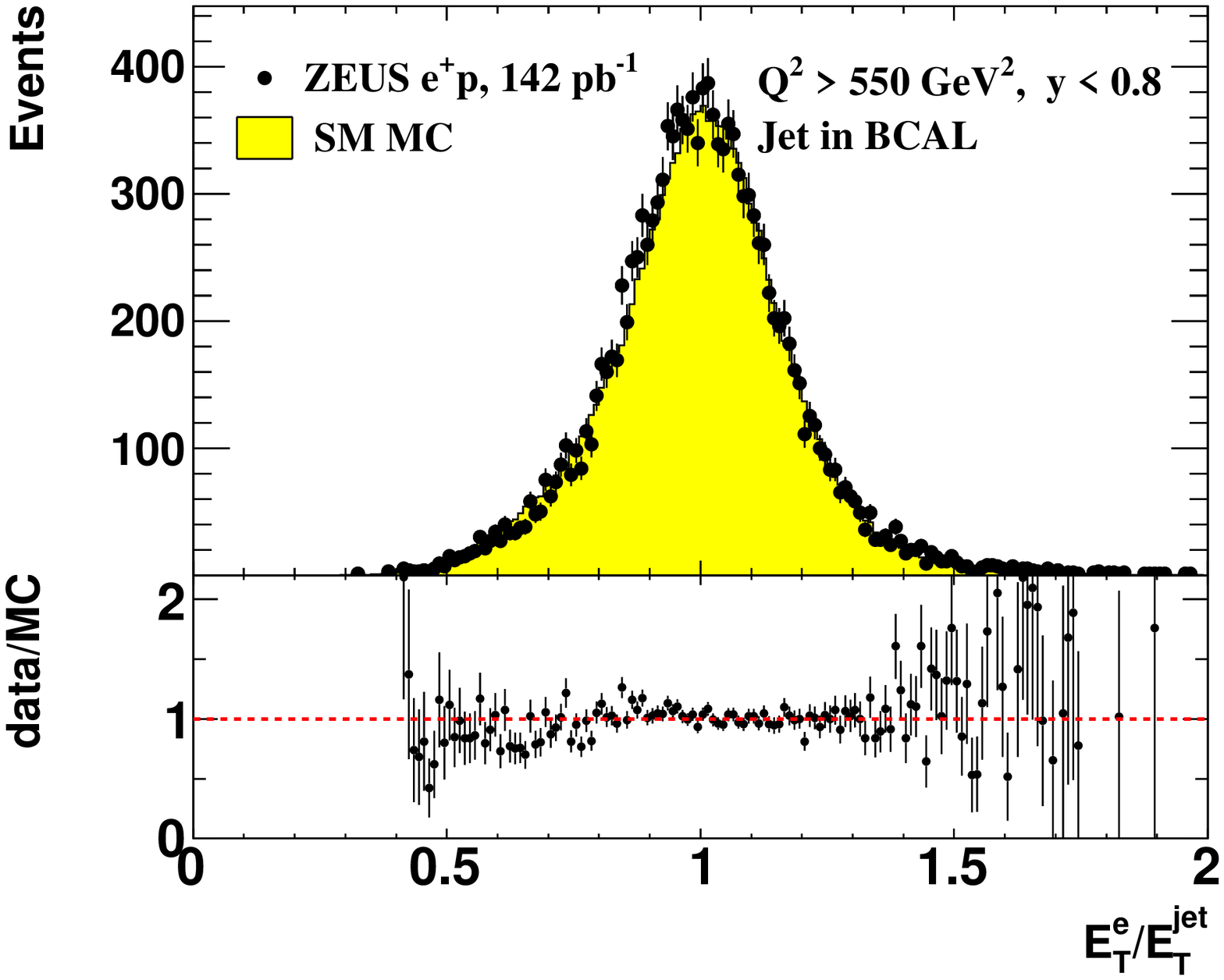}}
\caption{For $e^-p$ and $e^+p$ event samples with $Q^2>550 \units{GeV^2}$, distributions of (a), (b) the ratio of the measured electron energy, $E^{'}_{e}$, to the expected energy, $E_\mathrm{KP}$, for $y <$ 0.1, (c), (d)  of the transverse energy of the electron, $E_T^e$, to the transverse energy of the jet, $E_T^\mathrm{jet}$, in one-jet events with $y <$ 0.8 for jets in the FCAL and (e), (f) similarly for jets in the BCAL. In all upper panels, data samples are represented by dots while MC samples, normalised to the number of data events, are represented by histograms. The lower panels show the ratio of data to MC distributions.
}
\label{fig:KP+jetscale}
\end{center}
\end{figure}

\captionsetup[subfloat]{font={bf}, captionskip=-0.4cm, margin=4cm}

%/// kin properties
\begin{figure}[h!]
\begin{center}
\includegraphics[height=0.06\textheight]{zeusLabel.eps}\\
\subfloat[]{\includegraphics[height=0.22\textheight]{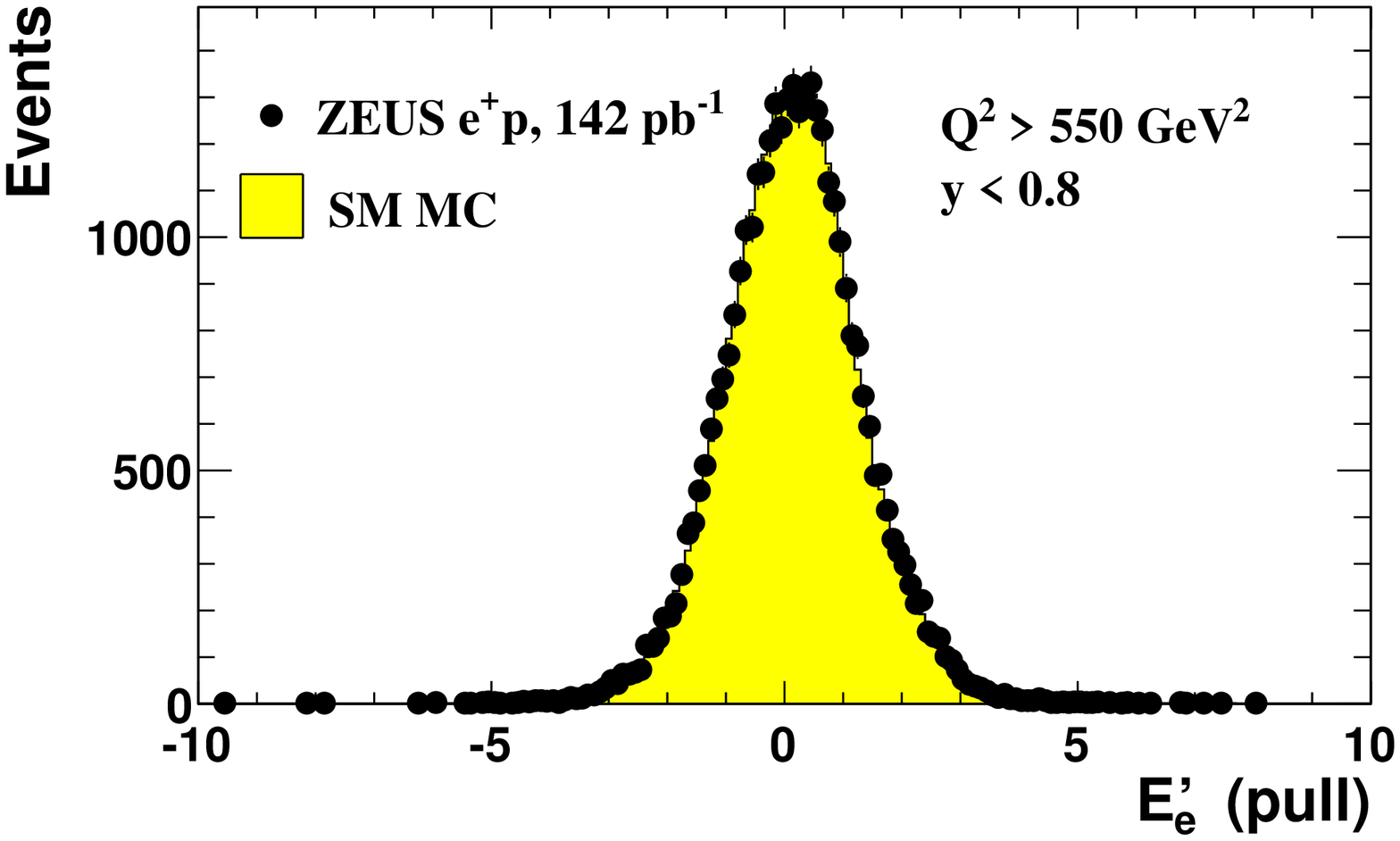}}
\subfloat[]{\includegraphics[height=0.22\textheight]{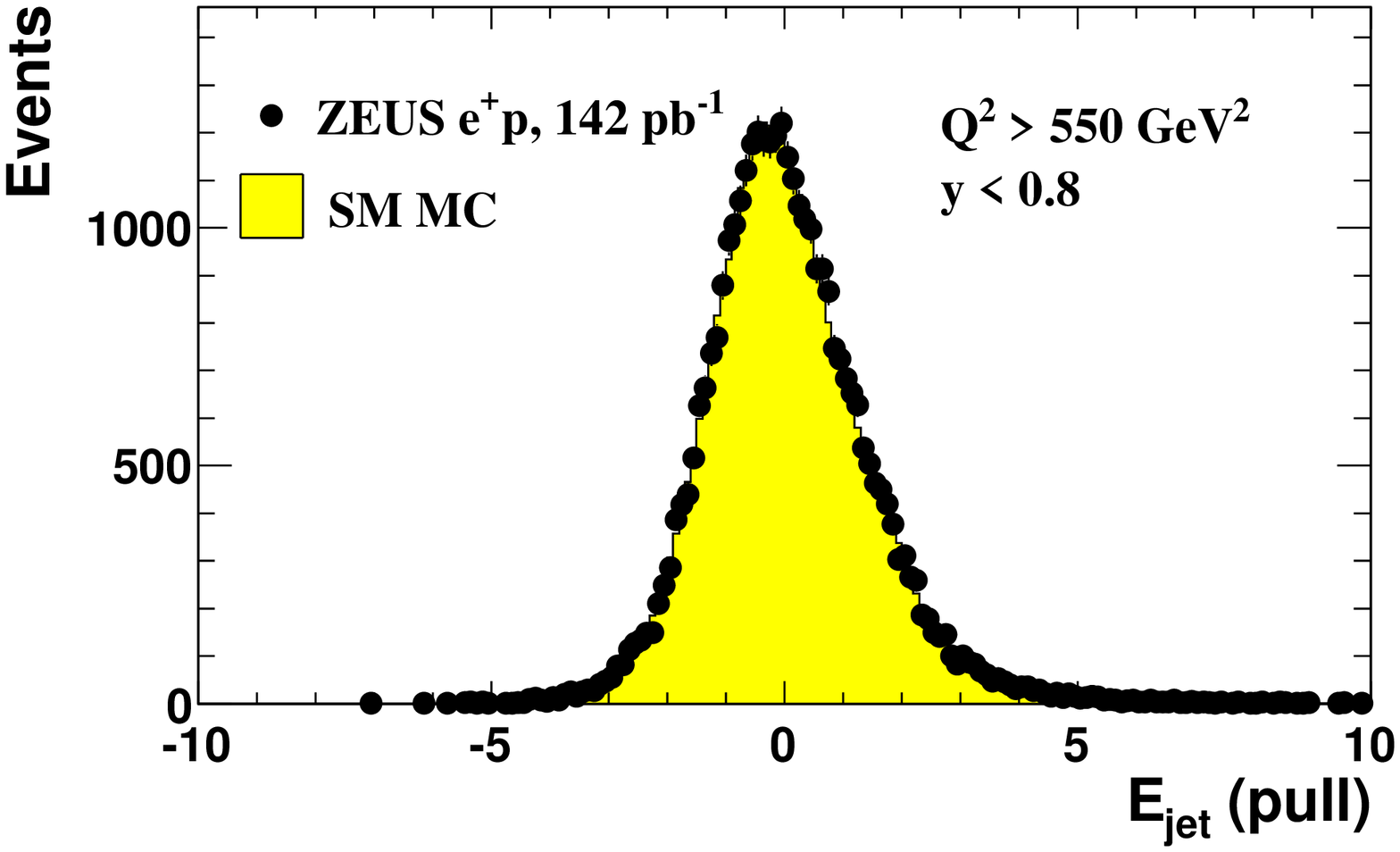}}\\
\subfloat[]{\includegraphics[height=0.22\textheight]{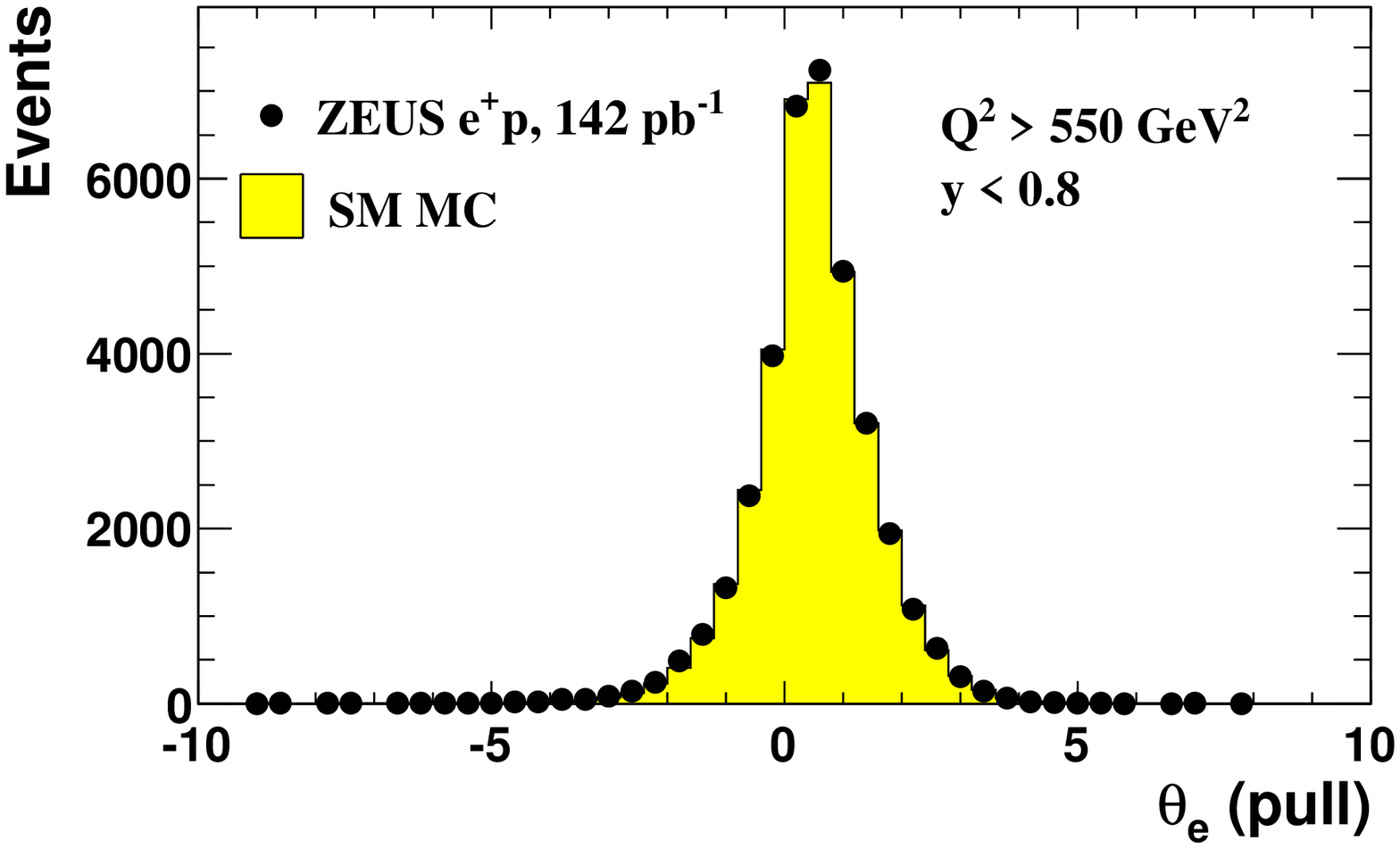}}
\subfloat[]{\includegraphics[height=0.22\textheight]{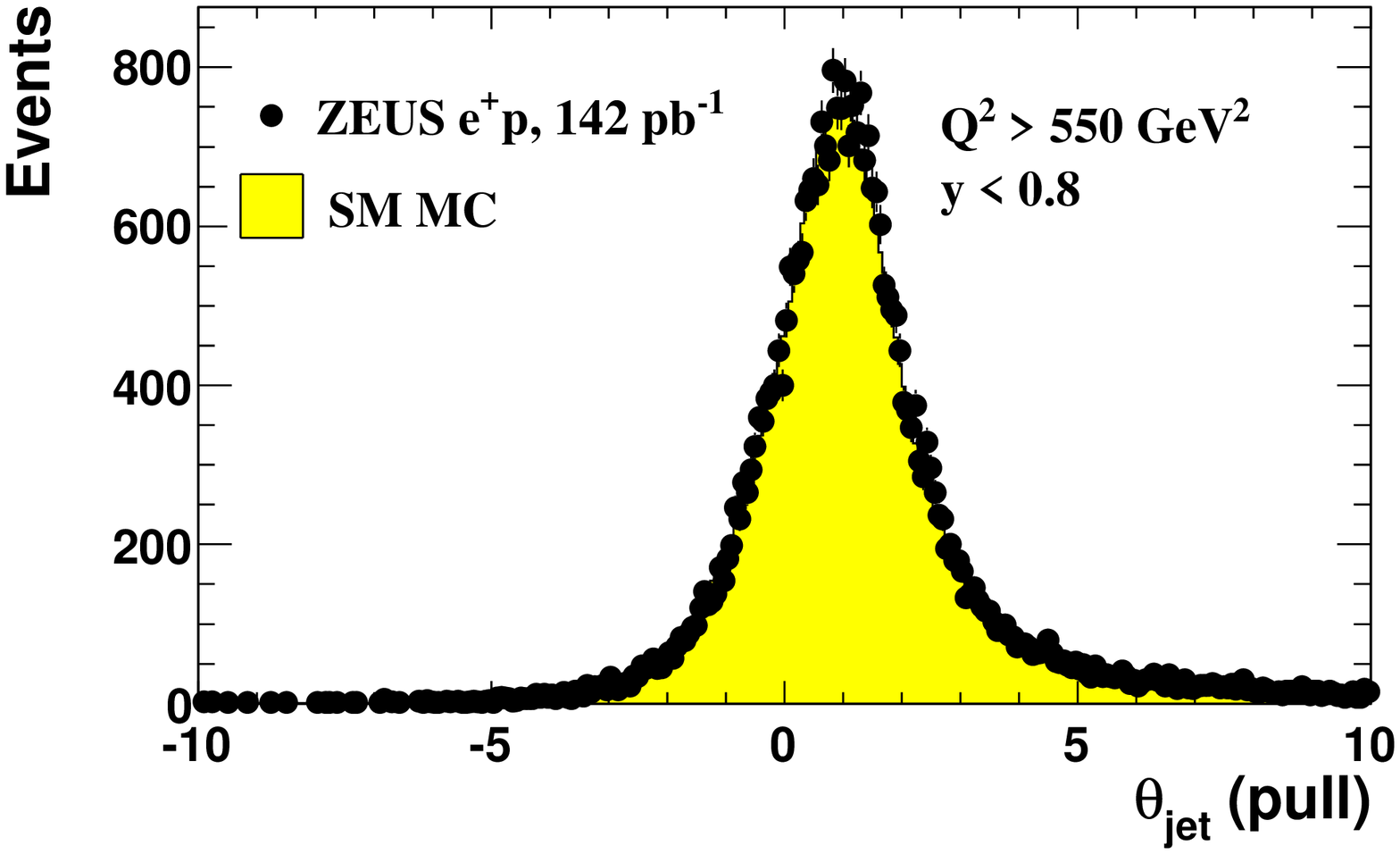}}
\caption{Distribution of the pulls obtained from the kinematic fit  for  (a) the electron energy, $E^{'}_e$ , (b) jet energy, $E_{\mathrm{jet}}$, (c) electron scattering angle, $\theta_e$, and  (d) jet angle, $\theta_{\mathrm{jet}}$, in data (dots) compared to the corresponding distribution in MC (histogram) for the $e^+p$ events with measured $Q^2>550 \units{GeV^2}$ and $y<0.8$. The MC distributions are normalised to the number of events in the data.}
\label{fig:batiep}
\end{center}
\end{figure}

% control plots ePp

\begin{figure}[h!]
\begin{center}
\includegraphics[height=0.06\textheight]{zeusLabel.eps}\\
\subfloat[]{\includegraphics[height=0.22\textheight]{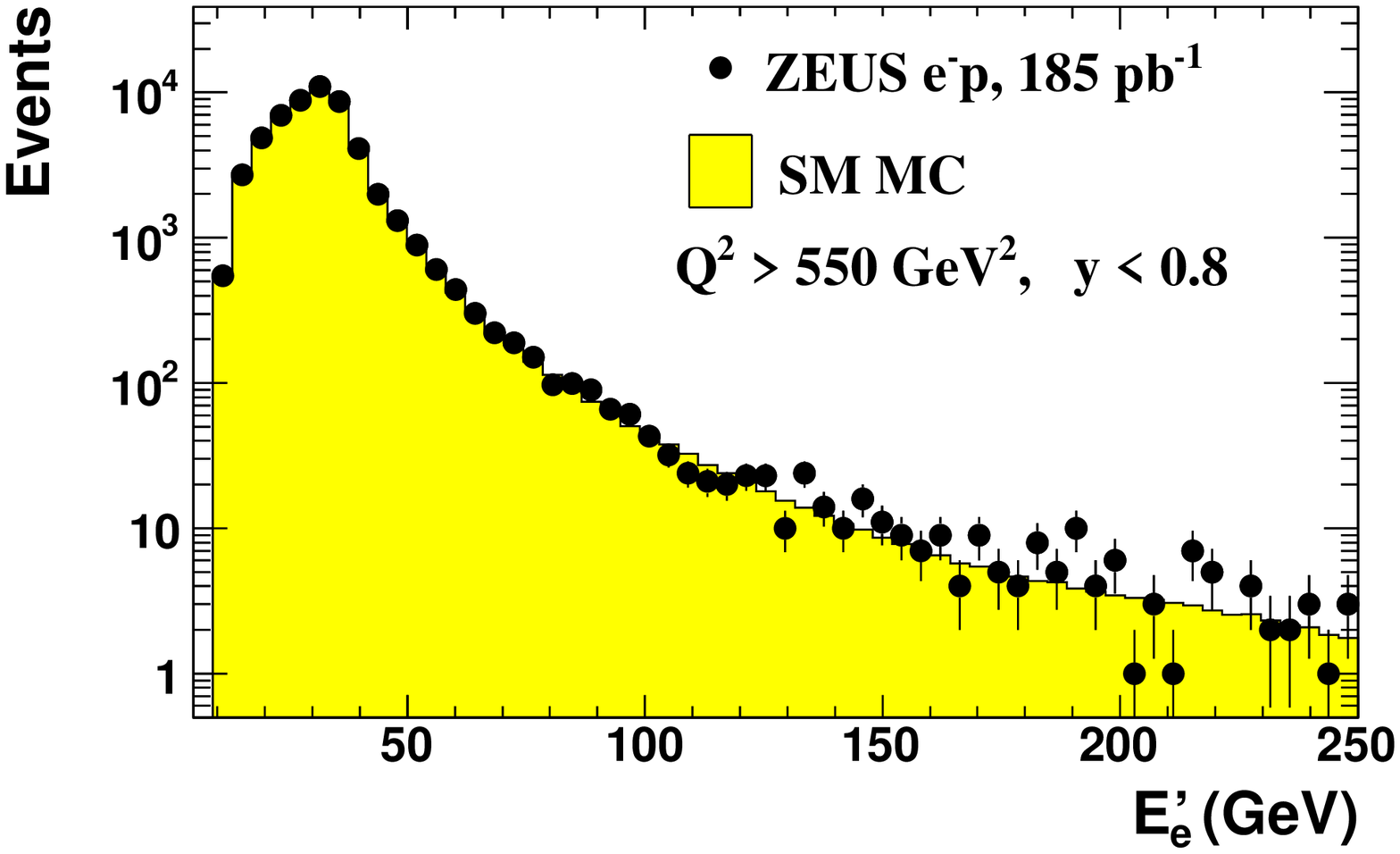}}%\\
\subfloat[]{\includegraphics[height=0.22\textheight]{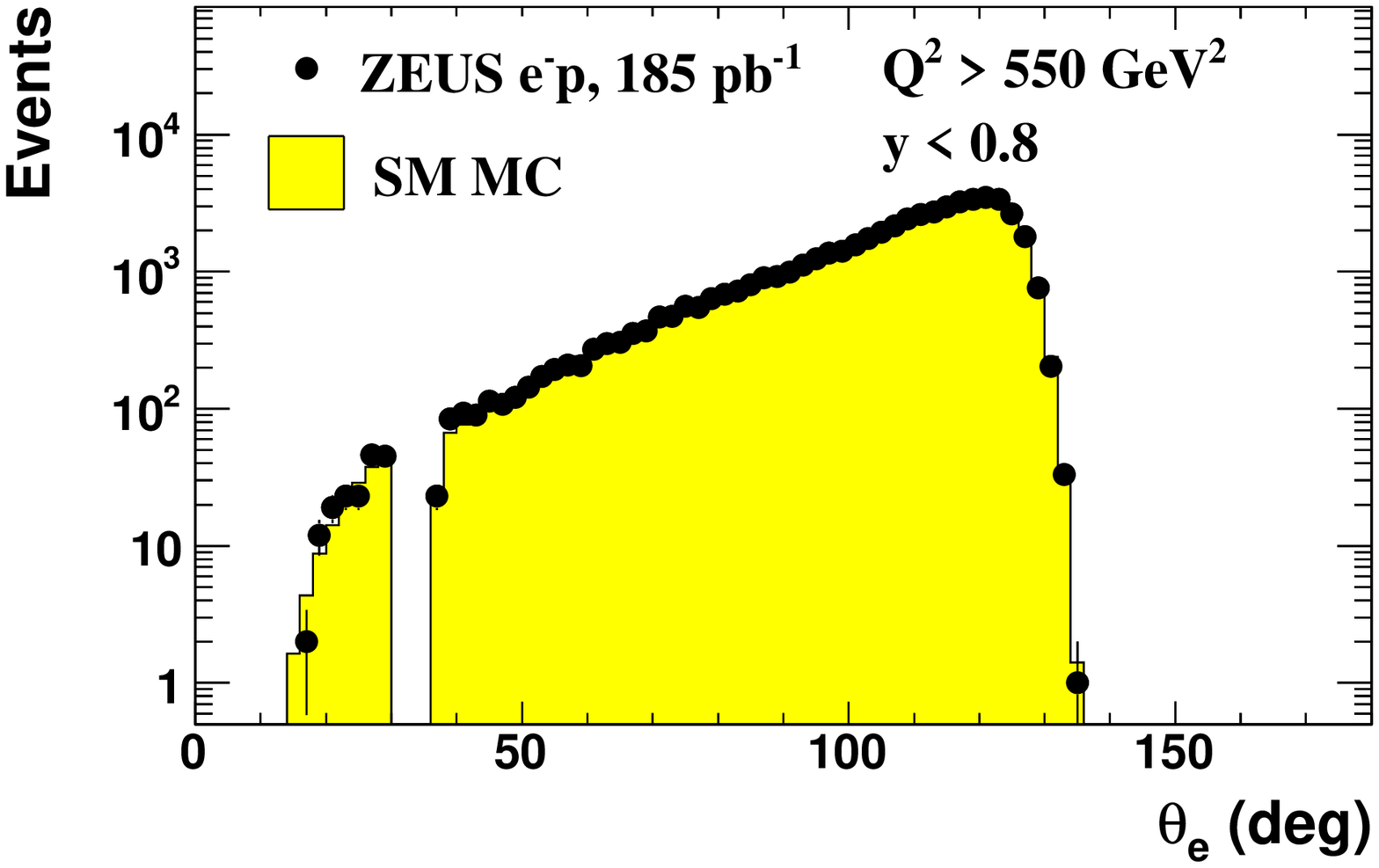}}\\
\subfloat[]{\includegraphics[height=0.22\textheight]{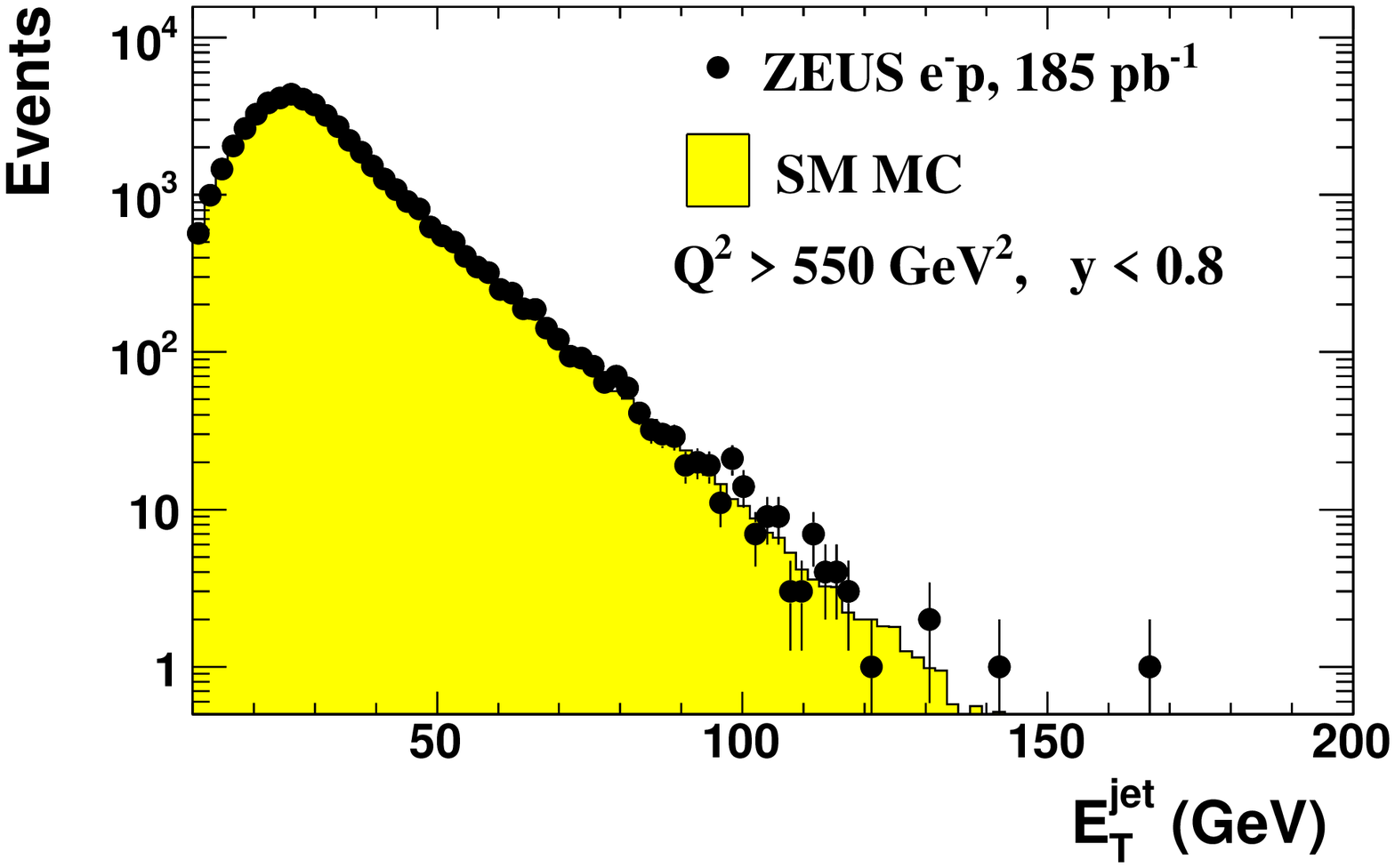}}
\subfloat[]{\includegraphics[height=0.22\textheight]{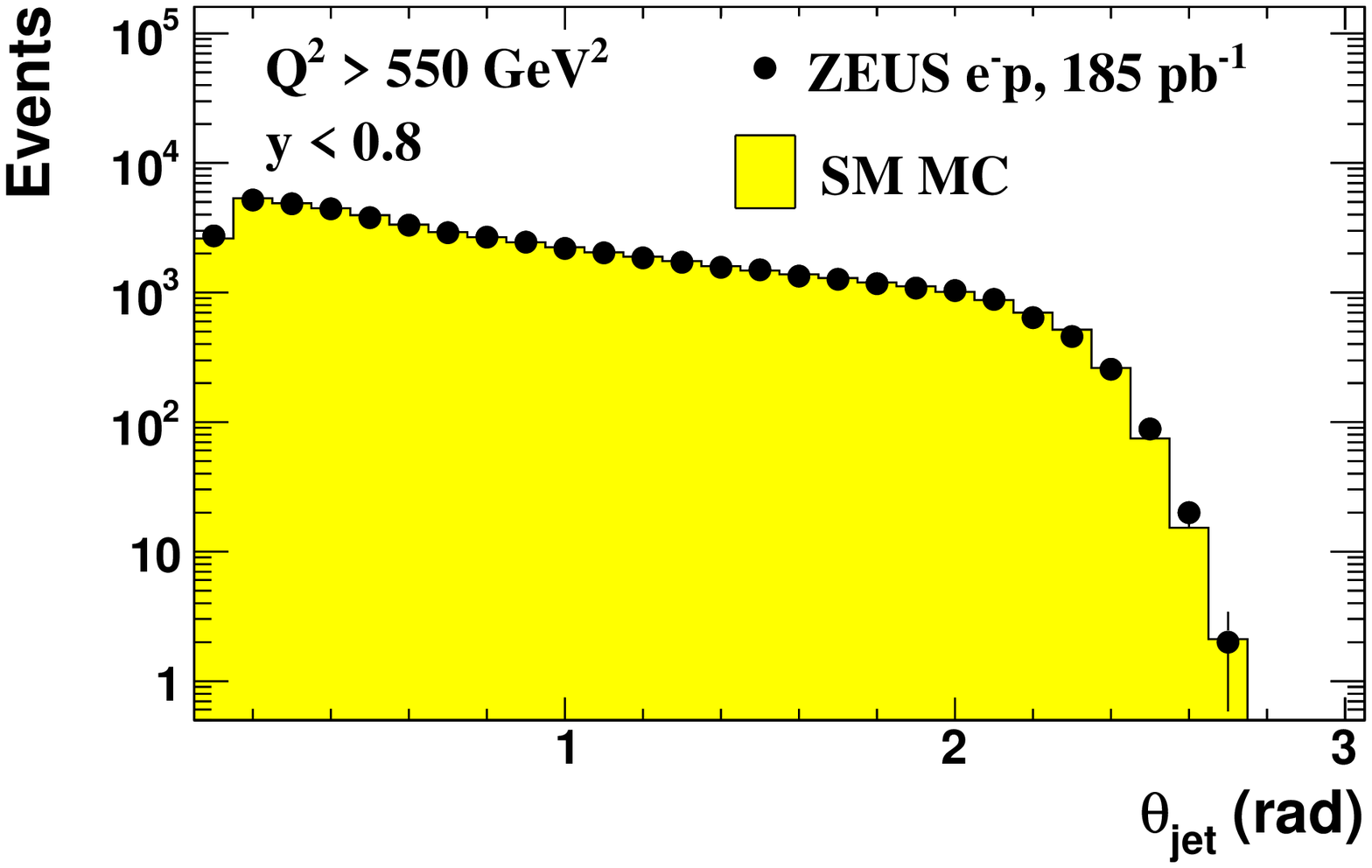}}
\caption{Distributions of (a) electron energy, $E^{'}_e$, (b) electron scattering angle, $\theta_e$,  (c) transverse energy of all jets, $E_T^\mathrm{jet}$, and (d) jet production angle, $\theta_\mathrm{jet}$, for $e^-p$ events with  $Q^2>550 \units{GeV^2}$ and $y<0.8$ in  data (dots) compared to the corresponding distributions in  MC (histograms). The MC distributions are normalised to the number of events in the data.}
\label{fig:mcmixem}
\end{center}
\end{figure}

% control plots eMp
\begin{figure}[h!]
\begin{center}
\includegraphics[height=0.06\textheight]{zeusLabel.eps}\\
\subfloat[]{\includegraphics[height=0.22\textheight]{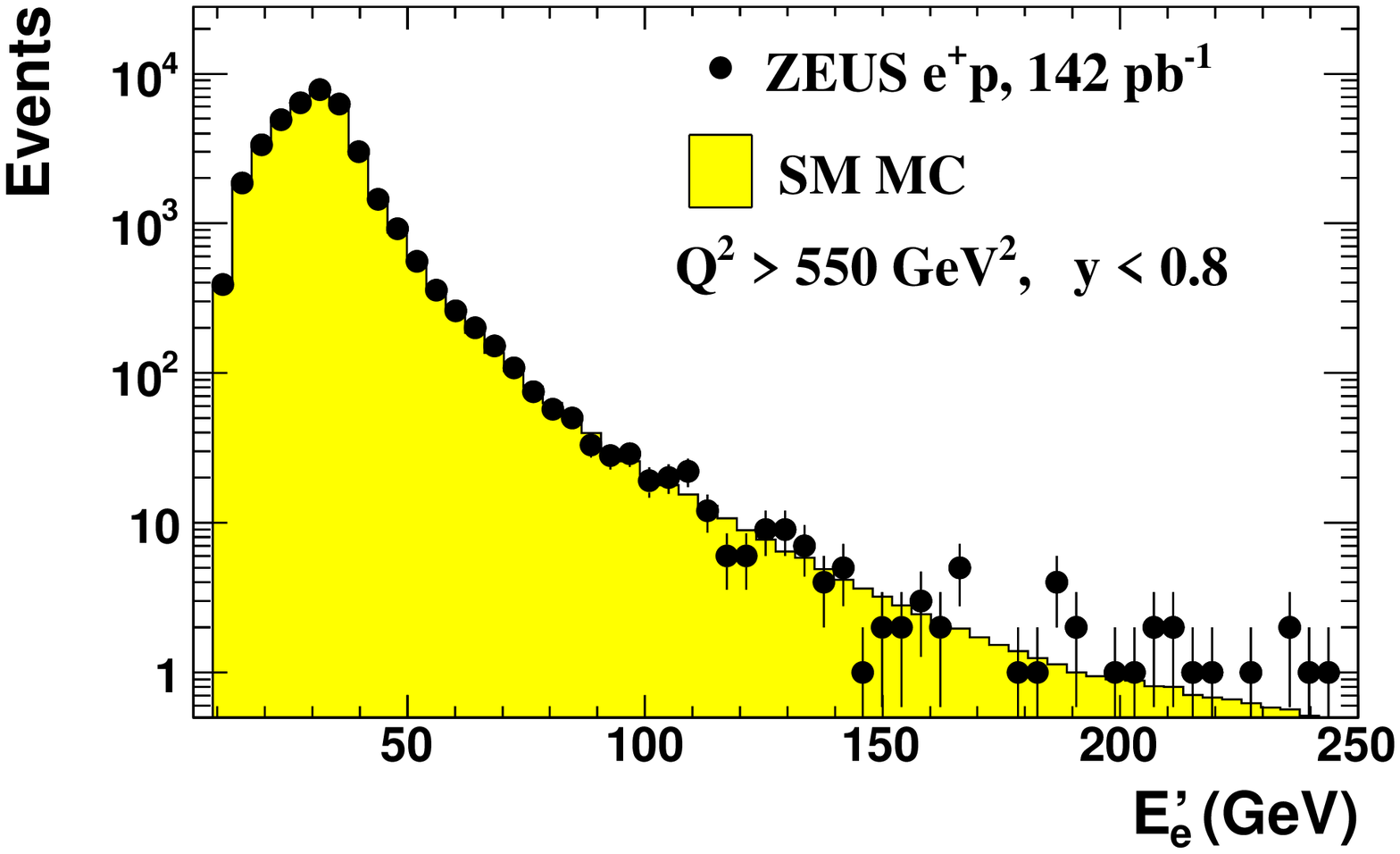}}%\\
\subfloat[]{\includegraphics[height=0.22\textheight]{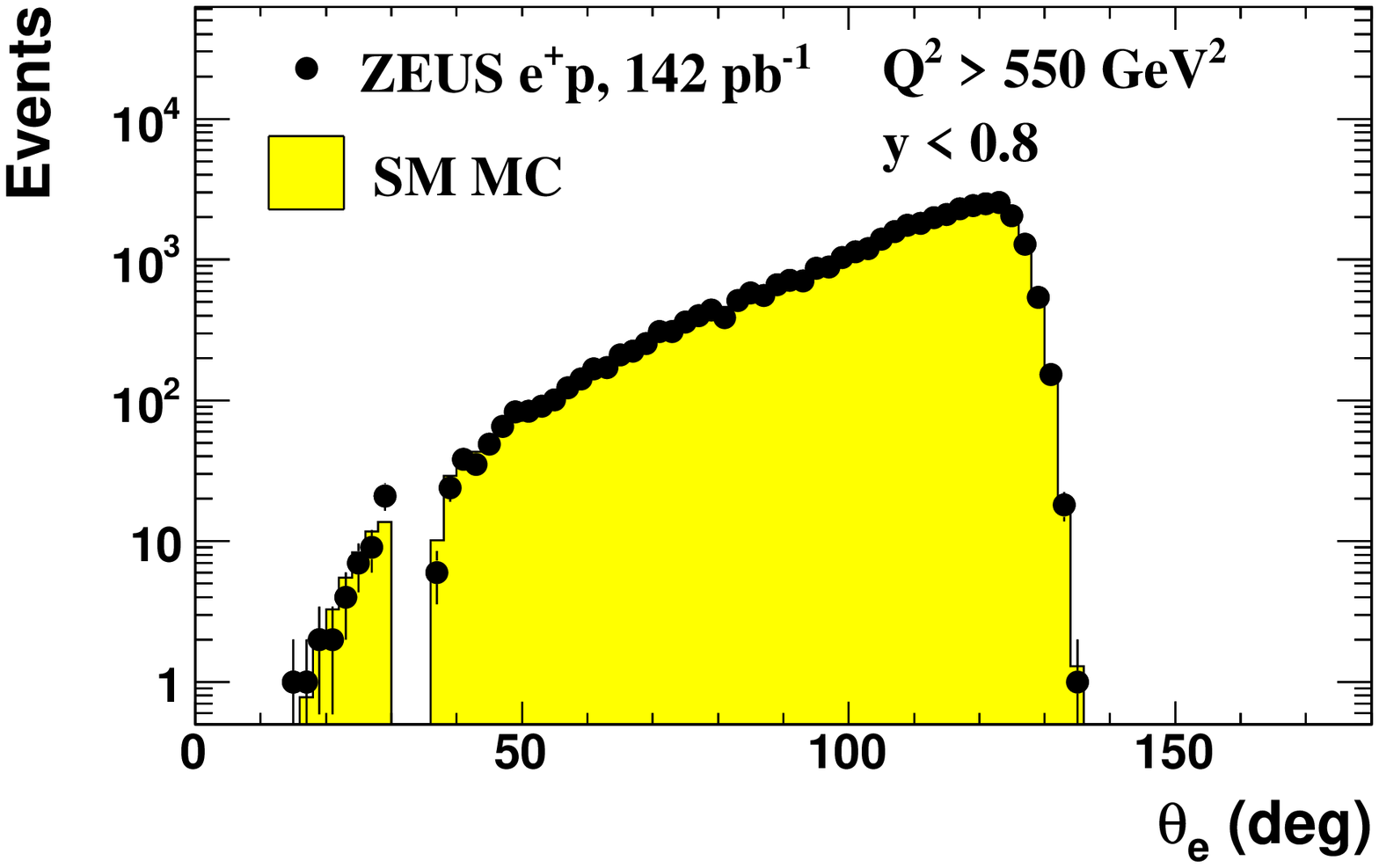}}\\
\subfloat[]{\includegraphics[height=0.22\textheight]{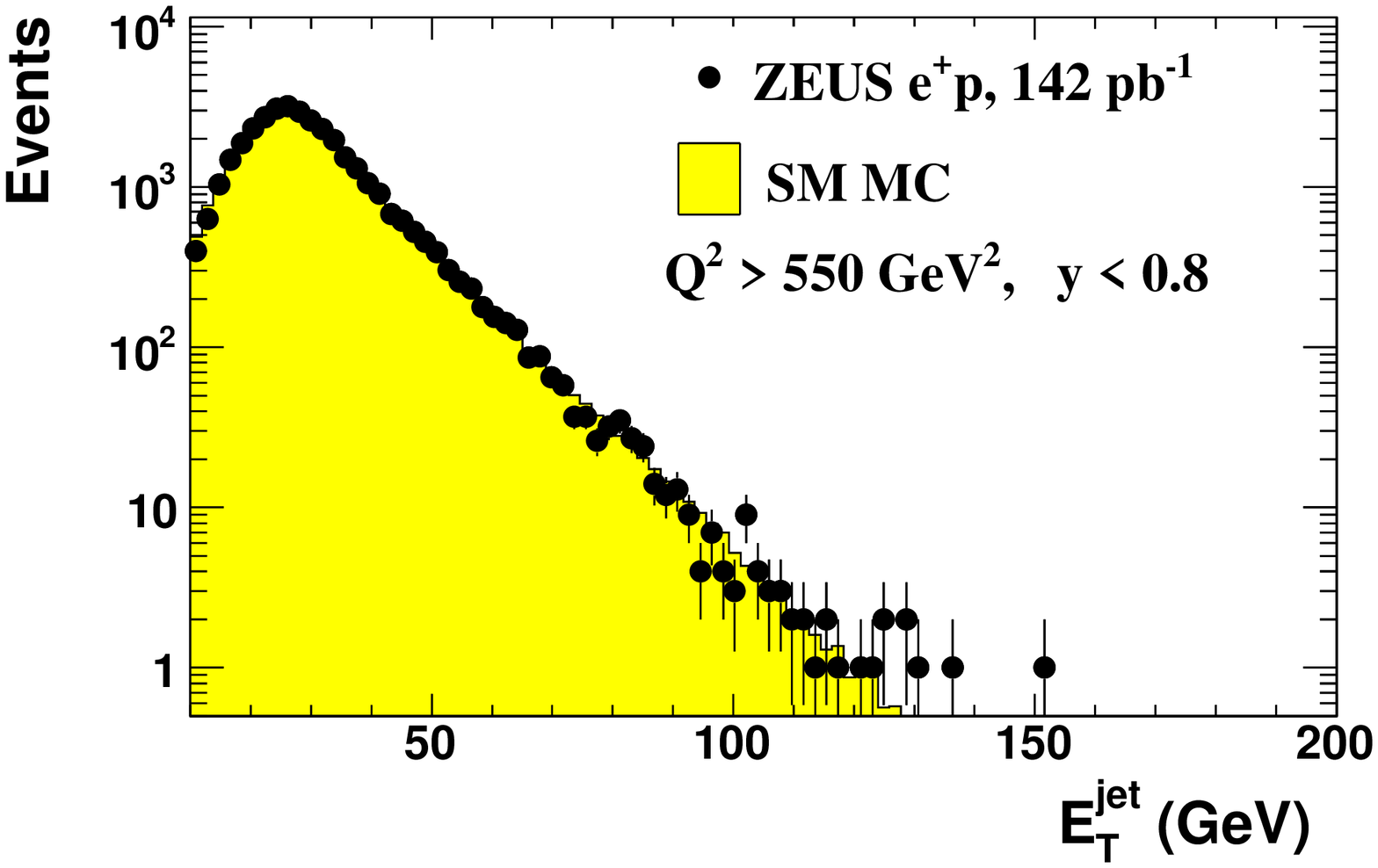}}
\subfloat[]{\includegraphics[height=0.22\textheight]{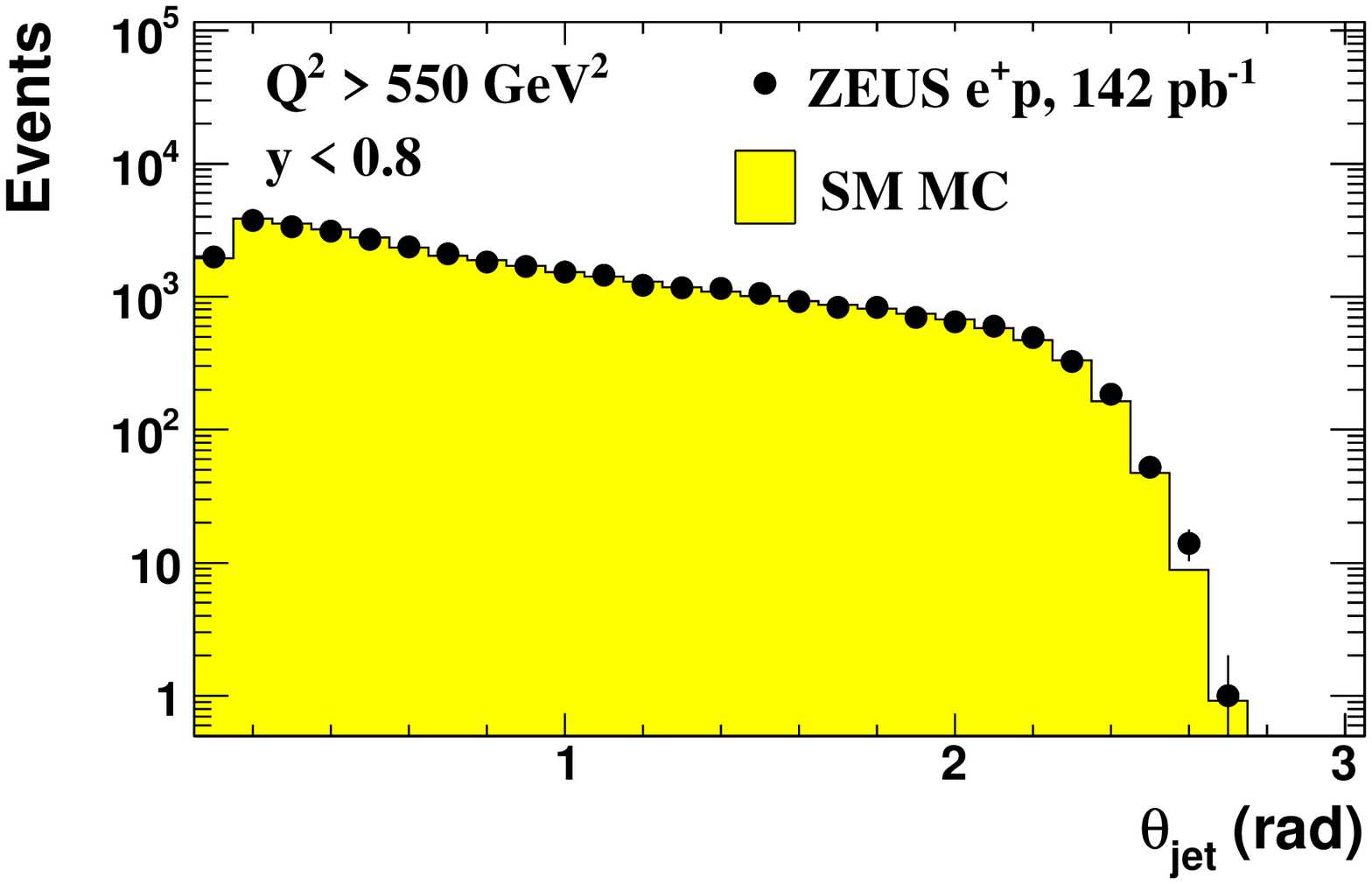}}
\caption{Distributions of (a) electron energy, $E^{'}_e$, (b) electron scattering angle, $\theta_e$,  (c) transverse energy of all jets, $E_T^\mathrm{jet}$, and (d) jet production angle, $\theta_\mathrm{jet}$, for $e^+p$ events with  $Q^2>550 \units{GeV^2}$ and $y<0.8$ in  data (dots) compared to the corresponding distributions in  MC (histograms). The MC distributions are normalised to the number of events in the data.
}
\label{fig:mcmixep}
\end{center}
\end{figure}

\begin{figure}[h!]
\begin{center}
\includegraphics[height=0.06\textheight]{zeusLabel.eps}\\
\subfloat[]{\includegraphics[height=0.22\textheight]{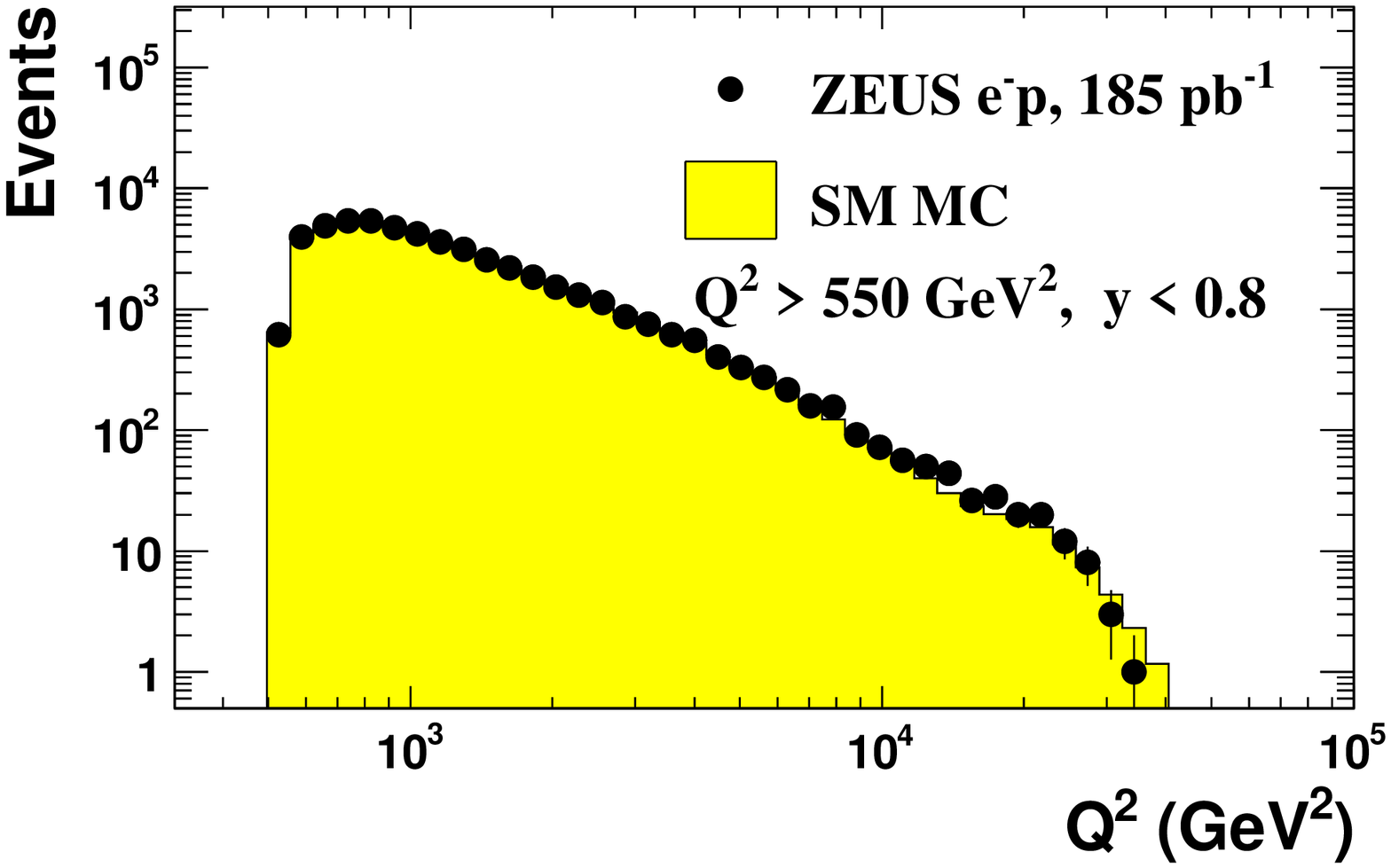}}
\subfloat[]{\includegraphics[height=0.22\textheight]{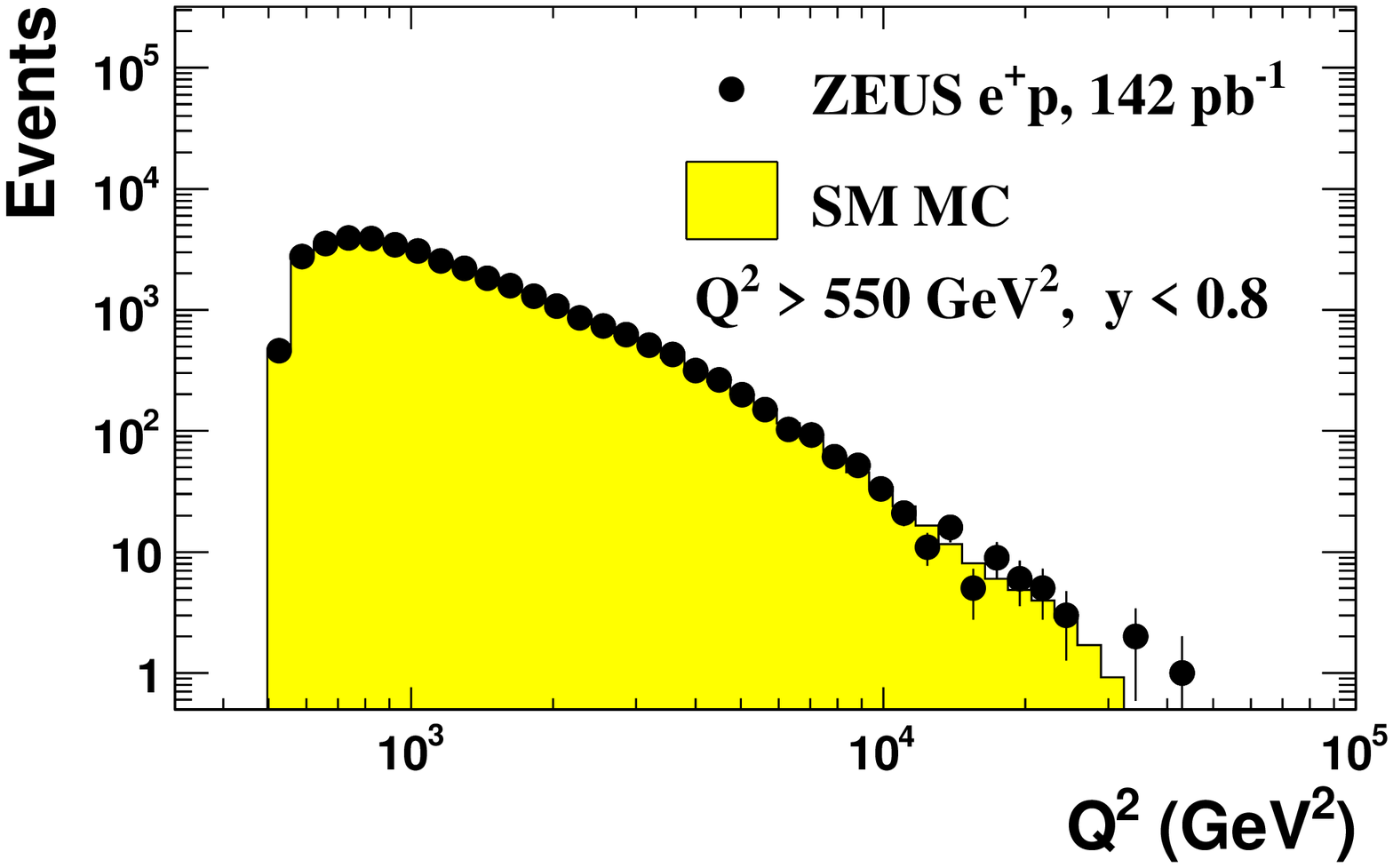}}\\
\subfloat[]{\includegraphics[height=0.22\textheight]{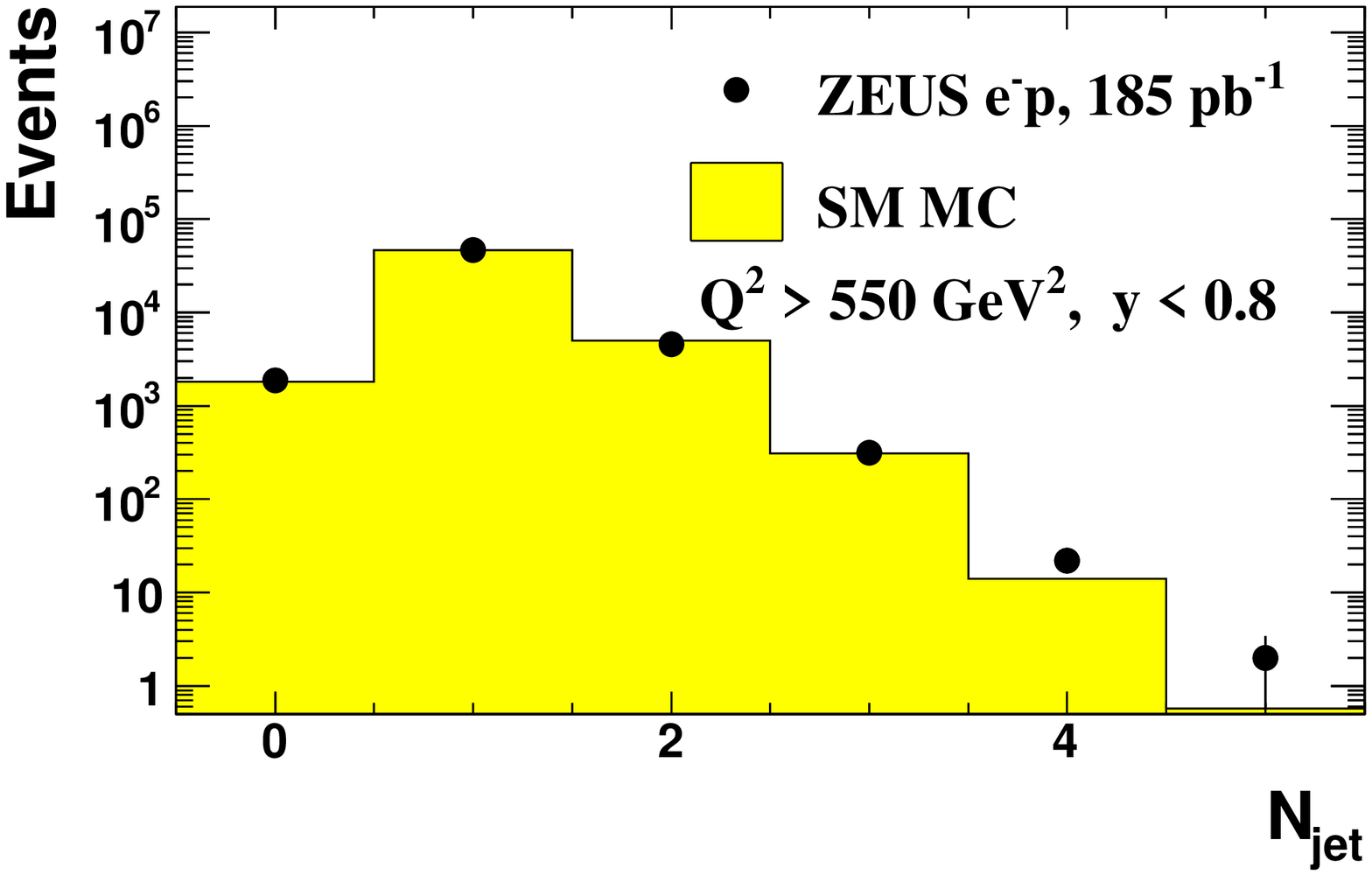}}
\subfloat[]{\includegraphics[height=0.22\textheight]{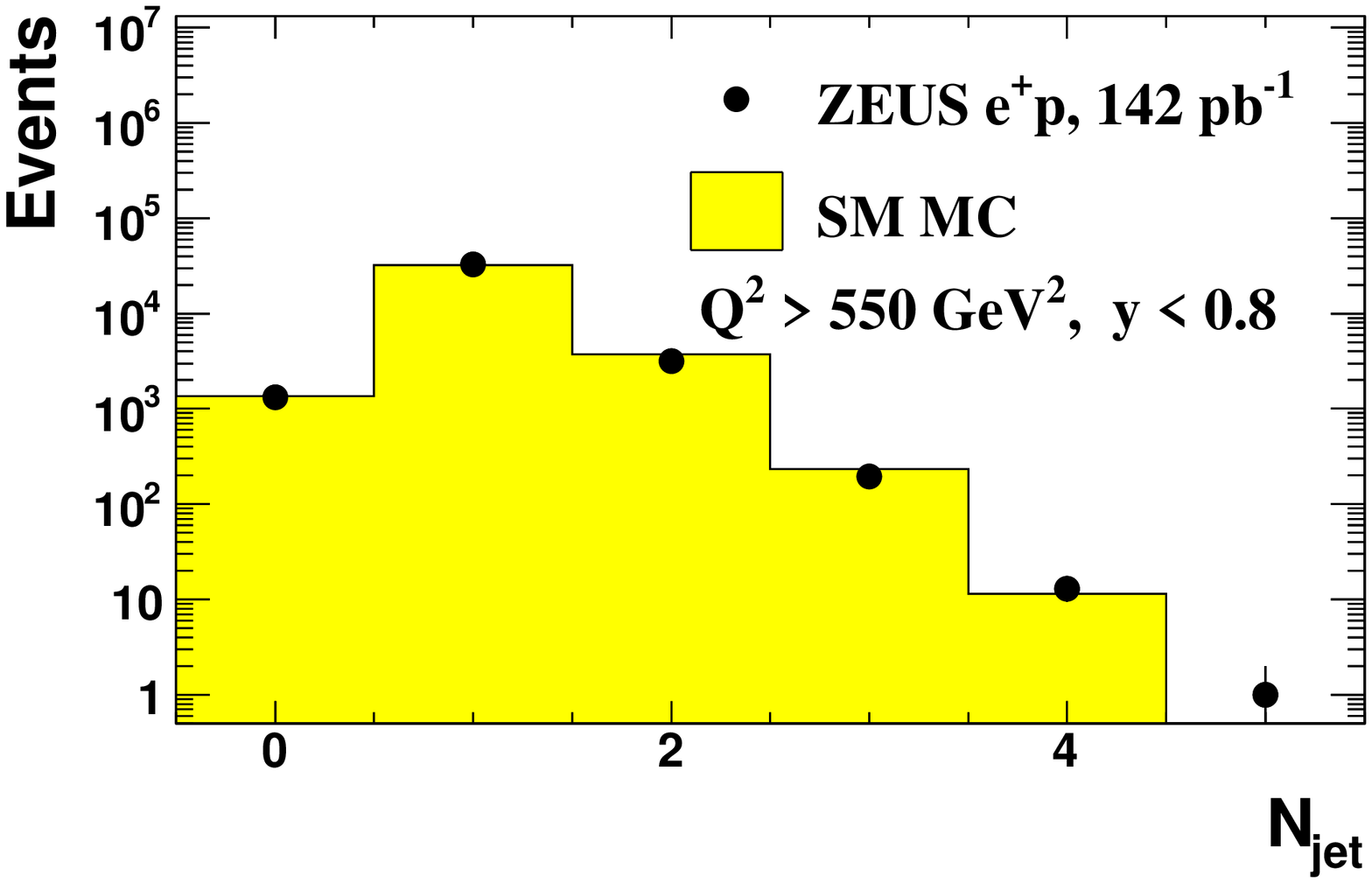}}\\
\subfloat[]{\includegraphics[height=0.22\textheight]{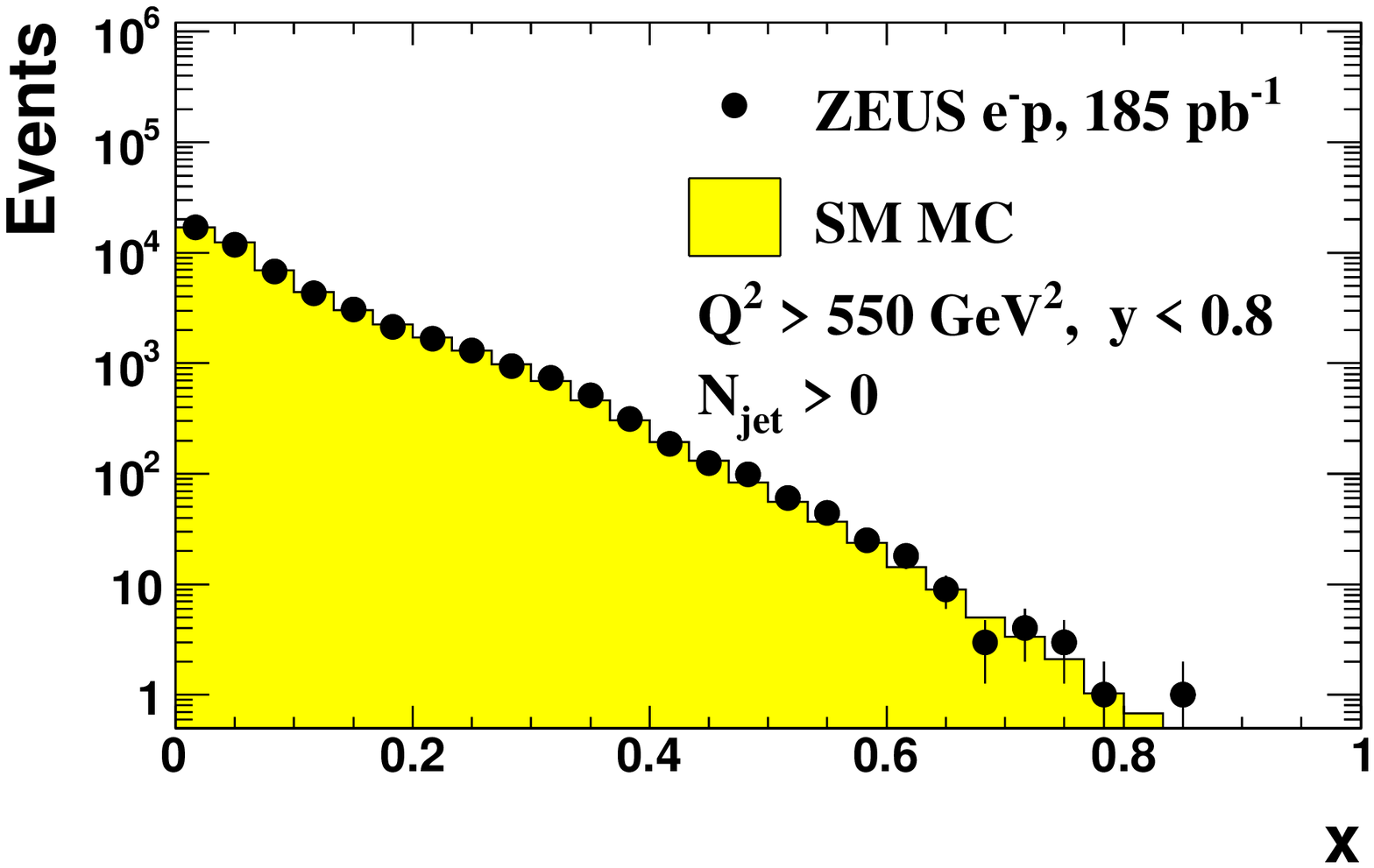}}
\subfloat[]{\includegraphics[height=0.22\textheight]{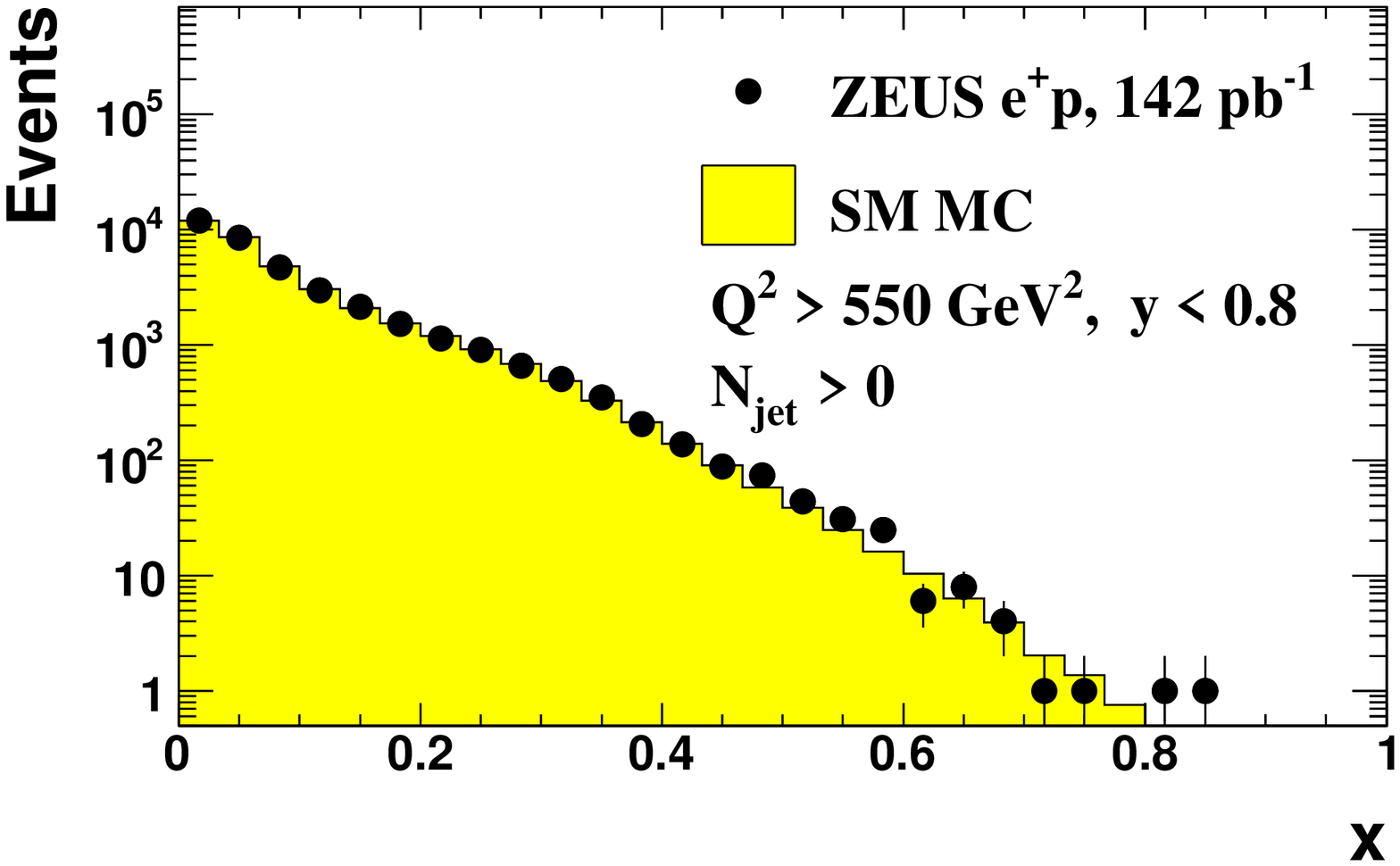}}
\caption{Distribution of  (a), (b) $ Q^2$,  (c), (d)  the jet multiplicity, $N_\mathrm{jet}$ and (e), (f)  $x$ for events with at least one jet for the $e^-p$ and $e^+p$  data samples (dots)  for $Q^2>550 \units{GeV^2}$ and $y<0.8$. The distributions are compared to MC expectations (histograms) normalised to the number of events in the data.  }
\label{fig:xq2mix}
\end{center}
\end{figure}

\begin{figure}[h!]
\begin{center}
\includegraphics[angle = 0.,scale = 0.80]{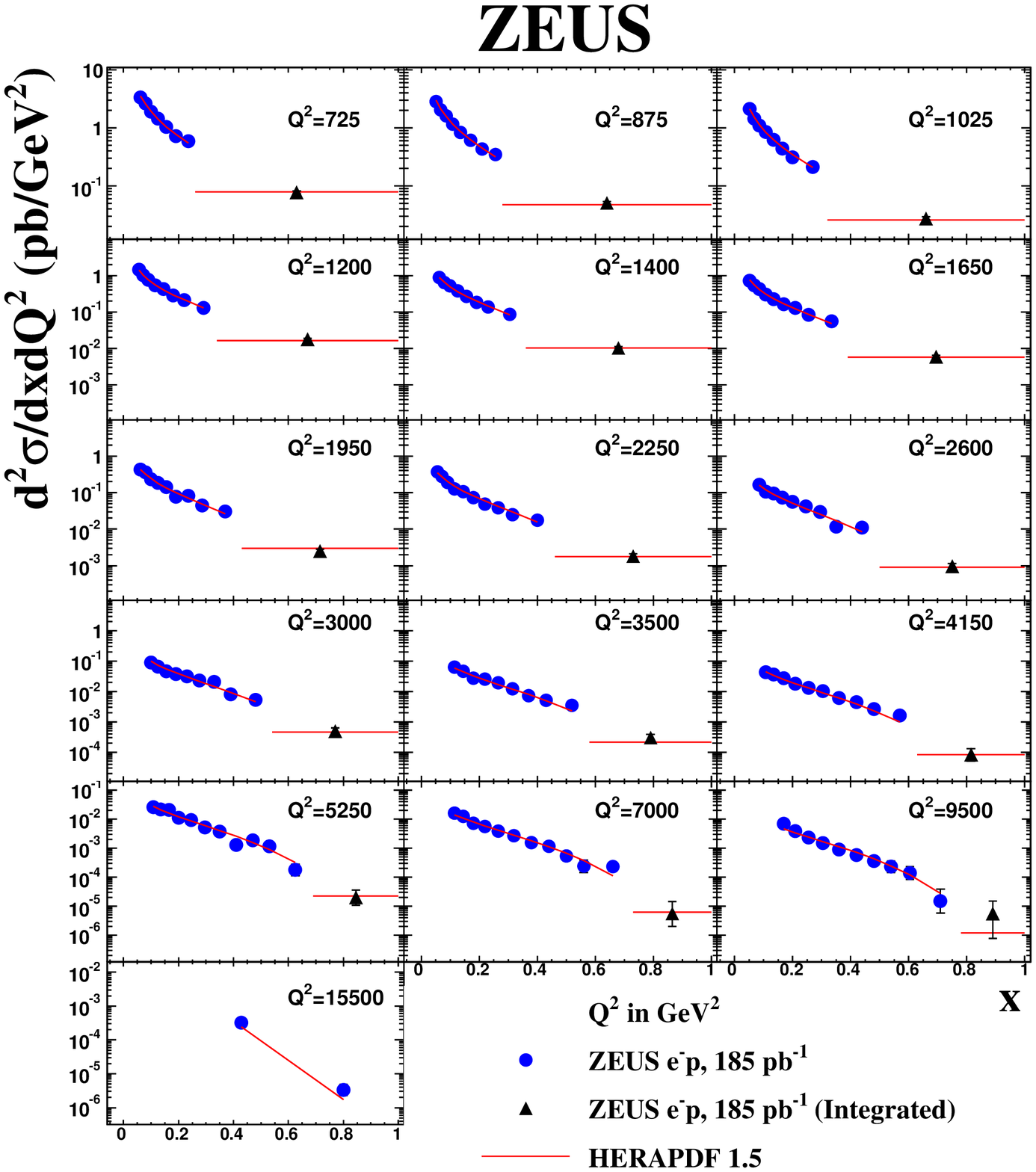} 
\caption{ The double-differential cross section for NC $e^-p$ scattering
at $\sqrt{s}=318\units{GeV}$
(dots)  as a function of $x$ and  the double-differential cross section integrated over $x$ divided by the bin width and placed at the centre of the bin
(triangles) for different values of $Q^2$ as shown, compared to the Standard Model
expectations evaluated using HERAPDF1.5  PDFs (line). The error bars
show the statistical and systematic uncertainties added in quadrature. }
\label{e-xsc}
\end{center}
\end{figure}

\begin{figure}[h!]
\begin{center}
\includegraphics[angle = 0.,scale = 0.80]{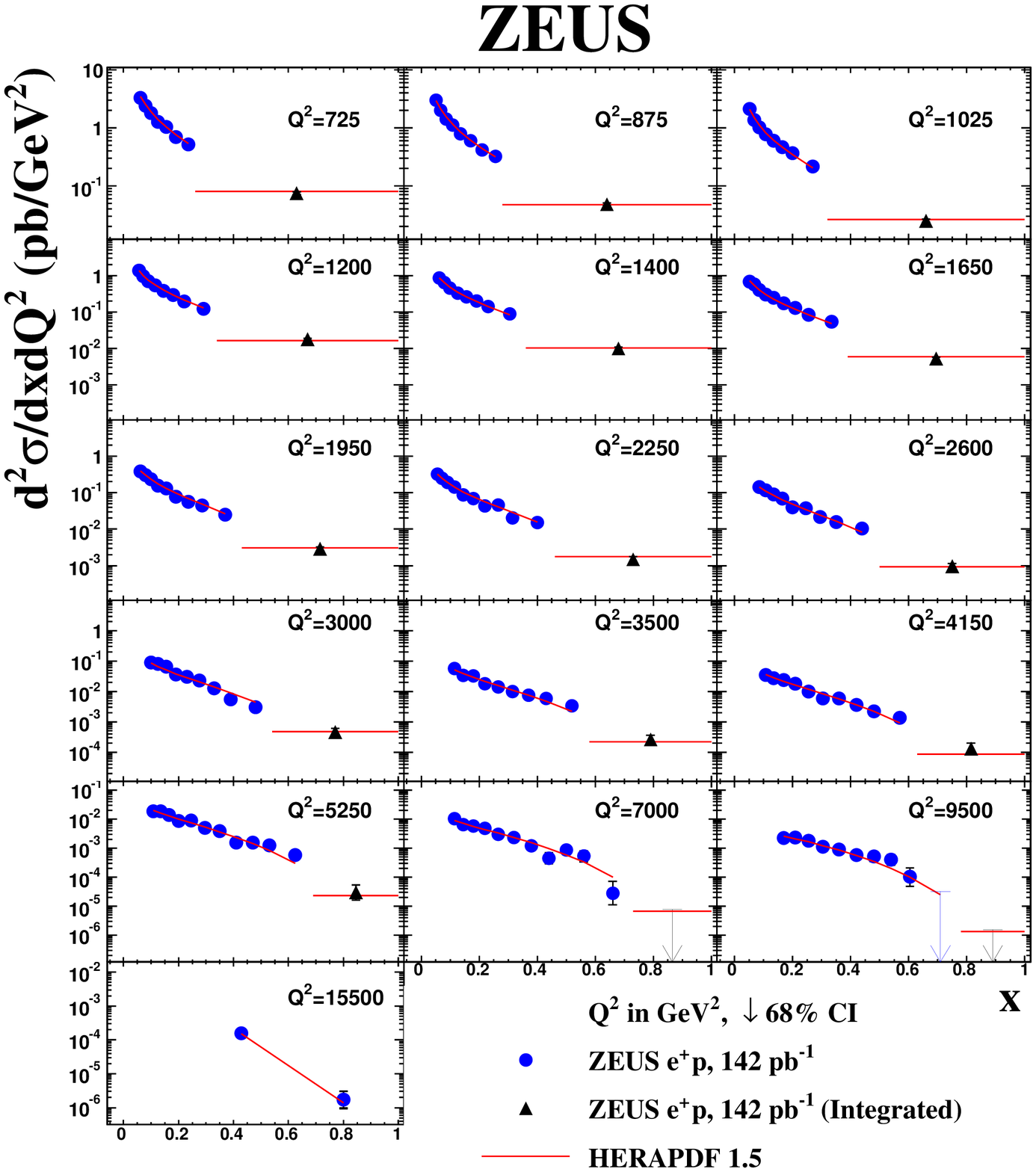} 
\caption{ The double-differential cross section for NC $e^+p$ scattering
at $\sqrt{s}=318\units{GeV}$
(dots)  as a function of $x$ and  the double-differential cross section integrated over $x$ divided by the bin width and placed at the centre of the bin
(triangles) for different values of $Q^2$ as shown, compared to the Standard Model
expectations evaluated using HERAPDF1.5  PDFs (line). The error bars
show the statistical and systematic uncertainties added in quadrature. For bins with zero measured events, a 68\% probability limit is given.}
\label{e+xsc}
\end{center}
\end{figure}

\begin{figure}[h!]
\begin{center}
\vspace{-1cm}
\hspace{-1.8cm}
\includegraphics[angle = 0.,scale = 0.80]{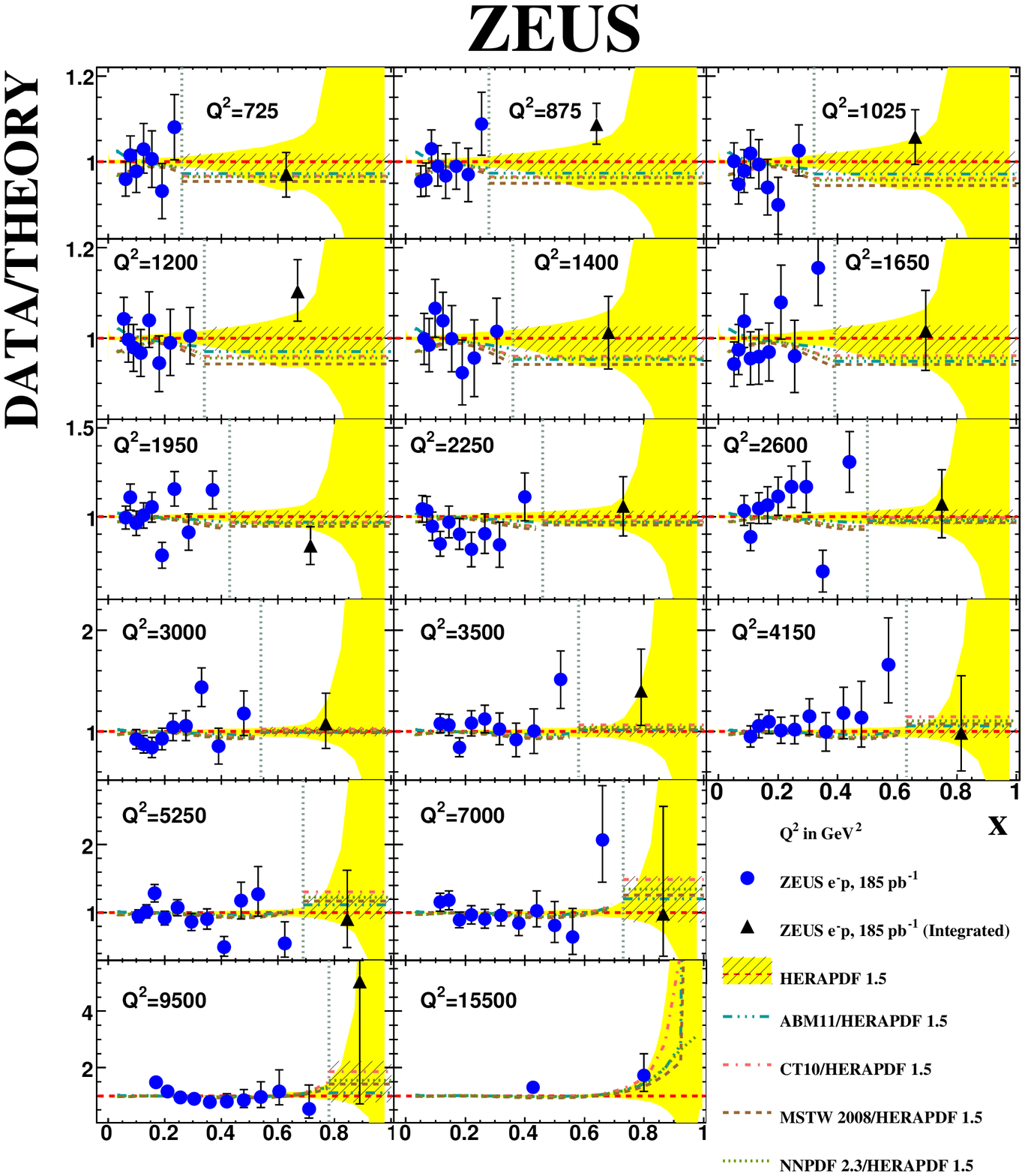} 
\caption{Ratio of the double-differential cross section for NC $e^-p$ 
scattering and  of the double-differential
cross section integrated over $x$ 
to the Standard Model expectation evaluated using the HERAPDF1.5 PDFs as a function of $x$ at different $Q^2$ values as described in the legend.  For HERAPDF1.5, the uncertainty is given as a band. The expectation for the integrated bin is also shown as a hatched box. 
The error bars show the statistical and systematic uncertainties added in quadrature.
The expectations of other commonly used PDF sets normalised to HERAPDF1.5 PDFs are also shown, as listed in the legend. Note that the scale on the $y$ axis changes with $Q^2$.  }
\label{e-rxsc}
\end{center}
\end{figure}

\begin{figure}[h!]
\begin{center}
\vspace{-1.8cm}
\includegraphics[angle = 0.,scale = 0.80]{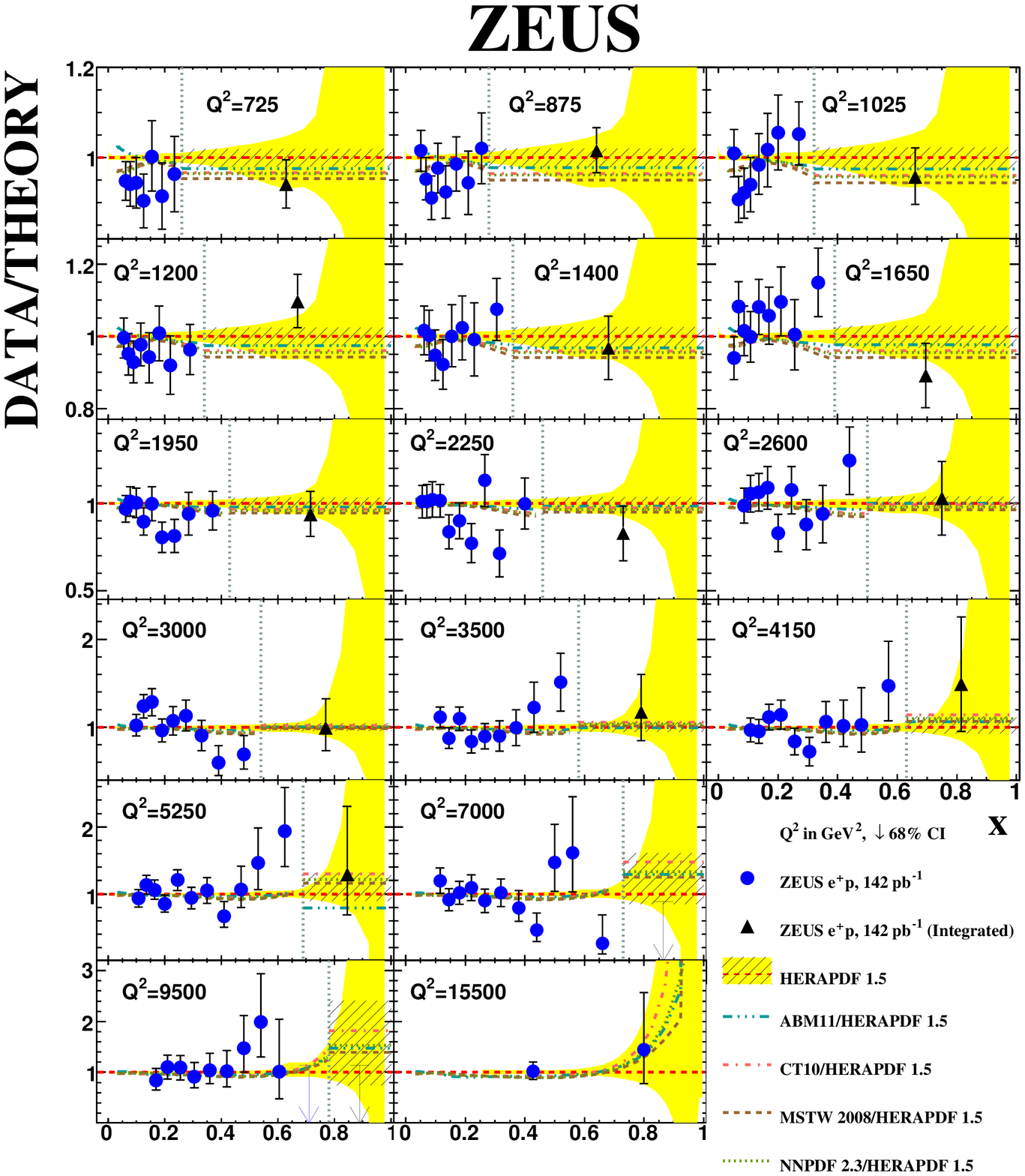} 
\caption{Ratio of the double-differential cross section for NC $e^+p$ 
scattering and  of the double-differential
cross section integrated over $x$ 
to the Standard Model expectation evaluated using the HERPDF1.5 PDFs as a function of $x$ at different $Q^2$ values as described in the legend.For HERAPDF1.5, the uncertainty is given as a band. The expectation for the integrated bin is also shown as a hatched box. 
The error bars show the statistical and systematic uncertainties added in quadrature.
For bins with zero measured events, a $68\%$ probability limit is given.  The expectations of other commonly used PDF sets normalised to HERAPDF1.5 PDFs are also shown, as listed in the legend.  Note that the scale on the $y$ axis changes with $Q^2$.}
\label{e+rxsc}
\end{center}
\end{figure}